\documentclass[11pt]{article}

\usepackage[dvips]{epsfig}
\usepackage{latexsym}
\usepackage{amssymb}
\usepackage{amsmath}
\usepackage{graphicx}
\usepackage{subfigure}
\usepackage{algorithmic}
\usepackage{algorithm}
\usepackage{bm}
\usepackage{enumitem}

\renewcommand{\baselinestretch}{1.15} 

\newcommand{\idc}{\bm{1}}
\newcommand{\p}{\mathbf{P}}
\newcommand{\q}{\mathbf{Q}}

\newcommand{\G}{\mathcal{G}}
\newcommand{\N}{\mathcal{N}}
\newcommand{\E}{\mathcal{E}}
\newcommand{\Ex}{\mathbb{E}}
\newcommand{\pr}{\mathbb{P}}
\newcommand{\rt}{\right}
\newcommand{\lt}{\left}

\newcommand{\Rn}{\mathbb{R}}

\newcommand{\var}{\mathrm{Var}}
\newcommand{\cov}{\mathrm{Cov}}

\def \endthm {~\hfill~{$\Box$}}

\newtheorem{theorem}{Theorem}

\newtheorem{proposition}{Proposition}

\newtheorem{remark}{Remark}

\newtheorem{proof}{Proof}

\begin{document}

\title{Beyond Random Walk and Metropolis-Hastings Samplers:
Why You Should Not Backtrack for Unbiased Graph Sampling}

\author{\vspace{3mm}
Chul-Ho Lee, Xin Xu, and Do Young Eun \\
Department of Electrical and Computer Engineering \\
North Carolina State University, Raleigh, NC 27695\\
Email: \{clee4, xxu5, dyeun\}@ncsu.edu }

\date{April, 2012}
\maketitle

\pagestyle{plain} \setcounter{page}{1} \pagenumbering{arabic}

\renewcommand{\baselinestretch}{1.1} 

\begin{abstract}
Graph sampling via crawling has been actively considered as a
generic and important tool for collecting uniform node samples so
as to consistently estimate and uncover various characteristics of
complex networks. The so-called simple random walk with
re-weighting (SRW-rw) and Metropolis-Hastings (MH) algorithm have been
popular in the literature for such unbiased graph sampling.
However, an unavoidable downside of their core random walks --
slow diffusion over the space, can cause poor estimation accuracy.
In this paper, we propose non-backtracking random walk with
re-weighting (NBRW-rw) and MH algorithm with delayed acceptance (MHDA) which
are \emph{theoretically guaranteed} to achieve, at \emph{almost no
additional cost}, not only unbiased graph sampling but also higher
efficiency (smaller asymptotic variance of the resulting unbiased
estimators) than the SRW-rw and the MH algorithm,
respectively. In particular, a remarkable feature of the MHDA is
its applicability for \emph{any} non-uniform node sampling like
the MH algorithm, but ensuring better sampling efficiency than the
MH algorithm.
We also provide simulation results to confirm our theoretical findings.
\end{abstract}

\vspace{2mm}


\renewcommand{\baselinestretch}{1.2} 

\section{Introduction}

Estimating various nodal and topological properties of complex
networks such as online social networks (OSNs), peer-to-peer (P2P)
networks, and the world wide web (WWW) has recently attracted much
attention from research community because of their ever-increasing
popularity and importance in our daily life. However, the
estimation of network characteristics is a non-trivial task, as
these networks are typically too large to measure, making a
complete picture of the network hard to obtain and even its size
unknown. It is thus infeasible to perform `independence sampling' which
obtains uniform node samples (for unbiased estimation) directly
and independently from such a large, unknown network. Instead,
graph crawling techniques -- graph sampling via crawling, have
been widely used for that purpose. In particular, random
walk-based graph sampling methods (or Markov chain samplers) have
become popular, as they are simple and implementable in a
distributed fashion and also able to provide \emph{unbiased} graph
sampling, unlike the breadth-first-search (BFS) and its variants
leading to unknown bias~\cite{GjokaJSAC11-2,Kurant11}.

In the literature, the most popular random walk-based graph
sampling methods are the so-called simple random walk with
re-weighting (SRW-rw)~\cite{WillingerInfocom09,GjokaJSAC11-2} and
Metropolis-Hastings (MH)
algorithm~\cite{Metropolis53,Hastings70,WillingerToN09,WillingerInfocom09,GjokaJSAC11-2,Hasan09}.
The former launches a simple random walk (SRW) over a graph $\G$, which moves from
a node to one of its neighbors chosen uniformly at random, to
collect random node samples, followed by a re-weighting process in
order to eliminate the bias caused by the non-uniform stationary
distribution of the SRW. The other method is to rely on a
Metropolis-Hastings random walk (MHRW) crawling over $\G$ -- a
random walk achieving a unform distribution constructed by the
famous MH algorithm~\cite{Metropolis53,Hastings70}, to obtain
uniform node samples.

\vspace{2.5mm}
\textbf{\textsf{Motivation and Contributions:}} While the
SRW-rw and MH algorithm ensure unbiased graph
sampling, the core components -- SRW and MHRW, suffer from their
slow diffusion over the space, which can in turn lead to poor
estimation accuracy. In particular, their fully random nature in
selecting the next node, when making a transition, often cause
them to go back to the previous node from where they just came. This produces many duplicate samples for a short to moderate
time span, thereby reducing estimation accuracy.
It is apparently desirable to avoid such backtracking transitions
whenever possible, so as to steer them toward `unvisited' places
(or to obtain new node samples), as long as such a modification does
not affect the unbiased estimation.

However, it is still uncertain how to achieve this \emph{at almost
no additional cost} and whether it really results in better estimation accuracy. We
provide affirmative answers for these questions. Specifically, we
propose non-backtracking random walk with re-weighting (NBRW-rw) and
MH algorithm with delayed acceptance (MHDA), and prove
that each of them guarantees not only \emph{unbiased} graph
sampling but also \emph{higher efficiency} (smaller asymptotic
variance of the estimators) than the SRW-rw and the
MH algorithm, respectively. A notable feature of our MHDA is
its generic purpose: the MHDA is theoretically guaranteed to enhance the
standard MH algorithm for constructing a random walk or a Markov
chain with \emph{any} arbitrarily given stationary distribution
under the constraints of graph structure. Thus, the MHDA is
applied, as `a special case', to construct a random walk crawling
over a graph $\G$ achieving a uniform stationary distribution,
leading to higher efficiency than the MHRW while ensuring the
unbiased estimation. To the best of our knowledge, this is the
first theoretical result to improve, with proven guarantee,
\emph{both} SRW-rw and the MH algorithm for unbiased
graph sampling.

\vspace{2.5mm}
\textbf{\textsf{Related Work:}} Very recently, there have been a
few attempts to improve the estimation accuracy against the SRW-rw (not the MH algorithm) through multiple
dependent random walks~\cite{Ribeiro10}, a random walk on a
weighted graph (with a priori estimate of network
information)~\cite{Kurant11}, and the addition of random jumps (to
anywhere in the graph)~\cite{Avrachenkov10}. The corresponding
Markov chains are time-reversible, whereas the main kernel of our
proposed methods is transforming `any' reversible Markov chain to
its related non-reversible chain which avoids backtracking
transitions and also achieves the same stationary distribution.
Thus, our work is complementary to their approaches.

On the other hand, there is a body of research works across many
disciplines for speeding up a random walk, or Markov chain, on a
graph $\G$ in terms of its mixing time, hitting time, and/or cover
time. The fastest mixing (reversible) Markov chain on a graph is
obtained in~\cite{Boyd04-SIAM} with complete knowledge of entire
graph. \cite{Chen99,Neal00} showed that certain `lifted'
(non-reversible) Markov chains converge to their stationary
distributions faster than their related reversible chain, and
\cite{Jung06,Dai07} subsequently applied this idea to design a
fast and efficient average consensus algorithm. It is, however,
still unknown how to construct such a `lifted' Markov chain in a
distributed or decentralized manner for a general graph.

\cite{Alon07,Berenbrink10,Ikeda09} recently undertook speeding up
a SRW based only on local information, but did not provide any
direct implication to the unbiased graph sampling. As the MH
algorithm is the most popular method of Markov Chain Monte Carlo
(MCMC) simulations or samplers, it has been an active research
topic to improve the MH algorithm in terms of the sampler
performance (asymptotic variance) in the MCMC literature
(e.g.,~\cite{Mira01,Green01,Wu05}). However, most works toward
more efficient MCMC samplers (including
\cite{Mira01,Green01,Wu05}) do not take into account
graph-topological constraints in that transition from node $i$ to
$j \! \ne \! i$ is allowed only when they are neighbors of each
other, and thus cannot be directly applicable to unbiased graph
sampling.

\vspace{2.5mm}
\textbf{\textsf{Organization:}} The rest of the paper is
organized as follows. We first provide an in-depth overview on
generic Markov chain samplers for unbiased graph sampling in
Section 2, and then briefly review the SRW-rw and
MH algorithm in Section 3. In Section 4, we present a general
recipe for the transformation of a time-reversible Markov to its
related non-reversible Markov chain, which forms a common building
block for our proposed NBRW-rw and MHDA. We then
explain the details of the NBRW-rw and MHDA, and
provide relevant analysis. In Section 5, we provide simulation
results obtained based on real graphs to support our theoretical
findings. We finally conclude in Section 6.

\section{Background on Markov Chain Samplers}

\subsection{Unbiased Graph Sampling}\label{se:unbiased}

Consider a \emph{connected, undirected} graph $\G = (\N,\E)$ with
a set of nodes (vertices) $\N =\{1,2,\ldots, n\}$ and a set of
edges $\E$. We assume that $ 3 \leq |\N| \!=\! n  < \infty$. We
also assume that the graph $\G$ has no self-loops and no
multi-edges. Let $N(i) \triangleq \{j \in \N : (i,j) \in \E\}$ be
the set of neighbors of node $i \in \N$, and $d(i) \triangleq
|N(i)|$ be the degree of node $i$.

\emph{Unbiased} graph (or node) sampling, via crawling, is to
consistently estimate nodal or topological properties of a target
graph $\G$\footnote{A target graph for sampling may be
time-varying due to node join/leave, which is beyond the scope of
this paper.} (e.g., an overlay network or an OSN) based upon
\emph{uniform} node samples obtained by a random walk (or possibly
multiple random walks) crawling over the graph $\G$. The goal here
is to unbiasedly estimate a proportion of the nodes with a
specific characteristic. Thus, the unbiased, uniform graph
sampling is, in principle, developing a random walk-based
``estimator" or a Markov chain sampler for the expectation of any
given, desired function $f$ with respect to a uniform
distribution, i.e.,
\begin{equation}
\Ex_{\bm{u}}(f) \triangleq  \sum_{i \in \N} f(i)\frac{1}{n},
\label{unbiased}
\end{equation}
where $\bm{u} \triangleq [u(1), u(2), \ldots, u(n)] = [1/n, 1/n,
\ldots, 1/n]$. Note that a nodal (or topological) characteristic
of interest can be specified by properly choosing a function $f$.
For example, for a target graph $\G$, if one is interested in
estimating its degree distribution (say, $\pr\{D_\G = d\}, d =
1,2,\ldots,n\!-\!1$), then choose a function $f$ such that $f(i) =
\idc_{\{d(i) = d\}}$ for $i \in \N$, i.e., $f(i) = 1$ if $d(i) =
d$, and $f(i) = 0$ otherwise.

We below review a basic Markov chain theory which serves as the
mathematical foundation for unbiased graph sampling via a
(Markovian) random walk crawling over a graph $\G$. Define a
random walk or a finite discrete-time Markov chain $\{X_t \!\in\!
\N, t \!=\! 0,1,\ldots\}$ on the nodes of the graph $\G$ with its
transition matrix $\p \triangleq \{P(i,j)\}_{i,j \in \N}$ in which
\begin{equation*}
P(i,j) = \pr\{X_{t+1} = j ~|~ X_t = i\},~~~i,j \in \N,
\end{equation*}
and $\sum_j P(i,j) = 1$ for all $i$. Each edge $(i,j) \in \E$ is
associated with a transition probability $P(i,j) \geq 0$ with
which the chain (or random walk) makes a transition from node $i$
to node $j$. We allow the chain to include self-transitions, i.e., $P(i,i) > 0$ for some $i$, although $\G$ has no
self-loops. Clearly, $P(i,j) = 0$ for all $(i,j) \not\in \E$ ($i
\ne j$). We then assume that the Markov chain $\{X_t\}$ is
irreducible, i.e., every node in $\N$ is reachable in finite time
with positive probability, such that the chain has a unique
stationary distribution $\bm{\pi} \triangleq [\pi(1), \pi(2),
\ldots, \pi(n)]$.

For any function $f: \N \!\to\! \Rn$, define an estimator
\begin{equation}
\hat{\mu}_t(f) \triangleq
\frac{1}{t}\sum^{t}_{s=1}f(X_s)\label{estimator}
\end{equation}
for the expectation of the function $f$ with respect to $\bm{\pi}$
which is given by
\begin{equation} \Ex_{\bm{\pi}}(f) \triangleq \sum_{i\in
\N}f(i)\pi(i).\label{limit}
\end{equation}
Then, the following Strong Law of Large Numbers (SLLN) (a.k.a.,
ergodic theorem) has been a fundamental basis for most of the
random walk-based graph sampling methods in the
literature~\cite{WillingerToN09,WillingerInfocom09,GjokaJSAC11-2,Hasan09,Ribeiro10,Avrachenkov10,Kurant11},
and more generally, MCMC
samplers~\cite{Peskun73,Mira01,Wu05,Jones04,Rosenthal04,Neal04}.

\begin{theorem}\cite{Jones04,Rosenthal04}\label{ergodic}
Suppose that $\{X_t\}$ is a finite, irreducible Markov chain with
its stationary distribution $\bm{\pi}$. Then, for any initial
distribution $\pr\{X_0 \!=\! i\}, i \!\in\! \N$, as $t \!\to\!
\infty$,
\begin{equation*}
\hat{\mu}_t(f) ~\to~ \Ex_{\bm{\pi}}(f)~~\text{almost surely
(a.s.)}
\end{equation*}
for any function $f$ with $\Ex_{\bm{\pi}}(|f|) < \infty$.
\endthm
\end{theorem}

The SLLN ensures that the estimator $\hat{\mu}_t(f)$ based on any
finite, irreducible Markov chain with the same $\bm{\pi}$ can
serve as a valid and unbiased approximation of
$\Ex_{\bm{\pi}}(f)$. In particular, the two popular random
walk-based graph sampling methods (or two different Markov chain
samplers for unbiased graph sampling) in the networking literature
-- SRW-rw~\cite{WillingerInfocom09,GjokaJSAC11-2} and
MH algorithm~\cite{WillingerToN09,WillingerInfocom09,GjokaJSAC11-2,Hasan09}
are built upon the SLLN to asymptotically guarantee the
\emph{unbiasedness} of their estimators for $\Ex_{\bm{u}}(f)$. We
will review in detail these two graph sampling methods in
Section~\ref{se:sampling}.

\subsection{Central Limit Theorem and Asymptotic Variance}

For a given graph $\G$, there are potentially many (finite)
irreducible Markov chains (or different random walks) preserving
the same stationary distribution $\bm{\pi}$, all of which can be
used to obtain asymptotically unbiased estimates of
$\Ex_{\bm{\pi}}(f)$, and also of $\Ex_{\bm{u}}(f)$, together with
proper re-weighting if $\bm{\pi} \ne \bm{u}$. One important
question would then be how to compare these `competing' Markov
chains, or rather, which one is `better' or more efficient than
the others as a Markov chain sampler for unbiased graph sampling.

Mixing time can perhaps be a criterion to compare several
irreducible, aperiodic Markov chains, all with the same stationary
distribution. The mixing time captures the notion of the speed of
convergence to the stationary distribution, and is typically
defined via the total variation distance: for an irreducible,
aperiodic Markov chain $\{X_t\}$ with its transition matrix $\p$
and stationary distribution $\bm{\pi}$, the mixing time can be
written as
\begin{equation*}
t_{mix}(\varepsilon) = \min\{t \geq 1: \max_{i\in \N}
\|P^t(i,\cdot) - \bm{\pi} \|_{TV} \leq \varepsilon \},
\end{equation*}
where the total variation distance is defined by $\|P^t(i, \cdot)
\!-\! \bm{\pi}\|_{TV} \!\triangleq\! \max_{A \subseteq \N}|P^t(i,A) \!-\!
\pi_A|$. Here, $P^t(i,A)$ denotes the $t$-step transition
probability from node (state) $i$ to subset $A \!\subseteq\! \N$, and
$\pi_A \!=\! \sum_{j \in A}\pi(j)$. The mixing time has been actively
studied in the literature, especially for irreducible, aperiodic,
\emph{time-reversible}\footnote{If the Markov chain $\{X_t\}$
satisfies the reversibility condition (or detailed balance
equation), i.e., $\pi(i) P(i,j) \!=\! \pi(j) P(j,i)$ for all
$i,j$, then the chain is called time-reversible.} Markov chains
(e.g.,~\cite{Boyd04-SIAM,Levin09}). In particular, it is now well
known that the mixing time of such a Markov chain (or the
asymptotic rate of convergence to its stationary distribution) is
mainly governed by the second largest eigenvalue modulus (SLEM) --
the second largest eigenvalue in absolute value, of its transition
matrix, and smaller SLEM leads to smaller (faster) mixing
time~\cite{Boyd04-SIAM,Levin09}.

If the speed of convergence to the stationary distribution is a
primary concern, then the mixing time is surely the right metric
to compare different Markov chains with the same stationary
distribution. However, this is not the case for the unbiased graph
sampling. Random walk-based graph sampling methods typically adopt
an initial burn-in period over which (initial) sampled values are
discarded to get rid of the dependence on the initial position
of a random walk~\cite{GjokaJSAC11-2}. After such a burn-in
period, the Markov chain (or random walk) will be close to its
stationary regime (well mixed), but many samples are still yet to
be obtained from this point onward.
Therefore, the primary concern should be, instead, the
\emph{efficiency} of the estimator $\hat{\mu}_t(f)$ in deciding
how many random samples are required to achieve a certain accuracy
of $\hat{\mu}_t(f)$ in regard to $\Ex_{\bm{\pi}}(f)$ (and
eventually to $\Ex_{\bm{u}}(f)$ after proper re-weighting if
necessary).

To that end, we define by $\sigma^2(f)$ the asymptotic variance of
the estimator $\hat{\mu}_t(f)$ based on an irreducible Markov
chain $\{X_t\}$ with its stationary distribution $\bm{\pi}$, which
is given by
\begin{equation}
\sigma^2(f) \triangleq \lim_{t \to \infty}
t\cdot\var\lt(\hat{\mu}_t(f)\rt) = \lim_{t \to \infty}\frac{1}{t} ~ \Ex\lt\{\lt[\sum^t_{s=1}(f(X_s)
- \Ex_{\bm{\pi}}(f))\rt]^2 \rt\}\label{asymptotic}
\end{equation}
for any function $f$ with $\Ex_{\bm{\pi}}(f^2) < \infty$,
where the initial position (state) $X_0$ is drawn from the
stationary distribution $\bm{\pi}$, i.e., $X_0 \sim \bm{\pi}$.
Note that the asymptotic variance $\sigma^2(f)$ is, in fact,
\emph{independent} of the distribution of the initial state
$X_0$~\cite{Peskun73,Rosenthal04}. We below explain how effective the
asymptotic variance $\sigma^2(f)$ can be, through its connection
to the Central Limit Theorem (CLT), in measuring the performance
of the estimator $\hat{\mu}_t(f)$.

Suppose first that the random samples $X_1,X_2,\ldots,X_t$ are
\textit{i.i.d.} and drawn directly from $\bm{\pi}$. Then, the
standard Central Limit Theorem (CLT) says that
\begin{equation*}
\sqrt{t}\cdot[\hat{\mu}_t(f) - \Ex_{\bm{\pi}}(f)]
~\stackrel{d}{\Longrightarrow}~ \mathrm{N}(0,\sigma^2(f)),
~\mbox{as} ~t \to \infty,
\end{equation*}
where $\stackrel{d}{\Longrightarrow}$ denotes convergence in
distribution and $\mathrm{N}(0,\sigma^2(f))$ is a Gaussian random
variable with zero mean and variance $\sigma^2(f) \!=\!
\var(f(X_1))$. That is, the distribution of $\hat{\mu}_t(f)$ is
asymptotically normal. For sufficiently large $t$, we also have
\begin{equation*}
\pr\lt\{\frac{\hat{\mu}_t(f) -
\Ex_{\bm{\pi}}(f)}{\sigma(f)/\sqrt{t}}
> x\rt\} \thickapprox
\frac{1}{\sqrt{2\pi}}\int^\infty_xe^{-\frac{y^2}{2}}dy,
\end{equation*}
which allows us to identify an approximate \emph{confidence
interval} for $\Ex_{\bm{\pi}}(f)$. For instance, for sufficiently
large $t$, we can be $95\%$ confident that $\Ex_{\bm{\pi}}(f)$ is
approximately between $\hat{\mu}_t(f) - 2(\sigma(f)/\sqrt{t})$ and
$\hat{\mu}_t(f) + 2(\sigma(f)/\sqrt{t})$. This clearly
demonstrates the importance of the asymptotic variance
$\sigma^2(f)$ in conjunction with the CLT in assessing the
accuracy of the estimator $\hat{\mu}_t(f)$.

Not only for the above case with \textit{i.i.d.} samples, the CLT
holds also for Markov chains, as given below.
\begin{theorem}\cite{Jones04,Rosenthal04}\label{clt}
For a finite, irreducible Markov chain $\{X_t\}$ with its
stationary distribution $\bm{\pi}$,
\begin{equation*}
\sqrt{t}\cdot[\hat{\mu}_t(f) - \Ex_{\bm{\pi}}(f)]
~\stackrel{d}{\Longrightarrow}~
\mathrm{N}(0,\sigma^2(f)),~\mbox{as} ~t \to \infty,
\end{equation*}
for any function $f$ with $\Ex_{\bm{\pi}}(f^2) < \infty$
regardless of any initial distribution, and $\sigma^2(f)$ is given
by (\ref{asymptotic}).
\endthm
\end{theorem}

\vspace{2mm}
\noindent Note that Theorems~\ref{ergodic}--\ref{clt} (SLLN and
CLT) do not require any assumption of
aperiodicity~\cite{Rosenthal04}. However, for simplicity, we do
not consider periodic Markov chains in our analysis throughout the
paper. In addition, we focus on \emph{bounded} functions $f$ (and
thus $\Ex_{\bm{\pi}}(f^2) < \infty$), which is typical in
graph sampling applications.

As shown above for \textit{i.i.d.} samples, the CLT allows one to
evaluate the asymptotic variance $\sigma^2(f)$ in order to decide
approximately how many (correlated) samples are required to
achieve a certain accuracy of the estimator $\hat{\mu}_t(f)$.
Hence, the asymptotic variance $\sigma^2(f)$ has been an important
criterion to rank the \emph{efficiency} among competing Markov
chains with the same $\bm{\pi}$ for the MCMC
samplers~\cite{Peskun73,Mira01,Wu05,Jones04,Rosenthal04,Neal04},
although quantifying $\sigma^2(f)$ may not be easy. In particular,
by noting that the asymptotic variance is independent of any
initial distribution for which the CLT holds, the efficiency
ordering over competing Markov chains with the same $\bm{\pi}$
(the smaller the asymptotic variance, the better the estimator
performance) is still in effect even when the competing Markov
chains are already in their stationary regimes (already `mixed').
Observe that from $X_0 \sim \bm{\pi}$, $X_t \sim \bm{\pi}$ for all
$t$ (the chain $\{X_t\}$ is in the stationary regime), and thus (\ref{asymptotic}) becomes
\begin{equation}
\sigma^2(f) = \var(f(X_0)) + 2\sum^\infty_{k =
1}\cov(f(X_0),f(X_k)),\label{asymptotic2}
\end{equation}
where $\cov(f(X_0),f(X_k)) = \Ex\{f(X_0)f(X_k)\} -
\Ex^2_{\bm{\pi}}(f)$ denotes the covariance between $f(X_0)$ and
$f(X_k)$. That is, even if the competing Markov chains are already in
their stationary regimes, the correlation structure over random
samples given by each of these Markov chains can vary and
significantly affect their asymptotic variances. Observe that
reducing the temporal correlation over random samples can lead to
smaller asymptotic variances. This intuition can be leveraged to
improve the existing Markov chain samplers for unbiased graph sampling.

Motivated by the effectiveness of the asymptotic variance with its
connection to the CLT, in this paper, we consider the asymptotic
variance as a primary performance metric, and develop two random
walk-based graph sampling methods, each of which guarantees the
unbiased graph sampling with smaller asymptotic variance than its
corresponding counterpart in the current networking literature.
Before going into details, we next briefly review the existing two
random walk-based graph sampling methods.

\section{Random Walk-based Graph Sampling}\label{se:sampling}

\subsection{Simple Random Walk with Re-weighting}\label{se:srw}

We first review the SRW-rw, a.k.a., respondent-driven
sampling~\cite{Salganik04}, which has been recently used
in~\cite{WillingerInfocom09,GjokaJSAC11-2} for unbiased
graph sampling. This method operates based upon a sequence of
(correlated) random samples obtained by a SRW, together with a
proper re-weighting process to ensure the unbiased sampling. It is
essentially a special case of the importance sampling (a Monte
Carlo method) applied for random samples generated by a Markov
chain~\cite{Bassetti06,Malefaki08,Salganik09}. While there are
similar variants of such method (e.g.,~\cite{Massoulie06}), the
main idea behind them is still to correct the sampling bias caused
by the stationary distribution of the SRW.

Consider a SRW on $\G$ that moves from a node to one of its
neighbors chosen uniformly at random (u.a.r.). Specifically, let
$\{X_t\}$ be the Markov chain representing the sequence of visited
nodes by the SRW, with its transition matrix $\p = \{P(i,j)\}_{i,j
\in \N}$ given by
\begin{equation}\label{srw}
P(i,j) =
\begin{cases}
\frac{1}{d(i)} &~\text{if}~ (i,j) \in \E, \\
0 &~\text{otherwise}.
\end{cases}
\end{equation}
It is well known that $\p$ is irreducible, and reversible with
respect to a unique stationary distribution $\bm{\pi}$ for which
$\pi(i) = d(i)/(2|\E|), i \in \N$~\cite{Aldous}.

Suppose that there are $t$ random samples $\{X_s\}^t_{s=1}$ from the
SRW. Then, for a function of interest $f$, choose a weight
function $w: \N \!\to\! \Rn$ such that
\begin{equation*}
w(i) = \frac{u(i)}{\pi(i)} = \frac{2|\E|}{n}\frac{1}{d(i)},~~ i
\in \N.
\end{equation*}
Observe that from the SLLN in Theorem~\ref{ergodic}, as $t \to
\infty$,
\begin{equation*}
\hat{\mu}_t(wf) = \frac{1}{t}\sum^t_{s=1}w(X_s)f(X_s) ~\to~
\Ex_{\bm{\pi}}(wf) = \Ex_{\bm{u}}(f) ~~\text{a.s.}
\end{equation*}
and thus the estimator $\hat{\mu}_t(w f)$ is unbiased for
$\Ex_{\bm{u}}(f)$. However, this estimator itself is not
practical, since $n$ and $|\E|$ are typically unknown a priori. Instead,
another estimator $\hat{\mu}_t(w f)/\hat{\mu}_t(w)$ is often used
as an unbiased estimator for $\Ex_{\bm{u}}(f)$. Indeed,
Theorem~\ref{ergodic} asserts that $\hat{\mu}_t(w f)$ and
$\hat{\mu}_t(w)$ converge to $\Ex_{\bm{u}}(f)$ and 1 almost
surely, as $t \to \infty$, respectively. This yields
\begin{equation*}
\frac{\hat{\mu}_t(wf)}{\hat{\mu}_t(w)} =
\frac{\sum^t_{s=1}w(X_s)f(X_s)}{\sum^t_{s=1}w(X_s)} ~\to~
\Ex_{\bm{u}}(f) ~~\text{a.s.}
\end{equation*}
Hence, the estimator $\hat{\mu}_t(w f)/\hat{\mu}_t(w)$ can be made
in such a way that we need to know $w(i)$ only up to a
multiplicative constant. That is, if we set $w(i) = 1/d(i)$, $i
\!\in\! \N$, then the estimator $\hat{\mu}_t(w f)/\hat{\mu}_t(w)$
remains intact, and is more practical as an unbiased estimator for
$\Ex_{\bm{u}}(f)$. Throughout this paper, we refer to the
estimator $\hat{\mu}_t(w f)/\hat{\mu}_t(w)$ with $w(i) = 1/d(i)$
($i \!\in\! \N$) as the unbiased estimator for $\Ex_{\bm{u}}(f)$
in the SRW-rw~\cite{WillingerInfocom09,GjokaJSAC11-2}.

As an example, for a target graph $\G$, choose a function $f$ such
that $f(i) = \idc_{\{d(i) = d\}}$ for $i \in \N$ in order to
estimate the degree distribution $\pr\{D_\G = d\}$. Then, for any
given $d$,
\begin{equation*}
\frac{\hat{\mu}_t(w f)}{\hat{\mu}_t(w)} =
\frac{\sum^t_{s=1}\idc_{\{d(X_s) =
d\}}/d(X_s)}{\sum^t_{s=1}1/d(X_s)} ~\longrightarrow~ \Ex_{\bm{u}}(f) = \sum_{i \in \N}\idc_{\{d(i) = d\}}\frac{1}{n}~~\text{a.s.},
\end{equation*}
implying that the estimator $\hat{\mu}_t(w f)/\hat{\mu}_t(w)$
yields a valid unbiased estimate of $\pr\{D_\G = d\}$.

\subsection{Metropolis-Hastings Algorithm}

The MH algorithm~\cite{Metropolis53,Hastings70} was developed to
construct a transition matrix $\p$ of a time-reversible Markov
chain $\{X_t\}$ with a given, desired stationary distribution
$\bm{\pi}$. Here, we only discuss the MH algorithm under the
topological constraints of a graph $\G$ in that transition from
node $i$ to $j \!\ne\! i$ is allowed only when they are neighbors
of each other. The MH algorithm is defined as follows. At the
current state $i$ of $X_t$, the next state $X_{t+1}$ is proposed
with a \emph{proposal} probability $Q(i,j)$, which is a state
transition probability of an arbitrary irreducible Markov chain on
the state space $\N$, where $Q(i,j)
> 0$ if and only if $Q(j,i)
> 0$, and $Q(i,j) = 0$ for all $(i,j) \not\in \E$ ($i \!\ne\! j$).
Let $\q \triangleq \{Q(i,j)\}_{i,j \in \N}$ be a proposal
(transition) matrix. The proposed state transition to $X_{t+1}
\!=\! j$ is accepted with an \emph{acceptance} probability
\begin{equation}
A(i,j) = \min\lt\{1, \frac{\pi(j) Q(j,i)}{\pi(i)
Q(i,j)}\rt\},\label{mh-accept}
\end{equation}
and rejected with probability $1 \!-\! A(i,j)$ in which case
$X_{t+1} \!=\! i$. Thus, the transition probability $P(i,j)$
becomes, for $i \!\ne\! j$,
\begin{equation}\label{mh}
P(i,j) = Q(i,j)A(i,j) = \min\lt\{Q(i,j),
Q(j,i)\frac{\pi(j)}{\pi(i)}\rt\},
\end{equation}
with $P(i,i) \!=\! 1 \!-\! \sum_{j \ne i}P(i,j)$, which ensures
that $\p$ is reversible with respect to $\bm{\pi}$. Note that the
uniqueness of $\bm{\pi}$ is granted due to the irreducibility of
$\q$ (so is $\p$) and the finite state space.

The MH algorithm, in addition to its popular applications for MCMC
simulation, has been also widely used as a means for unbiased
graph
sampling~\cite{WillingerToN09,WillingerInfocom09,GjokaJSAC11-2,Hasan09}.
Specifically, the MH algorithm has been applied to construct a MHRW
on $\G$ achieving a
uniform stationary distribution, i.e., $\bm{\pi} \!=\! \bm{u}$.
This is done with transition probabilities of a SRW as the
proposal probabilities, i.e., $Q(i,j) = 1/d(i)$ if $(i,j) \in \E$
and $Q(i,j) = 0$, otherwise. The resulting transition probability
of the MHRW on $\G$ becomes
\begin{equation}\label{mh-srw}
P(i,j) =
\begin{cases}
\min\lt\{\frac{1}{d(i)},\frac{1}{d(j)}\rt\}
& \text{if}~ (i,j) \in \E,\\
0 & \text{if}~ (i,j) \not\in \E, i \ne j,
\end{cases}
\end{equation}
and $P(i,i) \!=\! 1 \!-\! \sum_{j \ne i}P(i,j)$. Thus, $\p$ is
reversible with respect to $\bm{\pi} = \bm{u}$, implying that for
any function $f$, the estimator $\hat{\mu}_t(f)$ based upon random
samples by the MHRW is unbiased for $\Ex_{\bm{u}}(f)$. This
version of MH algorithm is summarized in Algorithm~\ref{alg1},
where $X_t \in \N$ denotes the location of the MHRW at time $t$,
and $d(X_t)$ denotes the degree of node $X_t$. Here, $X_0$ can be
arbitrarily chosen.

\begin{algorithm}[h!]
\caption{MH algorithm for MHRW (at time $t$)}\label{alg1}
\begin{algorithmic}[1]
\STATE Choose node $j$ u.a.r. from neighbors of $X_t$, i.e.,
$N(X_t)$ \STATE Generate $p \sim U(0,1)$ \IF{$p \leq
\min\lt\{1,\frac{d(X_t)}{d(j)}\rt\}$} \STATE $X_{t+1} \leftarrow
j$ \ELSE \STATE $X_{t+1} \leftarrow X_t$ \ENDIF
\end{algorithmic}
\end{algorithm}

\begin{remark}\label{remark1}
It is worth noting that the MH algorithm (Algorithm~\ref{alg1})
does not need to know the self-transition probabilities $P(i,i)$
explicitly, nor does it require all the neighbors' degree
information of the current node $X_t$ at each time $t$. Instead,
only the degree information of the randomly chosen neighbor $j$ is
enough for making decision whether or not to move to $j$.
\end{remark}

Recall that the above unbiased estimators are based on $t$ random,
consecutive samples obtained under SRW or MHRW, respectively.
Observe that the SRW, currently at node $i$ at time $s$ can
`backtrack' to the previously visited node with probability
$1/d(i)$, i.e., $X_{s+1} = X_{s-1}$, trapping the SRW
temporarily in a local region. The situation can be
worse for the regions in which nodes have small degrees (so higher
chance of backtracking). Similarly, the MHRW at node $i$ can also
backtrack to the previously visited node after staying at node $i$
for some random time. This slow `diffusion' of SRW/MHRW over the
space can, in turn, lead to highly duplicated random samples for a
short to moderate time duration, thereby increasing the variance
of the unbiased estimators. Recall that the asymptotic variance in
(\ref{asymptotic2}) involves covariance terms
$\cov(f(X_0),f(X_k))$. Thus, it would be beneficial for both SRW
and MHRW (or precisely, their variants) to avoid backtracking to
the previously visited node up to the extent possible in order to
reduce the temporal correlation over random consecutive samples,
while maintaining the same stationary distribution so that the
aforementioned mathematical framework for the unbiased estimation
remains intact. Thus motivated, for the rest of this paper, we
investigate how to achieve this at \emph{almost no additional
cost}, and rigorously prove that our proposed sampling methods
give smaller (no worse) asymptotic variance than the SRW (with
re-weighting) and MHRW-based ones, respectively.

\section{Avoid Backtracking To Previously Visited Node}\label{se:nonreversible}

In this section, we propose two random walk-based graph sampling
methods -- \textbf{(i) non-backtracking random walk with
re-weighting} and \textbf{(ii) MH algorithm with delayed
acceptance}, each of which \emph{theoretically guarantees}
unbiased graph sampling with smaller asymptotic variance than the
SRW-rw and the (original) MH algorithm,
respectively. In particular, our proposed sampling methods require
almost no additional cost, or more precisely, just remembering
where the underlying random walk came from, when compared to the
conventional methods. The reasoning behind the improvement of
asymptotic variance is to modify each of SRW and MHRW, when making
a transition from the current node to one of its neighbors, to
reduce bias toward the previous state (one of the neighbors of the
current node), while maintaining the same stationary distribution. Note that such
directional bias breaks the time-reversibility of the SRW and
MHRW. Thus, a common building block for our proposed sampling
methods will be, for a given reversible Markov chain with its
stationary distribution $\bm{\pi}$, to construct a
\emph{non-reversible} Markov chain preserving the same $\bm{\pi}$
while avoiding (to the extent possible) transitions that backtrack
to the state from which the chain just came. Our challenge here is
to construct such a non-reversible chain with only one-state memory and theoretical guarantee for higher efficiency
(smaller asymptotic variance). In what follows, we first explain a
basic setup for this transformation and several relevant issues,
and then present the details of our proposed methods.

\subsection{From Reversible To Non-reversible
Chains}\label{se:recipe}

Consider a generic random walk on $\G$, or a finite, irreducible,
\emph{time-reversible} Markov chain $\{X_t \!\in\! \N, t \!=\!
0,1,\ldots\}$, with its transition matrix $\p = \{P(i,j)\}_{i,j
\in \N}$ and stationary distribution $\bm{\pi} = [\pi(i), i \in
\N]$. Our goal here is to construct its related new random walk or
a finite, irreducible, \emph{non-reversible} Markov chain with the
same $\bm{\pi}$ which avoids backtracking to the previously
visited node, which in turn produces a smaller asymptotic variance
than the original reversible chain. An important requirement is
that this transformation should be done at no additional cost and
in a distributed manner. It is worth noting that there have been
other works~\cite{Chen99,Neal00} showing that certain
non-reversible Markov chains or lifted Markov chains mix
substantially faster than their related reversible chains. While
this concept has been also applied to design a fast and efficient
average consensus algorithm~\cite{Jung06,Dai07}, it is still
unknown how to construct such a non-reversible chain or lifted
Markov chain in a fully distributed or decentralized fashion, not
to mention how to do so for any arbitrarily given target
stationary distribution  $\bm{\pi}$.

\vspace{3mm} \textbf{\textsf{A general recipe for constructing a
non-reversible Markov chain in an augmented state space:}} Let
$X'_t \in \N$, $t \!=\! 0,1,2\ldots$, be the location of a new
random walk at time $t$. At the current node $X'_t$, the next node
$X'_{t+1}$ is decided based upon not only the current node $X'_t$
but also the previous node $X'_{t-1}$ so as to avoid backtracking.
Due to the dependency (memory) to the previous node, $\{X'_t\}_{t
\geq 0}$ itself cannot be a Markov chain on the state space $\N$,
regardless of the choice of transition matrix. This walk, however,
can still be made Markovian on an augmented state space instead,
defined by
\begin{equation}
\Omega \triangleq \{(i,j) ~:~ i, j \in \N~\text{s.t.}~P(i,j) > 0\}
~\subseteq~ \N \!\times\! \N \label{space}
\end{equation}
with $|\Omega| \!<\! \infty$, and $Z'_t \!\triangleq\! (X'_{t-1},
X'_t) \!\in\! \Omega$ for $t \!\geq\! 1$.
For notational simplicity, let $e_{ij}$ denote state
$(i,j) \!\in\! \Omega$. Note that $e_{ij} \ne e_{ji}$. It is also
possible that $e_{ii} \in \Omega$ for some $i$. A similar interpretation of a weighted random walk (or a reversible Markov chain) on the augmented state space can be also found in~\cite[Ch.3]{Aldous}, although its purpose is not for the construction of a related non-reversible chain.

Let $\bm{\p'} \triangleq \{P'(e_{ij}, e_{lk}) \}_{e_{ij},e_{lk}
\in \Omega}$ be the transition matrix of an irreducible Markov
chain $\{Z'_t \in \Omega, t \!=\! 1,2,\ldots\}$ on the state space
$\Omega$. Here, by definition, $P'(e_{ij}, e_{lk}) \!=\! 0$ for
all $j \!\ne\! l$. If the unique stationary distribution
$\bm{\pi'} \triangleq [\pi'(e_{ij}), e_{ij} \in \Omega]$ of the
chain $\{Z'_t\}$ is given by
\begin{equation}
\pi'(e_{ij}) = \pi(i)P(i,j),~~~~~ e_{ij} \in
\Omega,\label{stationary}
\end{equation}
implying that $\pi'(e_{ij}) \!=\! \pi'(e_{ji})$ from the
reversibility of the original chain $\{X_t\}$, then the
probability of the new random walk $\{X'_t\}$ being at node $j$ in
the \emph{steady-state} is the same as $\pi(j)$ for all $j$ (the
stationary distribution of the original reversible chain
$\{X_t\}$). To see this, note that
\begin{equation}
\sum_{i \in \N : e_{ij} \in \Omega} \pi'(e_{ij}) = \sum_{i \in \N}
\pi(i)P(i,j) = \pi(j), ~\forall j \in \N, \label{steady-state}
\end{equation}
where the first equality follows from $P(u,v) \!=\! 0$, $\forall (u,v)
\!\not\in\! \Omega$. In particular, for any original function of
interest $f: \N \!\to\! \Rn$, choose another function $g: \Omega
\!\to\! \Rn$ such that $g(e_{ij}) = f(j)$, and observe
\begin{equation*}
\Ex_{\bm{\pi'}}(g) = \sum_{e_{ij} \in \Omega}g(e_{ij})\pi'(e_{ij}) = \sum_{j \in \N}\sum_{i \in \N} f(j) \pi(i)P(i,j)=  \sum_{j \in \N}f(j) \pi(j) = \Ex_{\bm{\pi}}(f).
\end{equation*}
Then, the SLLN in Theorem~\ref{ergodic} gives
\begin{equation}
\frac{1}{t}\sum^t_{s=1}g(Z'_s) = \frac{1}{t}\sum^t_{s=1}f(X'_s)
~\longrightarrow~ \Ex_{\bm{\pi'}}(g) = \Ex_{\bm{\pi}}(f)
~~\text{a.s.},\label{ergodic-nonrev}
\end{equation}
i.e., $\sum^t_{s=1}g(Z'_s)/t$ is a valid unbiased estimator for
$\Ex_{\bm{\pi}}(f)$. We thus define, for any given function $f: \N
\!\to\! \Rn$,
\begin{equation}
\hat{\mu}'_t(f) \triangleq
\frac{1}{t}\sum^t_{s=1}f(X'_s)\label{est-nonrev}
\end{equation}
to be clearly distinguished from $\hat{\mu}_t(f)$ in
(\ref{unbiased}) defined based on the original chain $\{X_t\}$,
while $\hat{\mu}'_t(f)$ and $\hat{\mu}_t(f)$ are both unbiased
estimators for $\Ex_{\bm{\pi}}(f)$. In addition, the CLT in
Theorem~\ref{clt} implies
\begin{equation}
\sqrt{t}\cdot\lt[\frac{1}{t}\sum^t_{s=1}g(Z'_s) -
\Ex_{\bm{\pi'}}(g)\rt] ~=~ \sqrt{t}\cdot[\hat{\mu}'_t(f) -
\Ex_{\bm{\pi}}(f)] ~\stackrel{d}{\Longrightarrow}~ \mathrm{N}(0,{\sigma'}^2(f)),
\label{clt-nonrev}
\end{equation}
where ${\sigma'}^2(f)$ denotes the asymptotic variance of
$\hat{\mu}'_t(f)$ (and also of $\sum^t_{s=1}g(Z'_s)/t$).
Throughout the paper, we use the prime symbol ($'$) for any
notation related to a newly defined process (e.g., $\{X'_t\}$) to
differentiate it from its counterpart defined on the original
process (e.g., $\{X_t\}$).

While there are infinitely many different transition matrices
$\p'$ leading to the unbiased estimator $\hat{\mu}'_t(f)$ for
$\Ex_{\bm{\pi}}(f)$, our primary goal is, at (almost) no
additional cost and in a distributed manner, to find a transition
matrix $\p'$ that also guarantees smaller asymptotic variance.
Under a rather restricted setting, R. Neal gave a partial answer
to this in~\cite{Neal04} saying that less backtracking (rendering
the resulting Markov chain $\{Z'_t\}$ non-reversible) can result
in a smaller asymptotic variance. We restate his finding below.

\begin{theorem}\cite[Theorem 2]{Neal04}\label{nonbacktrack}
Suppose that $\{X_t\}$ is an irreducible, reversible Markov chain
on the state space $\N$ with transition matrix $\p \!=\!
\{P(i,j)\}$ and stationary distribution $\bm{\pi}$. Construct a
Markov chain $\{Z'_t\}$ on the state space $\Omega$ with
transition matrix $\p' \!=\! \{P'(e_{ij},e_{lk})\}$ in which the
transition probabilities $P'(e_{ij},e_{lk})$ satisfy the following
two conditions: for all $e_{ij},e_{ji},e_{jk},e_{kj} \in \Omega$
with $i \ne k$,
\begin{align}
P(j,i)P'(e_{ij},e_{jk}) &= P(j,k)P'(e_{kj},e_{ji}),\label{condition1}\\
P'(e_{ij},e_{jk}) &\geq P(j,k).\label{condition2}
\end{align}
Then, the Markov chain $\{Z'_t\}$ is irreducible and
non-reversible with a unique stationary distribution $\bm{\pi'}$
in which $\pi'(e_{ij}) = \pi(i)P(i,j)$, $e_{ij} \!\in\! \Omega$.
Also, for any function $f$, the asymptotic variance of
$\hat{\mu}'_t(f)$ is no greater than that of $\hat{\mu}_t(f)$,
i.e., ${\sigma'}^2(f) \leq \sigma^2(f)$.
\endthm
\end{theorem}

\begin{remark}
The condition in (\ref{condition1}) ensures that the resulting
transition matrix $\p'$ is stationary with respect to $\bm{\pi'}$
in (\ref{stationary}) and, in turn, leads to the unbiased
estimator $\hat{\mu}'_t(f)$ for $\Ex_{\bm{\pi}}(f)$. Together with
this condition, the condition in (\ref{condition2}) -- less
backtracking to the previously visited node, brings out the
improvement of asymptotic variance.
\end{remark}

Theorem~\ref{nonbacktrack} is quite versatile and provides a
guideline on how to choose the transition matrix $\p'$ of a Markov
chain $\{Z'_t\}$ leading to smaller asymptotic variance, and thus
will play an essential role in developing our graph sampling
methods and subsequent analysis. Despite this large degree of
freedom, it is still uncertain how to choose such a transition
matrix $\p'$ \emph{at no additional cost}. While R. Neal suggested
a procedure to find $\p'$, it generally poses significant cost,
especially for improving the MH algorithm, as admitted
in~\cite{Neal04}. (See pp. 9--10 therein.) Recall that the MH
algorithm (Algorithm~\ref{alg1}) for MHRW only needs the degree
information of a randomly chosen neighbor $j$ of the current node
$X_t$ to decide whether or not to move $j$, as mentioned in
Remark~\ref{remark1}.\footnote{More generally, in the MH algorithm
with any proposal matrix $\q$, it is often unnecessary to know
self-transition probabilities $P(i,i)$ explicitly, or does not
require summing the probabilities of rejection for all possible
proposals just to compute $P(i,i) \!=\! Q(i,i) \!+\! \sum_{j \ne
i}Q(i,j)(1 \!-\! A(i,j))$.} However, the procedure by R. Neal
necessitates the \emph{explicit knowledge} of all
$P(i,i)$'s~\cite{Neal04}. That is, the corresponding modified MHRW
would require all the neighbors' degree information of the current
node $X'_t$ at each time $t$ in order to choose the next node
$X'_{t+1}$. Imagine such modified MHRW crawling over an OSN (say, Facebook) and located at a certain user's
page. To simply decide where to go, the walk would have to visit
all his/her friends' pages first and collect all their degree
information (i.e., the number of friends) before making decision
to move. This is clearly impractical for our graph sampling
purpose. Therefore, in this paper, we set out to develop our own graph samplers with
higher efficiency \emph{without any such overhead}, by leveraging
Theorem~\ref{nonbacktrack} as a building block.

\subsection{Non-backtracking Random Walk with Re-weighting}\label{se:nbrw}

We first introduce non-backtracking random walk with re-weighting (NBRW-rw)
that ensures unbiased graph sampling, and then prove that NBRW-rw guarantees a smaller asymptotic variance than SRW-rw. The non-backtracking random walk (NBRW) is a discrete-time random walk which
`never' backtracks (thus named non-backtracking) to the previous
node (whenever possible) while preserving the same stationary
distribution as that of a SRW. Thus, the proposed sampling method
is to use an NBRW, instead of a SRW, to collect a sequence of
samples by crawling over a target $\G$, and at the same
time, to employ the same re-weighting process as is done for SRW
in order to eliminate sampling bias induced from its non-uniform
stationary distribution.

\begin{figure}[t!]
    \centering
    \includegraphics[width=2.5in]{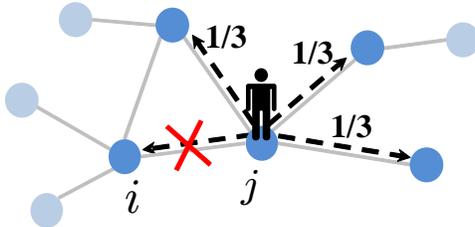}
    \vspace{-2mm}
    \caption{Illustrating the transitions of an NBRW over the nodes of
    $\G$. The walker is currently located at node $j$ (with $d(j)=4$) and just came from node $i$. From $j$, it will move to one of its
    neighbors except node $i$ with equal probability.}\label{fig:nbrw}
    \vspace{-0mm}
\end{figure}

Consider an irreducible, reversible Markov chain $\{X_t\}_{t \geq
0}$ (a sequence of visited nodes) by the SRW with its transition
matrix $\p \!=\! \{P(i,j)\}_{i,j \in \N}$ given by (\ref{srw}),
and stationary distribution $\pi(i) \!=\! d(i)/(2|\E|)$, $i
\!\in\! \N$. Then, the NBRW is defined as follows. A
(discrete-time) random walk at the current node $j$ with $d(j)
\geq 2$ moves to the next node $k$, chosen u.a.r. from the
neighbors of node $j$ except the previous node $i$. If the current
node $j$ has only one neighbor ($d(j) = 1$), the walk always
returns to the previous node $i$.
Figure~\ref{fig:nbrw} depicts this non-backtracking nature of the
NBRW in its transitions over the nodes of $\G$. Here, an initial
position of the NBRW can be arbitrarily chosen. The NBRW initially
moves from the initial position to one of its neighbors with equal
probability due to the absence of its `previous node', and then
proceeds as defined above thereafter.

Let $X'_t \!\in\! \N$, $t \!=\! 0,1,2,\ldots$, be the location of
an NBRW. As before, we construct a Markov chain $\{Z'_t \!=\!
(X'_{t-1},X'_t)\}_{t \geq 1}$ with its transition matrix $\p'
\!=\! \{P'(e_{ij},e_{lk})\}_{e_{ij},e_{lk} \in \Omega}$ given by,
for all $e_{ij}, e_{jk} \in \Omega$ with $i \ne k$ ($d(j) \geq
2$),
\begin{equation}
P'(e_{ij},e_{jk}) = \frac{1}{d(j) - 1} ~>~ \frac{1}{d(j)} =
P(j,k),
\end{equation}
implying that $P'(e_{ij},e_{ji}) = 0$. Also, $P'(e_{ij},e_{ji}) =
1$ for any $j$ with $d(j) = 1$. All other elements of $\p'$ are
zero. Clearly, $\p'$ satisfies the conditions in
(\ref{condition1})--(\ref{condition2}). From
Theorem~\ref{nonbacktrack}, the Markov chain $\{Z'_t\}$ is
irreducible and non-reversible with a unique stationary
distribution
\begin{equation}
\pi'(e_{ij}) = \pi(i)P(i,j) = \frac{1}{2|\E|},~~~e_{ij} \in
\Omega.
\end{equation}
That is, the probability of the NBRW being at node $j$ in the
steady-state is the same as $\pi(j)$. See (\ref{steady-state}).
From (\ref{ergodic-nonrev})--(\ref{est-nonrev}) and
Theorem~\ref{nonbacktrack}, we also know that for any given
function $f$ of interest, $\hat{\mu}'_t(f)$ and $\hat{\mu}_t(f)$
are both unbiased estimators for $\Ex_{\bm{\pi}}(f)$, and the
asymptotic variance of $\hat{\mu}'_t(f)$ (based on the random
samples by the NBRW) is no larger than that of $\hat{\mu}_t(f)$
(by the SRW), i.e., ${\sigma'}^2(f) \!\leq\! \sigma^2(f)$.

However, both unbiased estimators $\hat{\mu}'_t(f)$ and
$\hat{\mu}_t(f)$ are for $\Ex_{\bm{\pi}}(f)$, not
$\Ex_{\bm{u}}(f)$. It is unclear whether such improvement for the
asymptotic variance remains true even after a proper re-weighting
to obtain \emph{unbiased} samples. As explained in
Section~\ref{se:srw}, the SRW-rw is to use the
estimator $\hat{\mu}_t(wf)/\hat{\mu}_t(w)$ with $w(i) \!=\!
1/d(i)$ $(i \!\in\! \N)$ in order to consistently estimate
$\Ex_{\bm{u}}(f)$. Since the stationary distribution of the NBRW
remains the same as that of the SRW, we can also use the estimator
$\hat{\mu}'_t(wf)/\hat{\mu}'_t(w)$ with the same weight function
$w$, as a valid approximation of $\Ex_{\bm{u}}(f)$. Let
$\sigma^2_{\textsf{W}}(f)$ and ${\sigma'}^2_{\textsf{W}}(f)$ denote the
asymptotic variances of the estimators
$\hat{\mu}_t(wf)/\hat{\mu}_t(w)$ and
$\hat{\mu}'_t(wf)/\hat{\mu}'_t(w)$, respectively. To proceed, we
need the following.
%
\begin{theorem}[Slutsky's theorem]~\cite[pp.332]{Ash00}\\
Let $\{A_t\}$ and $\{B_t\}$ be the sequences of random variables.
If $A_t ~\stackrel{d}{\Longrightarrow}~ A$, and $B_t$ converges in
probability to a non-zero constant $b$, then $A_t/B_t
~\stackrel{d}{\Longrightarrow}~ A/b$. \endthm
\end{theorem}

Now we state our main result.

\begin{theorem}\label{nbrw-asymptotic}
For any function $f: \N \!\to\! \Rn$, the asymptotic variance of
$\hat{\mu}'_t(wf)/\hat{\mu}'_t(w)$ is no larger than that of
$\hat{\mu}_t(wf)/\hat{\mu}_t(w)$, i.e.,
${\sigma'}^2_{\textsf{W}}(f) \leq \sigma^2_{\textsf{W}}(f)$, where
the weight function $w$ is given by $w(i) = 1/d(i)$, $i \in \N$.
\endthm
\end{theorem}

\vspace{2mm} \textbf{Proof:}
Since the estimator $\hat{\mu}_t(wf)/\hat{\mu}_t(w)$ remains
invariant up to a constant multiple of $w$, without loss of
generality, we can set $w(i) = u(i)/\pi(i) = 2|\E|/(n d(i))$.
For any given $f$, observe that
\begin{align}
\sqrt{t} \lt[\frac{\hat{\mu}_t(wf)}{\hat{\mu}_t(w)} -
\Ex_{\bm{u}}(f)\rt] & = \sqrt{t}
\lt[\frac{\sum^t_{s=1}w(X_s)f(X_s)}{\sum^t_{s=1}w(X_s)} -
\Ex_{\bm{u}}(f)\rt] \nonumber\\
& =
\frac{t}{\sum^t_{s=1}w(X_s)}\sqrt{t}\lt[\frac{\sum^t_{s=1} w(X_s)(f(X_s)
-\Ex_{\bm{u}}(f))}{t}\rt]. \label{fff}
\end{align}
Define another function $h: \N \to \Rn$ such that
\begin{equation*}
h(i) \triangleq w(i)(f(i) - \Ex_{\bm{u}}(f)),~~~~ i \in \N,
\end{equation*}
implying $\Ex_{\bm{\pi}}(h) = \sum_{i \in \N}h(i)\pi(i) = 0$.
Then, from Theorems~\ref{ergodic} and \ref{clt}, we have, as $t
\to \infty$,
\begin{align*}
\frac{1}{t}\sum^t_{s=1}w(X_s) \to 1~\text{a.s.},~\text{and}~~
\sqrt{t}\lt[\frac{1}{t}\sum^t_{s=1}h(X_s)\rt]
\stackrel{d}{\Longrightarrow} \mathrm{N}(0,\sigma^2(h)).
\end{align*}
Since almost sure convergence implies convergence in
probability~\cite{Ash00}, by Slutsky's theorem, from (\ref{fff}),
we have
\begin{equation*}
\sqrt{t} \lt[\frac{\hat{\mu}_t(wf)}{\hat{\mu}_t(w)} -
\Ex_{\bm{u}}(f)\rt] ~\stackrel{d}{\Longrightarrow}~
\mathrm{N}(0,\sigma^2(h)),~\mbox{as}~ t \!\to\! \infty.
\end{equation*}
Together with (\ref{ergodic-nonrev}) and (\ref{clt-nonrev}),
following the same lines above, we similarly have
\begin{equation*}
\sqrt{t} \lt[\frac{\hat{\mu}'_t(wf)}{\hat{\mu}'_t(w)} -
\Ex_{\bm{u}}(f)\rt] ~\stackrel{d}{\Longrightarrow}~
\mathrm{N}(0,{\sigma'}^2(h)),~\mbox{as}~ t \!\to\! \infty.
\end{equation*}
Hence, for a given $f$, the asymptotic variance of the estimator
$\hat{\mu}_t(wf)/\hat{\mu}_t(w)$ is nothing but
$\sigma^2_{\textsf{W}}(f) \!=\! \sigma^2(h)$. Similarly,
${\sigma'}^2_{\textsf{W}}(f) \!=\! {\sigma'}^2(h)$. Therefore,
since Theorem~\ref{nonbacktrack} says that for any function $f$,
${\sigma'}^2(f) \leq \sigma^2(f)$, we also have
${\sigma'}^2_{\textsf{W}}(f) \leq \sigma^2_{\textsf{W}}(f)$. That
is, the asymptotic variance of $\hat{\mu}'_t(wf)/\hat{\mu}'_t(w)$
is no larger than that of $\hat{\mu}_t(wf)/\hat{\mu}_t(w)$.
\hfill $\Box$
\vspace{2mm}

\begin{remark}
In~\cite{Alon07}, the NBRW was originally considered for regular
graphs with $d(i) = d > 3, ~\forall i \in \N$, and shown to lead to
faster mixing rate (i.e., faster rate of convergence to its
stationary distribution) than that of the SRW. In contrast, for
any general (connected, undirected, not necessarily regular) graph
$\G$, we show that the NBRW-rw ensures not only the
unbiased graph sampling but also smaller asymptotic variance than
the SRW-rw.
\end{remark}

\subsection{Metropolis-Hastings Algorithm with Delayed Acceptance}\label{se:new-mh}

We turn our attention to improving the MH algorithm. For
any given, desired stationary distribution $\bm{\pi}$, we propose
Metropolis-Hastings algorithm with delayed acceptance (MHDA), which
theoretically guarantees smaller asymptotic variance than the
(generic) MH algorithm with proposal matrix $\q$ that
constructs a reversible Markov chain with arbitrary $\bm{\pi}$. In
particular, we demonstrate that MHDA can be applied, as a
special case, to construct a (non-Markovian) random walk on a
graph $\G$ which not only achieves a uniform stationary
distribution $\bm{\pi} \!=\! \bm{u}$ for unbiased graph sampling,
but leads to higher efficiency than MHRW by the MH algorithm
(Algorithm~\ref{alg1}). We emphasize that the only additional
overhead here is remembering the previously visited node (one
of the neighbors of the current node) from which the random walk
came.

\begin{figure}[t!]
    \centering
    \includegraphics[width=2.2in]{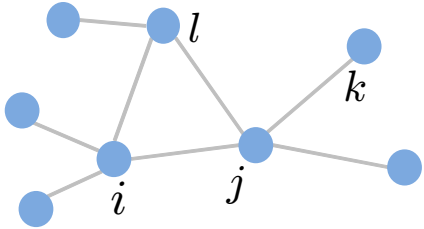}
    \vspace{-1mm}
    \caption{An example graph $\G$}\label{fig:example}
    \vspace{-0mm}
\end{figure}

\vspace{3mm} \textbf{\textsf{Interpreting a reversible MH Markov chain as
a semi-Markov chain:}} Consider an irreducible, reversible Markov
chain $\{X_t \!\in\! \N\}_{t \geq 0}$ by the MH algorithm with its
transition matrix $\p \!=\! \{P(i,j)\}_{i,j \in \N}$ given by
(\ref{mh}), and any arbitrarily given target stationary
distribution $\bm{\pi}$. Recall that the MH algorithm is nothing
but a repetition of proposing a state transition with proposal
probability $Q(i,j)$ that is then accepted with an acceptance
probability $A(i,j)$ in (\ref{mh-accept}) or rejected otherwise.
Observe that the process $\{X_t\}$, after entering into state
(node) $i$, stays at state $i$ for a geometrically distributed
time duration with mean $1/(1-P(i,i))$, and then moves to another
state $j \in N(i)$. Formally, define a Markov chain $\{\tilde{X}_m
\!\in\! \N\}_{m \geq 0}$ with its transition matrix $\tilde{\p} \!
\triangleq \! \{\tilde{P}(i,j)\}_{i,j \in \N}$ given by, for $j
\ne i$,
\begin{equation}
\tilde{P}(i,j) = \frac{P(i,j)}{1 - P(i,i)} =
\frac{Q(i,j)A(i,j)}{\sum_{j \ne i}Q(i,j)A(i,j)}= \frac{\min\{Q(i,j),Q(j,i)\pi(j)/\pi(i)\}}{\sum_{j \ne
i}\min\{Q(i,j),Q(j,i)\pi(j)/\pi(i)\}},\label{embedded}
\end{equation}
with $\tilde{P}(i,i) = 0$. It is not difficult to see that the
chain $\{\tilde{X}_m\}$ is irreducible, and reversible with
respect to a unique stationary distribution $\bm{\tilde{\pi}}
\!\triangleq\! [\tilde{\pi}(i), i \!\in\!\N]$, given by
\begin{equation*}
\tilde{\pi}(i) \propto \pi(i)(1-P(i,i)), ~~~i \in\N.
\end{equation*}
Also, we define a function $\gamma: \N \to \Rn$ such that, for $i
\in \N$,
\begin{equation}
\gamma(i) \triangleq 1 - P(i,i) = \sum_{j \ne
i}\min\{Q(i,j),Q(j,i)\pi(j)/\pi(i)\}, \label{gamma}
\end{equation}
and define a sequence $\{\xi_m\}_{m \geq 0}$ for which $\xi_m$
depends solely on $\{\tilde{X}_m\}_{m \geq 0}$ and is
geometrically distributed with parameter $\gamma(\tilde{X}_m)$. It
thus follows that $\Ex\{\xi_m | \tilde{X}_m \!=\! i\} \!=\!
1/\gamma(i)$, $i \!\in\!N$.
The process $\{X_t\}$ can now be interpreted as a
\emph{semi-Markov} chain with \emph{embedded} Markov chain
$\{\tilde{X}_m\}$ and respective sojourn times $\{\xi_m\}$. Suppose
that the random walk by the MH algorithm (or the process
$\{X_t\}$) enters node $j$ of a graph $\G$, depicted in
Figure~\ref{fig:example}, at time $t\!=\!1$ ($X_1 \!=\! j$). If we
consider a sample path $(X_1,X_2,\ldots, X_7) \!=\!
(j,j,j,i,i,j,k)$, then we have corresponding sequences
$(\tilde{X}_1,\tilde{X}_2,\tilde{X}_3,\tilde{X}_4) \!=\!
(j,i,j,k)$ and $(\xi_1,\xi_2,\xi_3) \!=\! (3,2,1)$.
Note that the standard definition of a semi-Markov process allows
the sojourn time $\xi_m$ to depend on both $\tilde{X}_m$ and
$\tilde{X}_{m+1}$ (and so we are dealing with a special case).
From the theory of semi-Markov processes (e.g.,~\cite{Ross96a}),
one can easily recover the stationary distribution $\bm{\pi}$ as
\begin{equation}
\pi(i) \propto \tilde{\pi}(i)/\gamma(i),~~~ i \in \N.
\label{semi-stationary}
\end{equation}
The above interpretation has been similarly given in the MCMC
literature~\cite{Malefaki08,Douc11}. In particular, it is known
that, for any given function $f: \N \!\to\! \Rn$,
\begin{equation}
\hat{\mu}_{m, \textsf{MH}}(f) \triangleq \frac{\sum^m_{l=1}\xi_l
f(\tilde{X}_l)}{\sum^m_{l=1}\xi_l}\label{new}
\end{equation}
converges almost surely to $\Ex_{\bm{\pi}}(f)$, as $m \to \infty$,
and thus $\hat{\mu}_{m, \textsf{MH}}(f)$ is also an unbiased
estimator for $\Ex_{\bm{\pi}}(f)$~\cite{Malefaki08}. This
definition of $\hat{\mu}_{m, \textsf{MH}}(f)$ enables more
tractable analysis on its asymptotic variance, denoted as
$\sigma^2_{\textsf{MH}}(f)$, by connecting it to its counterpart
in the importance sampling for Markov
chains~\cite{Malefaki08,Douc11}. Note that for sufficiently large
$t$ (also $m$), the (original) unbiased estimator $\hat{\mu}_t(f)
\!=\! \sum^t_{s=1}f(X_s)/t$ can be written as $\hat{\mu}_{m,
\textsf{MH}}(f)$ plus some negligible term (after setting the same
initial point $\tilde{X}_1 \!=\! X_1$), because it is always
possible to find $m$ such that $\sum^m_{l=1}\xi_l \leq t <
\sum^{m+1}_{l=1}\xi_l$. Also, in the limit $t,m \!\to\! \infty$,
$\hat{\mu}_t(f)$ and $\hat{\mu}_{m, \textsf{MH}}(f)$ are the same.
We thus focus on estimators in the form of $\hat{\mu}_{m,
\textsf{MH}}(f)$ in our subsequent analysis.

Consider a sequence of pairs $(\tilde{X}_m, \xi_m)$. From the
success in the NBRW-rw, one may ask what if the
reversible embedded Markov chain $\{\tilde{X}_m\}_{m \geq 0}$ is
replaced by a related stochastic process $\{\tilde{X}'_m \!\in\!
\N\}_{m \geq 0}$, or more precisely, a non-reversible Markov chain
$\{(\tilde{X}'_{m-1},\tilde{X}'_{m})\}_{m \geq 1}$ on the
augmented state space $\Omega$, which avoids backtracking
transitions to the extent possible, while preserving the same
stationary distribution $\bm{\tilde{\pi}}$. Another question can
be whether this transformation guarantees that the estimator in
(\ref{new}) based on $(\tilde{X}'_m, \xi_m)$ remains
\emph{unbiased} for $\Ex_{\bm{\pi}}(f)$ and also have \emph{higher
efficiency} than the original one. Our answer is in the
affirmative, and this is the reasoning behind the improvement of
our proposed MHDA over the standard MH algorithm. We stress here
that, in contrast to the NBRW, backtracking transitions in the
process $\{\tilde{X}'_m\}$ should be \emph{avoided only up to the extent
possible}\footnote{If $\{\tilde{X}'_m\}$ is made purely
non-backtracking just like we did for NBRW, then we lose
unbiasedness for the resulting MH-based estimator in general.} so
as to maintain the arbitrarily given original stationary
distribution $\bm{\tilde{\pi}}$. Thus, the extension from the MH
algorithm to our proposed MHDA becomes necessarily more involved
than the case of NBRW.

\vspace{3mm} \textbf{\textsf{Description of MHDA:}} Let $X'_t \in \N$,
$t = 0,1,2,\ldots$, be the position of a random walk (or the
state of a stochastic process). We also define the augmented state
space $\Omega$ in (\ref{space}) based on the reversible embedded
Markov chain $\{\tilde{X}_m\}$, where $\tilde{P}(i,i) = 0$ and so
$e_{ii} \not\in \Omega$ for all $i$.

MHDA is described as follows. Suppose that node $i$ is the
previous node from which the walk came. MHDA first operates
just like the MH algorithm. At the current node (state) $X'_t = j
\ne i$, the next node $X'_{t+1}=k \in N(j)$ is proposed with
probability $Q(j,k)$ ($j \ne k$). Then, the proposed
transition to $k$ is accepted with probability $A(j,k)$ in
(\ref{mh-accept}), and rejected with probability $1\!-\!A(j,k)$ in
which case $X'_{t+1} \!=\! j$. Here, in contrast to the MH
algorithm, MHDA renders the accepted transition to $X'_{t+1}
\!=\! k$ temporarily \emph{pending}, and applies another procedure
to proceed with the actual transition to $k$.\footnote{If the
transition to $k \!=\! j$ was accepted after a proposal with
$Q(j,j) \!>\!0$, then the MHDA accepts the transition as in the MH
algorithm without any further action.}

\begin{figure}[t!]
    \centering
    \vspace{-0mm}
    \hspace{-0mm}\subfigure[MH algorithm]{\includegraphics[width=2.4in]{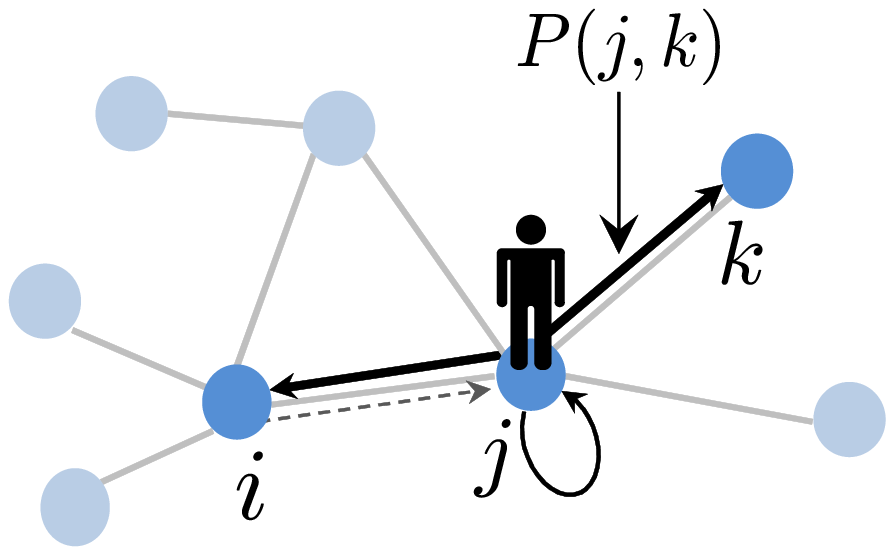}}
    \hspace{0mm}\subfigure[MHDA]{\includegraphics[width=2.8in]{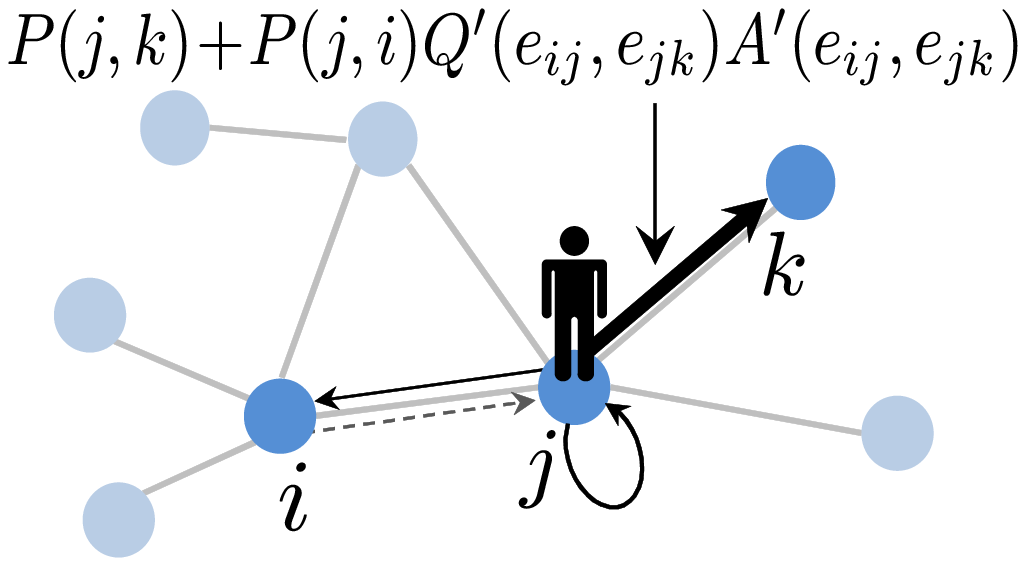}}
    \hspace{-0mm}
    \vspace{-0mm}
    \caption{Illustrating a difference between MH algorithm and MHDA: a walker moves from node $j$ to node
    $k$ with probability $P(j,k)$ in the MH algorithm, but with larger probability
    $P(j,k) + P(j,i)Q'(e_{ij},e_{jk})A'(e_{ij},e_{jk})$ in the MHDA.}\label{fig:difference}
    \vspace{-0mm}
\end{figure}

Specifically, for the accepted transition to $k$, if $k \ne i$,
then the `actual' transition takes place, i.e., $X'_{t+1} \!=\!
k$, which happens with probability $P(j,k) \!=\! Q(j,k)A(j,k)$ as
in the MH algorithm. On the other hand, if $k = i$, then the
transition to node $i$ is \emph{delayed} (thus named `delayed
acceptance').
The next node $X'_{t+1}$ is again proposed with another proposal
probability $Q'(e_{ij},e_{jk})$, which is a transition probability
of an arbitrary Markov chain on the state space $\Omega$, where
$Q'(e_{ij},e_{lk}) \!=\! 0$ for all $j \!\ne\! l$, and
$Q'(e_{ij},e_{jk}) \!>\! 0$ if and only if $Q'(e_{kj},e_{ji})
\!>\! 0$. The (second) proposed transition to $X'_{t+1} \!=\! k$
is accepted with another acceptance probability
$A'(e_{ij},e_{jk})$, and rejected with probability $1 \!-\!
A'(e_{ij},e_{jk})$ in which case $X'_{t+1} \!=\! i$
(backtracking occurs). That is, transition probability $P(j,i) \!=\!
Q(j,i)A(j,i)$ in the MH algorithm is leveraged to create another
transition opportunity from $j$ to $k \!\ne\! i$ in the MHDA.
So, the transition from $j$ to $k \!\ne\! i$ occurs with larger probability $P(j,k) \!+\!
P(j,i)Q'(e_{ij},e_{jk})A'(e_{ij},e_{jk})$ than the MH
algorithm (w.p. $P(j,k)$). This is also illustrated in Figure~\ref{fig:difference} where the thickness of arrows represents the corresponding transition probabilities
from node $j$ to other node (including
self-transition). The new acceptance probability $A'(e_{ij},e_{jk})$ will be
specified shortly.

In summary, under MHDA, the walker stays at each node for the
same random amount of time as it would be under the MH algorithm,
while reducing the bias toward the previous node when making
transitions to one of its neighbors.

\vspace{3mm} \textbf{\textsf{Analysis of MHDA:}} Let $\tilde{X}'_m$, $m
\!\geq\! 0$, be the sequence of nodes visited by the walk, which
moves over $\G$ according to the MHDA. The process $\{\tilde{X}'_m\}$ is clearly different from the
reversible, embedded Markov chain $\{\tilde{X}_m\}$ for the MH
algorithm. Also, let $\xi'_m$, $m \!\geq\! 0$, be the respective
sojourn time at node $\tilde{X}'_m$. Note that the MHDA behaves
differently from the MH algorithm (performs the additional
procedure) only when a proposed transition from node $j$ to node
$k \ne j$ (occurring with probability $Q(j,k)$) is accepted with
probability $A(j,k)$ in the MH algorithm. Thus, $\xi'_m$ is also
geometrically distributed with parameter $\gamma(\tilde{X}'_m)$.
See (\ref{gamma}) for $\gamma(\cdot)$. That is, given that
$\tilde{X}'_{m_1} \!=\! \tilde{X}_{m_2} \!=\! i$, the sojourn
times $\xi'_{m_1}$ and $\xi_{m_2}$ have identical distributions.
Therefore, the MHDA, similar to the MH algorithm, can be also
characterized by a sequence of the pairs $(\tilde{X}'_m,\xi'_m)$.
As an example, if the random walk by the MHDA (or
the process $\{X'_t\}$) enters node $i$ in
Figure~\ref{fig:example} at time $t\!=\!1$ ($X'_1 \!=\! i$) and
$(X'_1,X'_2,\ldots, X'_7) \!=\! (i,i,l,j,j,j,k)$, then we consequently have
$(\tilde{X}'_1,\tilde{X}'_2,\tilde{X}'_3,\tilde{X}'_4) \!=\!
(i,l,j,k)$ and $(\xi'_1,\xi'_2,\xi'_3) \!=\! (2,1,3)$.
We define, for any given $f: \N \!\to\! \Rn$,
\begin{equation}
\hat{\mu}'_{m, \textsf{MHDA}}(f) \triangleq
\frac{\sum^m_{l=1}\xi'_l f(\tilde{X}'_l)}{\sum^m_{l=1}\xi'_l}.
\end{equation}
We prove below that $\hat{\mu}'_{m, \textsf{MHDA}}(f)$ converges
almost surely to $\Ex_{\bm{\pi}}(f)$, implying that
$\hat{\mu}'_{m, \textsf{MHDA}}(f)$ is an unbiased estimator for
$\Ex_{\bm{\pi}}(f)$. We also prove that, after showing the CLT
holds for $\hat{\mu}'_{m, \textsf{MHDA}}(f)$, the asymptotic
variance of $\hat{\mu}'_{m}(f)$, denoted as
${\sigma'}^2_{\textsf{MHDA}}(f)$, is smaller than its counterpart
${\sigma}^2_{\textsf{MH}}(f)$ for the MH algorithm.

To this end, we first explain how to properly choose the new
acceptance $A'(e_{ij},e_{jk})$ so that the process
$\{\tilde{X}'_m\}$ has the same stationary distribution as that of
the reversible embedded chain $\{\tilde{X}_m\}$, while, at the
same time, the process $\{\tilde{X}'_m\}$ reduces
backtracking transitions. Instead of the
process $\{\tilde{X}'_m\}$, we deal with its related non-reversible Markov
chain defined on the augmented state space $\Omega$ by consulting
the general recipe for this purpose in Section~\ref{se:recipe}.
Recall the state space $\Omega$ in (\ref{space}) obtained from the
transition matrix $\tilde{\p} \!=\! \{\tilde{P}(i,j)\}$ of the
chain $\{\tilde{X}_m\}$. We define $\tilde{Z}'_m \!\triangleq\!
(\tilde{X}'_{m-1},\tilde{X}'_m) \!\in\! \Omega$ for $m \!\geq\!
1$, and $\tilde{\p}' \!\triangleq\!
\{\tilde{P}'(e_{ij},e_{lk})\}_{e_{ij},e_{lk} \in \Omega}$ to be
the transition matrix of a Markov chain $\{\tilde{Z}'_m\}_{m\geq
1}$. For instance, consider a
sample path $(\tilde{X}'_1,\tilde{X}'_2,\tilde{X}'_3,\tilde{X}'_4)
\!=\! (i,l,j,k)$ in the above example. We have
$(\tilde{Z}'_2,\tilde{Z}'_3,\tilde{Z}'_4) \!=\! ((i,l),(l,j),(j,k))$.
If the chain $\{\tilde{Z}'_m\}$ has a unique stationary
distribution $\bm{\tilde{\pi}'} \!\triangleq\! [\tilde{\pi}'(e_{ij}),
e_{ij} \!\in\! \Omega]$ given by
\begin{equation}
\tilde{\pi}'(e_{ij}) = \tilde{\pi}(i)\tilde{P}(i,j),~~~~~ e_{ij}
\in \Omega,\label{stationary2}
\end{equation}
implying that $\tilde{\pi}'(e_{ij}) \!=\! \tilde{\pi}'(e_{ji})$
from the reversibility of the embedded chain $\{\tilde{X}_m\}$,
then the \emph{steady-state} probability of the process
$\{\tilde{X}'_m\}$ being at node $j$ is the same as
$\tilde{\pi}(j)$ for all $j$. From the description of MHDA,
observe that, for all $e_{ij},e_{jk} \!\in\! \Omega$ with $i \ne
k$ ($d(j) \geq 2$),
\begin{equation}
\tilde{P}'(e_{ij},e_{jk}) = \tilde{P}(j,k) +
\tilde{P}(j,i)Q'(e_{ij},e_{jk})A'(e_{ij},e_{jk}),\label{new-embedded}
\end{equation}
while $\tilde{P}'(e_{ij},e_{ji}) \!=\! 1 \!-\! \sum_{k \ne
i}\tilde{P}'(e_{ij},e_{jk})$, as is also shown in
Figure~\ref{fig:difference}(b). Note that
$\tilde{P}'(e_{ij},e_{jk})$ specifies the next node of the random
walk by MHDA, \emph{given that} the walk has to move from the
current node to one of its neighbors (its sojourn time is over).
Thus, $\tilde{P}(j,k)$ and $\tilde{P}(j,i)$ are used here instead
of $P(j,k)$ and $P(j,i)$, respectively. In addition, for any $j$
with $d(j) \!=\! 1$, we have $\tilde{P}'(e_{ij},e_{ji}) \!=\!
\tilde{P}(j,i) \!=\! 1$, $(i,j) \in \E$, since $Q'(e_{ij},e_{ji}) \!=\! 1$ (due to the stochastic matrix $\{Q'(e_{ij},e_{lk}\}$) and $A'(e_{ij},e_{ji}) \!=\!1$ which is shown below.

Among many possible choices for the acceptance probability
$A'(e_{ij},e_{jk})$ in the MHDA, we have the following.

\begin{proposition} \label{proposition1}
For any given $\{Q'(e_{ij},e_{lk})\}$, suppose that the acceptance
probability $A'(e_{ij},e_{jk})$ is given by
\begin{equation}
A'(e_{ij},e_{jk}) =
\min\lt\{1,\frac{P^2(j,k)Q'(e_{kj},e_{ji})}{P^2(j,i)Q'(e_{ij},e_{jk})}\rt\}.\label{accept}
\end{equation}
Then, the resulting transition matrix $\tilde{\p}'$, and $\tilde{\p}$ satisfy
conditions (\ref{condition1})--(\ref{condition2}). $\!\!\!$
\endthm
\end{proposition}

\vspace{2mm} \textbf{Proof:} See Appendix A. \hfill $\Box$
\vspace{1mm}

From Theorem~\ref{nonbacktrack} and
Proposition~\ref{proposition1}, the Markov chain
$\{\tilde{Z}'_m\}$ with its transition matrix $\tilde{\p}'$ as in
(\ref{new-embedded}) and (\ref{accept}),
is irreducible and non-reversible with a unique stationary
distribution $\bm{\tilde{\pi}'}$ in (\ref{stationary2}). This also
implies that the process $\{\tilde{X}'_m\}$ has the same
stationary distribution $\bm{\tilde{\pi}}$, as explained before.
We now present our main result.

\begin{theorem}\label{mhda-asymptotic}
Consider a given, desired stationary distribution $\bm{\pi} \!=\!
[\pi(i), i \!\in\!\N]$. Under the MHDA with any given
$\{Q'(e_{ij},e_{lk})\}$ and its corresponding $A'(e_{ij},e_{jk})$
in (\ref{accept}), for any given function $f: \N \!\to\! \Rn$, as
$m \to \infty$, $\hat{\mu}'_{m, \textsf{MHDA}}(f)$ converges
almost surely to $\Ex_{\bm{\pi}}(f)$, and also the asymptotic
variance of $\hat{\mu}'_{m, \textsf{MHDA}}(f)$ is no larger than
that of $\hat{\mu}_{m, \textsf{MH}}(f)$, i.e.,
${\sigma'}^2_{\textsf{MHDA}}(f) \!\leq\!
{\sigma}^2_{\textsf{MH}}(f)$.
\endthm
\end{theorem}

\vspace{2mm} \textbf{Proof:} See Appendix B. \hfill $\Box$

\vspace{1mm}

\vspace{2mm} \textbf{\textsf{An application of MHDA for unbiased graph
sampling:}} We explain how MHDA can be applied for
unbiased graph sampling applications. In particular, we present
how to construct a (discrete-time) random walk by MHDRA, named
Metropolis-Hastings Random walk with Delayed Acceptance (MHRW-DA),
on $\G$ that achieves the uniform stationary distribution, i.e., $\bm{\pi}
\!=\! \bm{u}$. The MHRW-DA here operates as an extension of
Algorithm~\ref{alg1} with the following choice of
$\{Q'(e_{ij},e_{lk})\}$: for all $e_{ij}, e_{jk} \in \Omega$ with
$i \ne k$ ($d(j) \geq 2$),
\begin{equation}
Q'(e_{ij},e_{jk}) = 1/(d(j) - 1),
\end{equation}
implying that $Q'(e_{ij},e_{ji}) \!=\! 0$. Also,
$Q'(e_{ij},e_{ji}) \!=\! 1$ for any $j$ with $d(j) \!=\! 1$. All
other elements are zero. While $\{Q'(e_{ij},e_{lk})\}$ is the same
as the transition matrix of NBRW, a `Metropolizing' step, which is
done with $A'(e_{ij},e_{jk})$ in (\ref{accept}), must follow in order to ensure that
the stationary distribution is uniform and the resulting estimator is unbiased. In other words, $A'(e_{ij},e_{jk})$ in
(\ref{accept}) becomes
\begin{equation*}
A'(e_{ij},e_{jk}) = \min\!\lt\{1,
\min\lt\{\frac{1}{d(j)^2},\frac{1}{d(k)^2}\!\rt\}\Big/ \min\lt\{\frac{1}{d(j)^2},\frac{1}{d(i)^2}\!\rt\}\rt\}.
\end{equation*}
This version of the MHDA is summarized in Algorithm~\ref{alg2},
where $X'_t \in \N$ is the location of  MHRW-DA at time $t$
and $Y_t \in \N$ indicates the previous node from which the
MHRW-DA came ($Y_t \ne X'_t$). Here, $X'_0$ can be chosen
arbitrarily. Since there is no notion of `previous node' $Y_0$ at
time $t=0$, MHRW-DA initially behaves the same as MHRW
until it moves from the initial position to one of its neighbors,
and then proceeds as described in Algorithm~\ref{alg2} thereafter.

\begin{algorithm}[h!]
\caption{MHDA for MHRW-DA (at time $t$)}\label{alg2}
\begin{algorithmic}[1]
\STATE Choose node $i$ u.a.r. from neighbors of $X'_t$, i.e.,
$N(X'_t)$ \STATE Generate $p \sim U(0,1)$ \IF{$p \leq
\min\lt\{1,\frac{d(X'_t)}{d(i)}\rt\}$} \IF{$Y_t = i$ and $d(X'_t)
> 1$} \STATE Choose node $k$ u.a.r. from $N(X'_t) \setminus \{i\}$
\STATE Generate $q \sim U(0,1)$ \IF{$q \leq \min\!\lt\{1,
\min\!\lt\{1,\lt(\frac{d(X'_t)}{d(k)}\rt)^2
\rt\}\max\!\lt\{1,\lt(\frac{d(i)}{d(X'_t)}\rt)^2\rt\}\rt\}$} \STATE
$X'_{t+1} \leftarrow k$ and $Y_{t+1} \leftarrow X'_t$ \ELSE \STATE
$X'_{t+1} \leftarrow i$ and $Y_{t+1} \leftarrow X'_t$ \ENDIF \ELSE
\STATE $X'_{t+1} \leftarrow i$ and $Y_{t+1} \leftarrow X'_t$
\ENDIF \ELSE \STATE $X'_{t+1} \leftarrow X'_t$ and $Y_{t+1}
\leftarrow Y_t$ \ENDIF
\end{algorithmic}
\end{algorithm}

Theorem~\ref{mhda-asymptotic} states that the MHDA works for any
given stationary distribution $\bm{\pi}$, while allowing us to
freely choose the new proposal probabilities
$\{Q'(e_{ij},e_{lk})\}$ as desired. Thus, Algorithm~\ref{alg2} for
MHRW-DA is nothing but a `special case' of the MHDA.
Theorem~\ref{mhda-asymptotic} asserts that MHRW-DA produces unbiased samples
with higher efficiency than the corresponding
MHRW (Algorithm~\ref{alg1}). Again, we emphasize that the only
additional overhead for MHRW-DA, compared to the MHRW, is
\emph{remembering where it came from}, $Y_t$. Note that the
degree of the previous node $Y_t$ is already known and can easily
be retrieved, while the degree information of another randomly
chosen neighbor is also necessary anyway even in the MH
algorithm (to decide whether or not to move there).

\section{Simulation Results}

In this section, we present simulation results to support our
theoretical findings. To this end, we use the following real-world
network datasets~\cite{snap}: 

\begin{itemize}[itemsep=0ex,leftmargin=1em]
\item \textbf{AS-733} -- an undirected graph of autonomous systems (ASs) composed of 6474 nodes and 13233 edges, where nodes represent ASs and edges exist according to AS-AS peering relationships.
\item \textbf{HEP-TH} -- a collaboration network among authors who submit papers to High Energy Physics-Theory category in the e-print arXiv, forming an undirected graph with 9877 nodes and 51971 edges, where nodes represent authors and edges exist between authors if coauthoring a paper.
\item \textbf{Road-PA} -- a road network of Pennsylvania, forming an undirected graph with 1088092 nodes and 3083796 edges, where nodes represent intersections and endpoints and edges represent the roads connecting them.
\item \textbf{Web-Google} -- a directed web graph with 875713 nodes and 5105039 edges, where nodes represent web pages and directed edges represent hyperlinks between them. For our simulation, we use an undirected version of this web graph.
    \vspace{-2.5mm}
\end{itemize}
To ensure graph
connectivity, we also use the largest connected component (LCC) of each
graph, where the LCC sizes of the AS-733, HEP-TH, Road-PA, and Web-Google graphs are 6474, 8638, 1087562, and 855802, respectively. Here, the average degrees of AS-733, HEP-TH, Road-PA, and Web-Google
graphs are 4.09, 5.75, 2.83, and 10.03, while their maximum degrees
are 1459, 65, 9, and 6332, respectively.

As a test case, we consider the estimation of the degree
distribution of each graph -- $\pr\{D_\G \!=\! d\}$ (pdf) and
$\pr\{D_\G \!>\! d\}$ (ccdf), to evaluate and compare our proposed
NBRW-rw and MHRW-DA (MHDA in Algorithm~\ref{alg2})
against SRW-rw and MHRW (MH algorithm in
Algorithm~\ref{alg1}), respectively. As mentioned before, to
estimate $\pr\{D_\G \!=\! d\}$, we just need to choose a function
$f(i) \!=\! \idc_{\{d(i) = d\}}$, $i \!\in\!\N$, for the
corresponding estimators. Similarly, we choose $f(i) \!=\!
\idc_{\{d(i) > d\}}$ for $\pr\{D_\G \!>\! d\}$.
To measure the estimation accuracy, we use the following
normalized root mean square error
(NRMSE)~\cite{Avrachenkov10,Ribeiro10,Kurant11},
$\sqrt{\Ex\{(\hat{x}(t) - x)^2\}}/x$,
where $\hat{x}(t)$ is the estimated value out of $t$ samples and
$x$ is the (ground-truth) real value. ($x= \lim_{t\to\infty}
\hat{x}(t)$ from unbiasedness.)
In all simulations, an initial position of each random walk is
drawn from its stationary distribution as similarly used
in~\cite{Avrachenkov10}, unless otherwise specified. In practical implementations, one can
employ a `burn-in' period to drive the random walk close to its
steady-state~\cite{GjokaJSAC11-2}. Each data point reported here for AS-733 and HEP-TH graphs
is obtained from $10^4$ independent simulations, while, for Road-PA and Web-Google graphs, the data points are based on $10^5$ and $5\cdot10^5$ simulations, respectively.

\begin{figure}[t!]
    \centering
    \vspace{-0mm}
    \hspace{-0mm}\subfigure[SRW-rw vs. NBRW-rw]{\includegraphics[width=3in,height=2.3in]{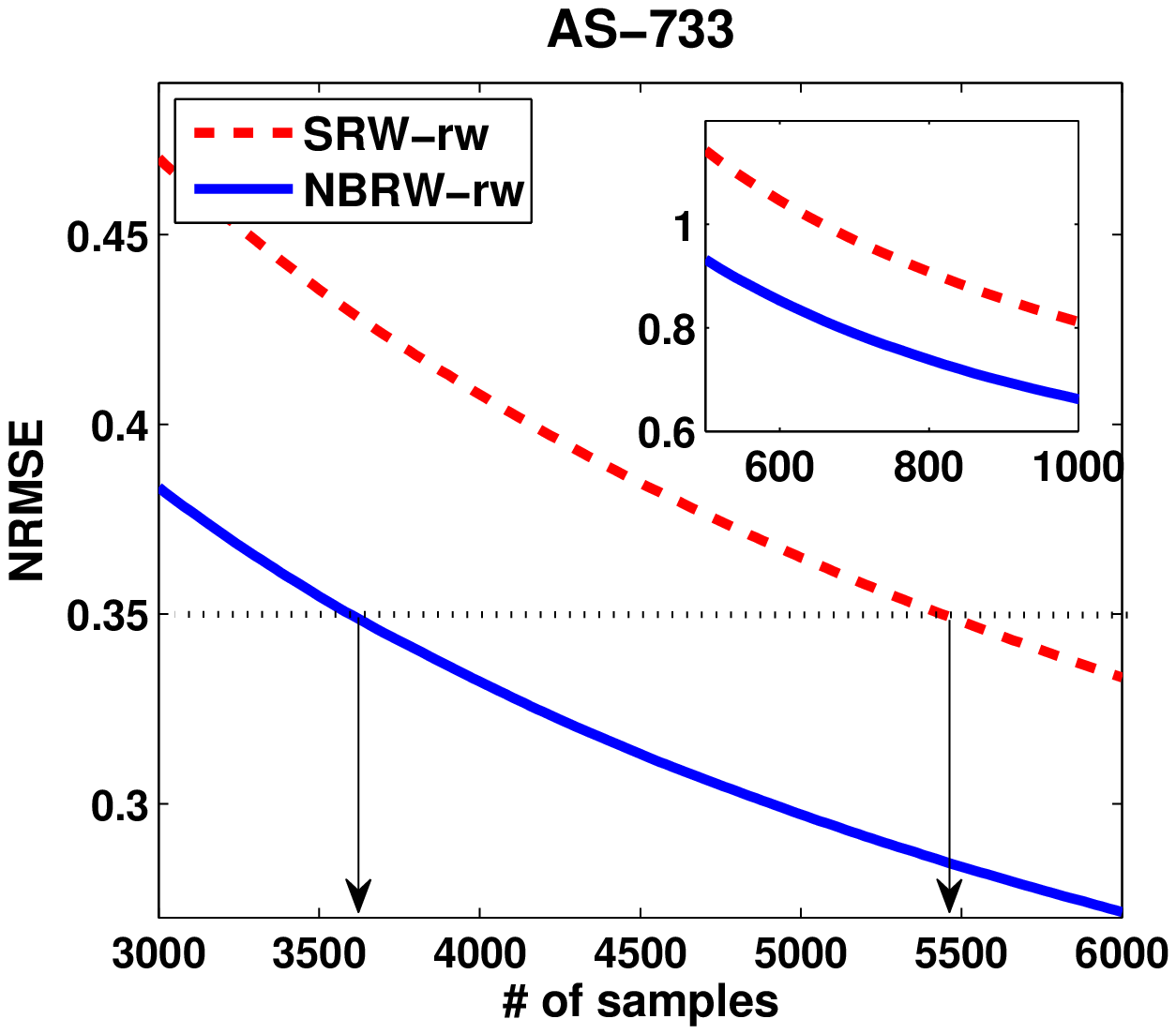}}
    \hspace{-0mm}\subfigure[MHRW vs. MHRW-DA]{\includegraphics[width=3in,height=2.3in]{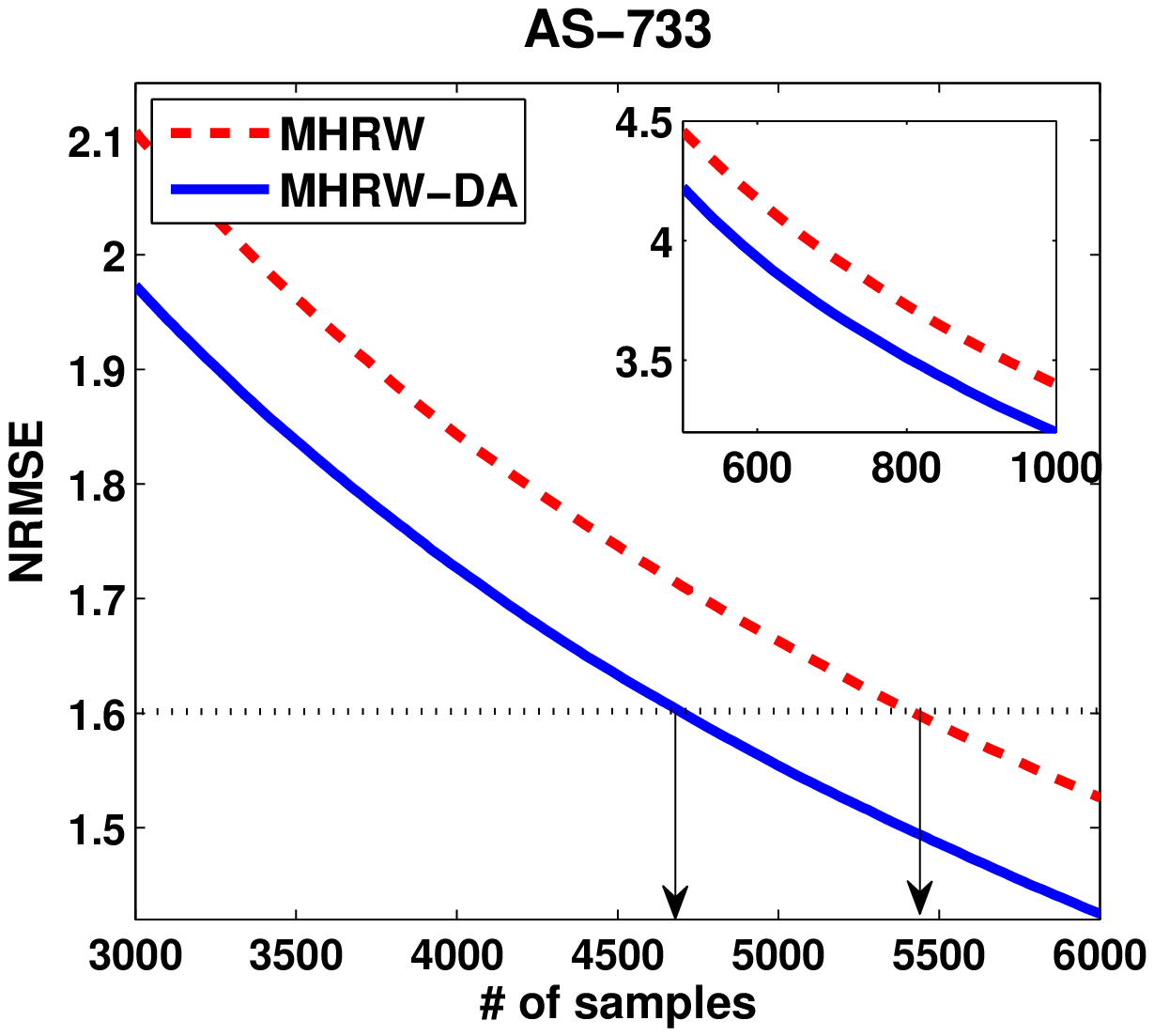}}
    \hspace{-0mm}
    \vspace{-0mm}
    \caption{AS-733 graph. NRMSE (averaged over all possible degrees $d$) of the estimator of $\pr\{D_\G = d\}$, when we vary the number of samples; the insets are for smaller number of samples.} \label{fig:nrmse-as}
    \vspace{-0mm}
\end{figure}
\begin{figure}[t!]
    \centering
    \vspace{-0mm}
    \hspace{-0mm}\subfigure[SRW-rw vs. NBRW-rw]{\includegraphics[width=3in,height=2.3in]{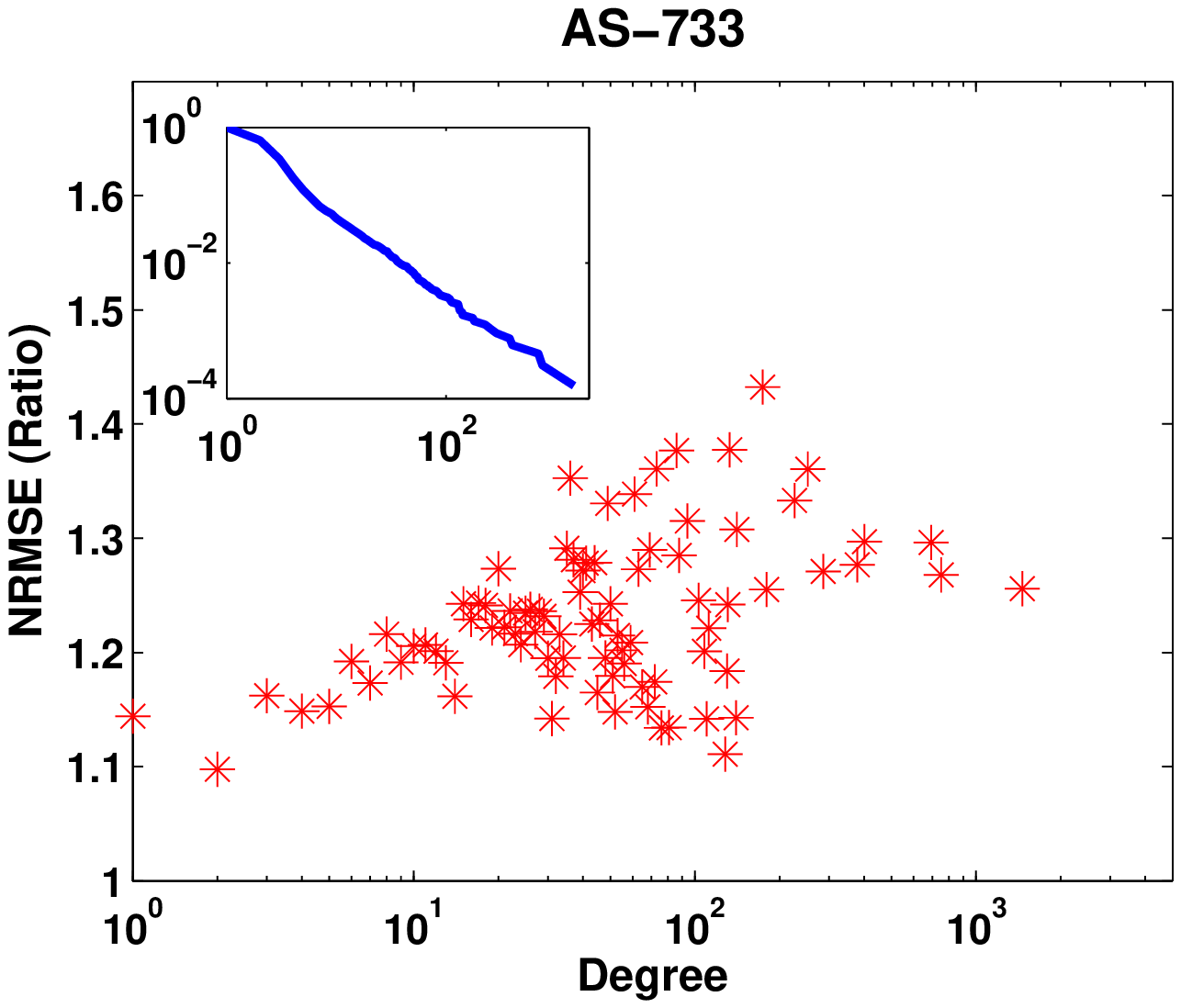}}
    \hspace{-0mm}\subfigure[MHRW vs. MHRW-DA]{\includegraphics[width=3in,height=2.3in]{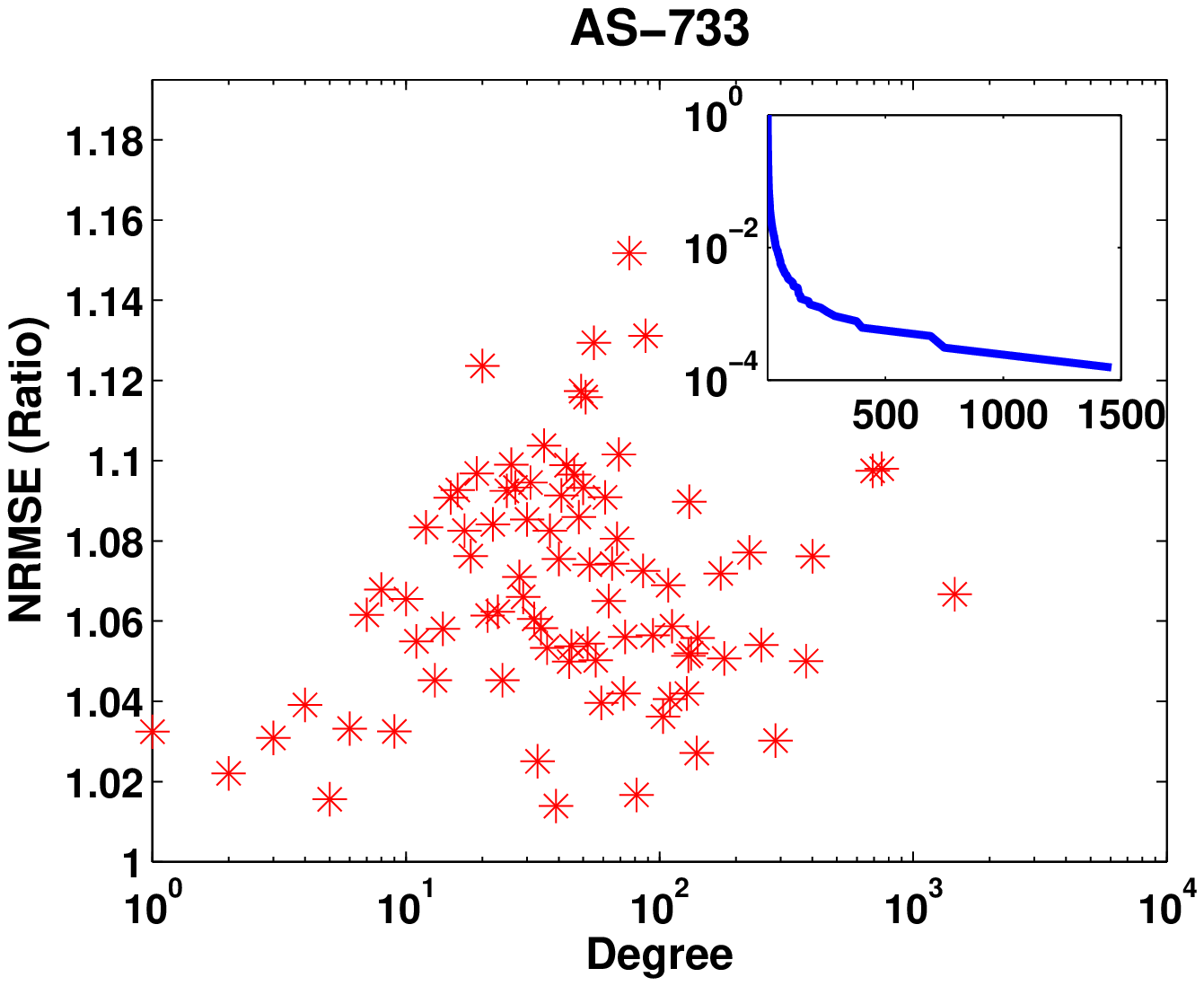}}
    \hspace{-0mm}
    \vspace{-0mm}
    \caption{AS-733 graph. NRMSE ratio (per degree $d$) when estimating $\pr\{D_\G \!=\! d\}$ with $10^4$ samples; the insets represent the `actual' degree distribution (ccdf) in (a) log-log scale, (b) semi-log scale.} \label{fig:perdegree-pdf-as}
    \vspace{-0mm}
\end{figure}
\begin{figure}[t!]
    \centering
    \vspace{-0mm}
    \hspace{-0mm}\subfigure[SRW-rw vs. NBRW-rw]{\includegraphics[width=3in,height=2.3in]{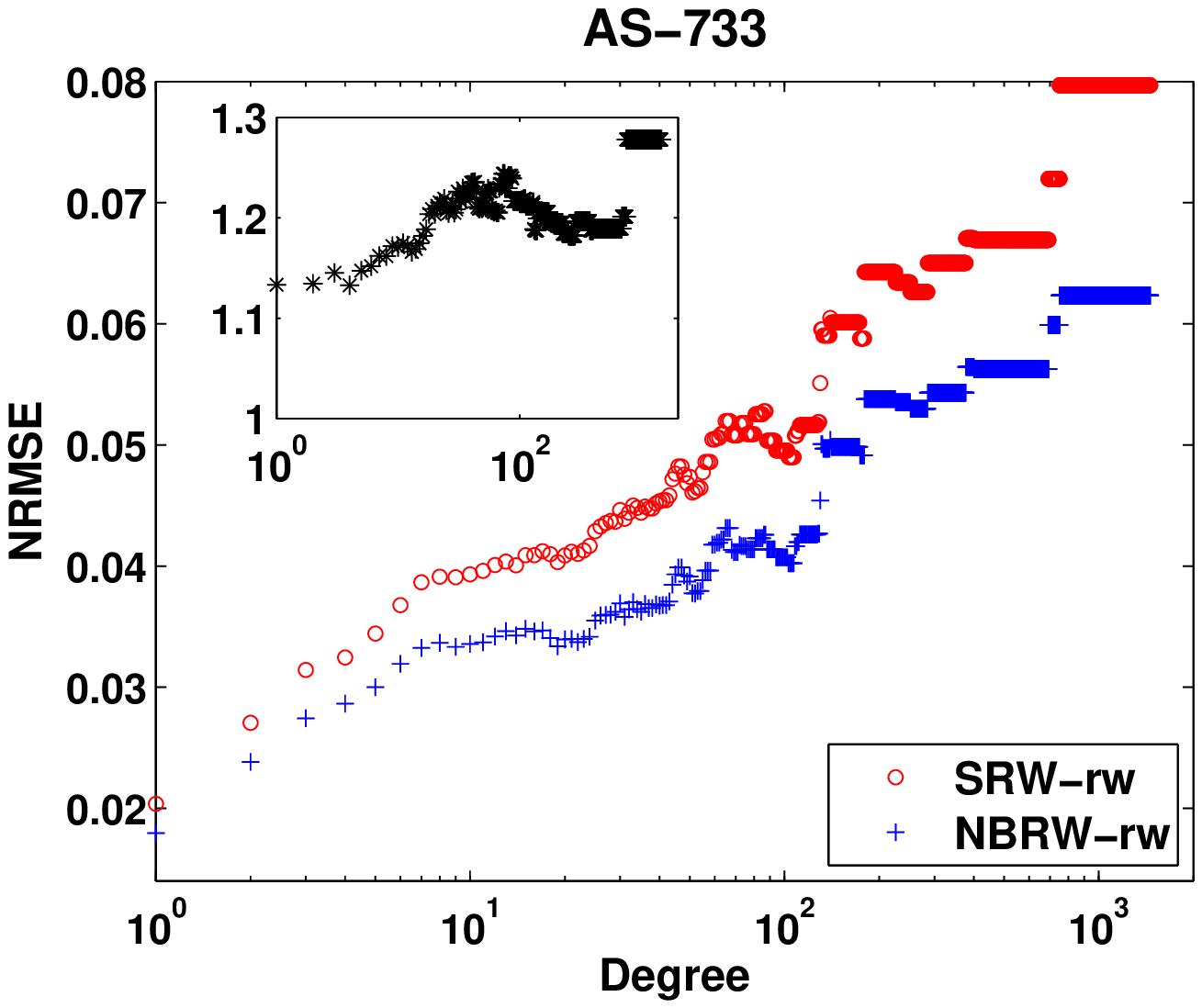}}
    \hspace{-0mm}\subfigure[MHRW vs. MHRW-DA]{\includegraphics[width=3in,height=2.3in]{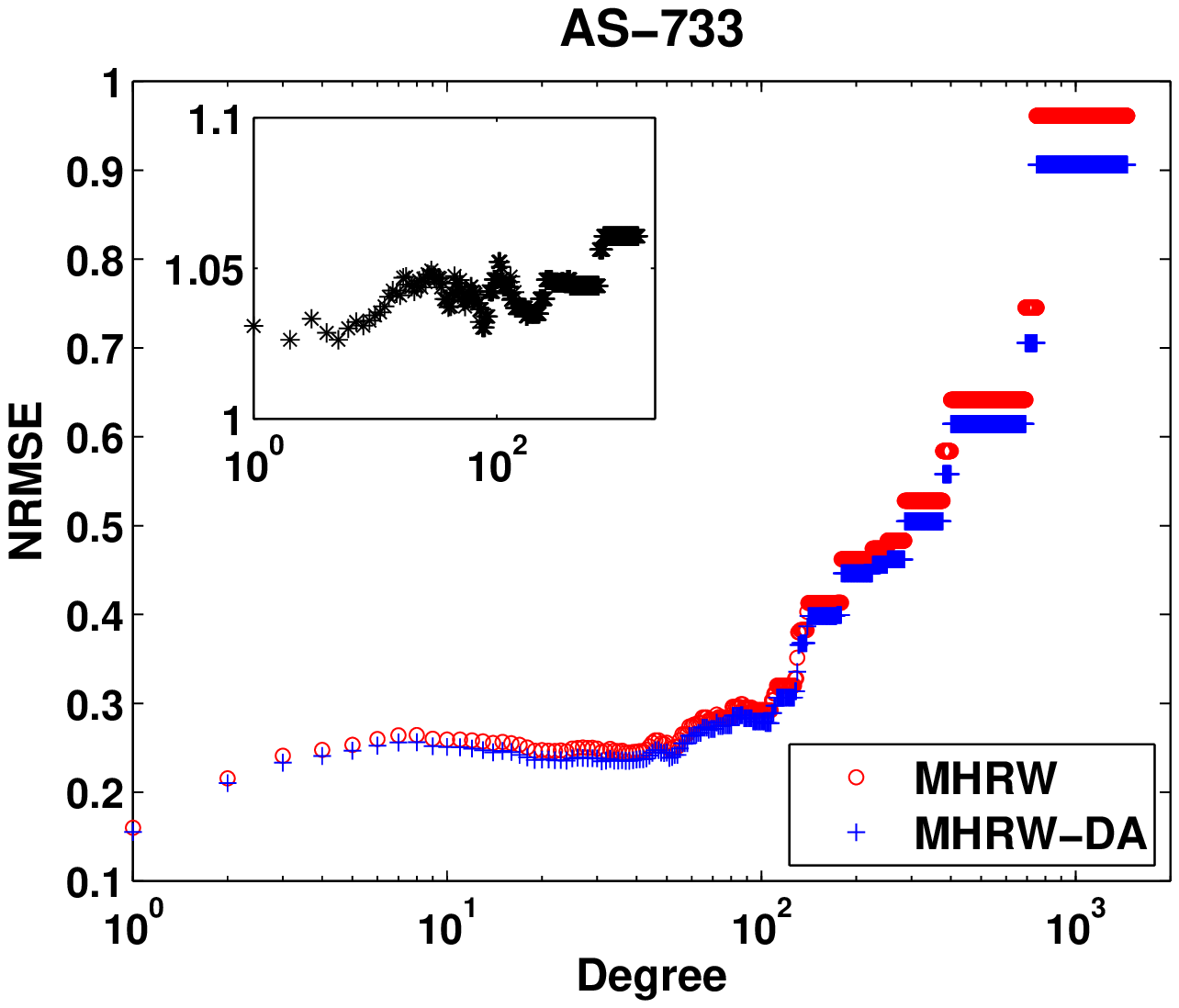}}
    \hspace{-0mm}
    \vspace{-0mm}
    \caption{AS-733 graph. NRMSE (per degree $d$) when estimating $\pr\{D_\G > d\}$ with $10^4$ samples; the insets show NRMSE ratio.} \label{fig:perdegree-ccdf-as}
    \vspace{-0mm}
\end{figure}
\begin{figure}[t!]
    \centering
    \vspace{-0mm}
    \hspace{-0mm}\subfigure[SRW-rw vs. NBRW-rw]{\includegraphics[width=3in,height=2.3in]{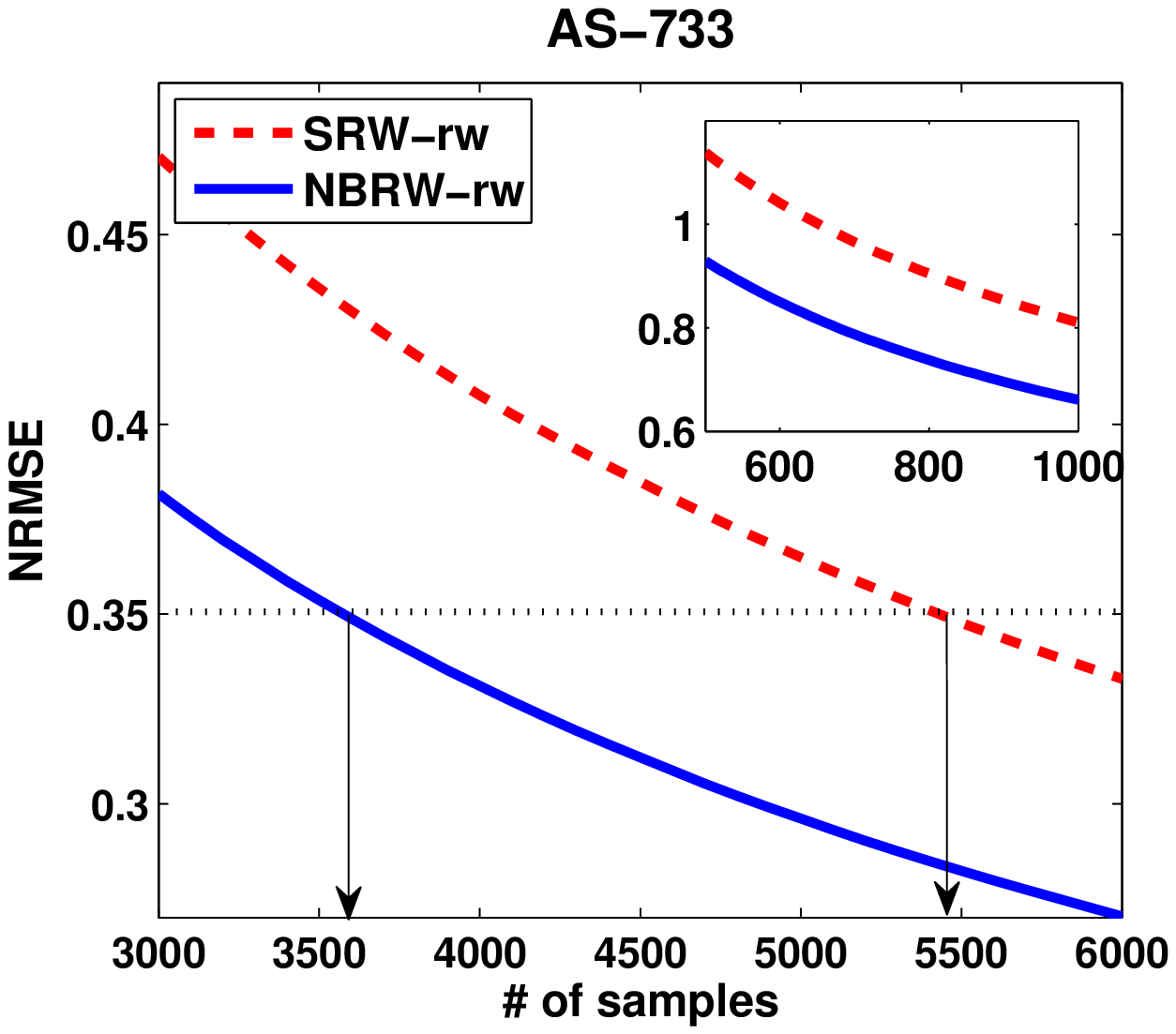}}
    \hspace{-0mm}\subfigure[MHRW vs. MHRW-DA]{\includegraphics[width=3in,height=2.3in]{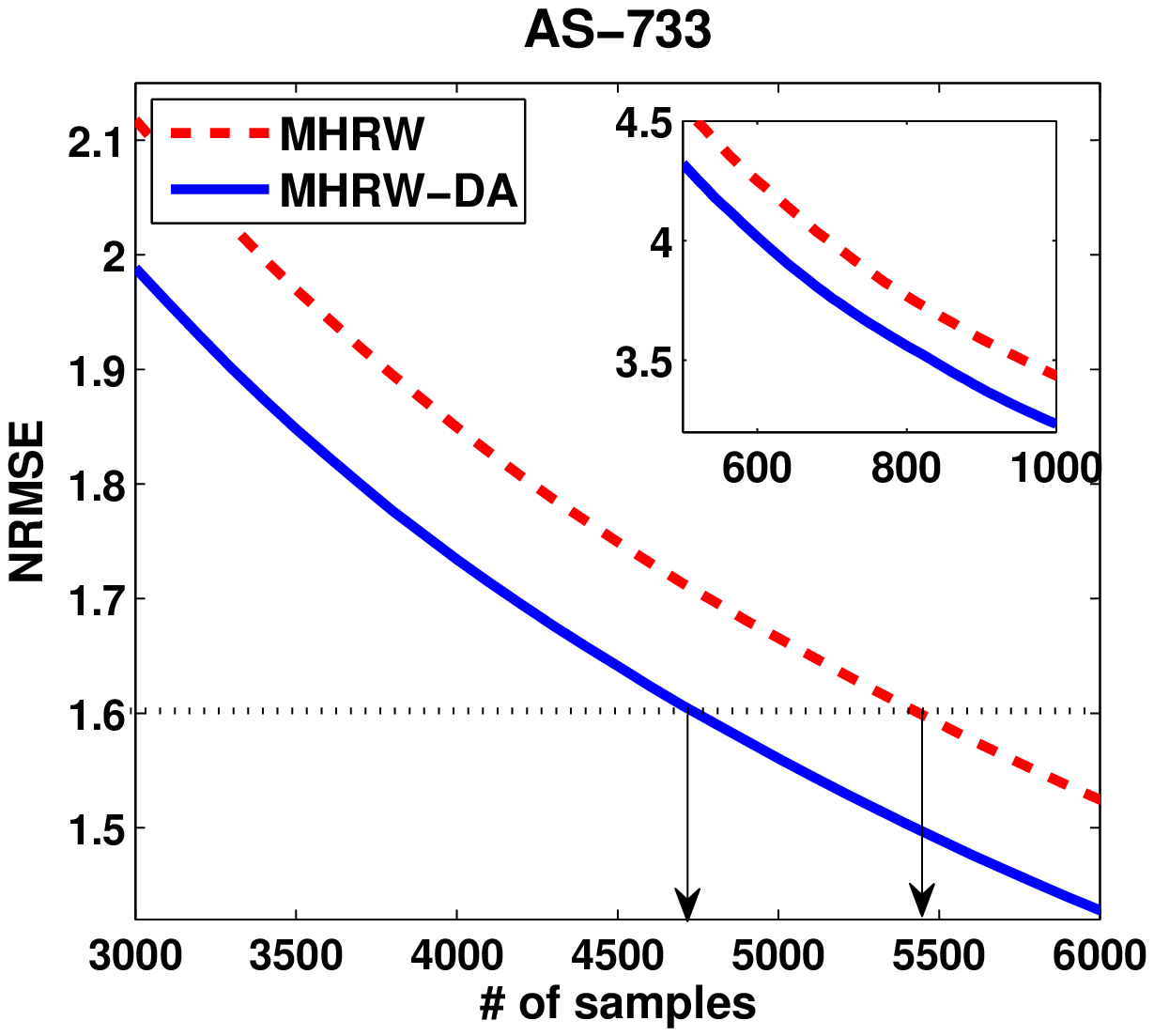}}
    \hspace{-0mm}
    \vspace{-0mm}
    \caption{AS-733 graph. NRMSE (averaged over all possible $d$) of the estimator of $\pr\{D_\G = d\}$, when each random walk does not start in the stationary regime.} \label{fig:nrmse-nonstationary-as}
    \vspace{-0mm}
\end{figure}

We first present the simulation results for AS-733 graph whose `actual' degree distribution is almost a `power-law' as depicted in Figure~\ref{fig:perdegree-pdf-as} (insets).  Figure~\ref{fig:nrmse-as} shows that NBRW-rw (resp.
MHRW-DA) outperforms SRW-rw (resp. MHRW) in terms
of the required number of samples (cost) to achieve the same level
of estimation error when estimating $\pr\{D_\G \!=\! d\}$, as expected from our theoretical results. Here, the
NBRW-rw (resp. MHRW-DA) brings out about 35\% (resp. 14\%) cost saving on average, when compared to the SRW-rw (resp. MHRW). In addition, we plot, in
Figure~\ref{fig:perdegree-pdf-as}, the NRMSE ratio of SRW-rw (resp. MHRW) to the case of NBRW-rw (resp. MHRW-DA) for every degree $d$ when estimating
$\pr\{D_\G \!=\! d\}$ with $10^4$ samples. It clearly shows the
improvement of our proposed methods for each degree $d$ (all data
points are above one). We also provide the NRMSE curve (with its ratio), in
Figure~\ref{fig:perdegree-ccdf-as}, for the comparison between  NBRW-rw (resp. MHRW-DA) and SRW-rw (resp. MHRW)
when estimating $\pr\{D_\G \!>\! d\}$ with
$10^4$ samples, which is again clearly consistent with our theoretical
findings. In addition, we conduct another simulation to see the impact of non-stationary start for each random walk on the sampling accuracy, for which an initial position of each SRW and NBRW is drawn from a uniform distribution, while the initial position for MHRW and MHRW-DA is chosen with a probability proportional to node degree.
Under this setting, we measure NRMSE of the estimator of $\pr\{D_\G \!=\! d\}$, and observe that NBRW-rw and MHRW-DA still outperform SRW-rw and MHRW, respectively, as shown in Figure~\ref{fig:nrmse-nonstationary-as}. Note that there is not much difference between the stationary start and non-stationary start cases. (See Figures~\ref{fig:nrmse-as} and~\ref{fig:nrmse-nonstationary-as}.)

\begin{figure}[t!]
    \centering
    \vspace{-0mm}
    \hspace{-0mm}\subfigure[SRW-rw vs. NBRW-rw]{\includegraphics[width=3in,height=2.3in]{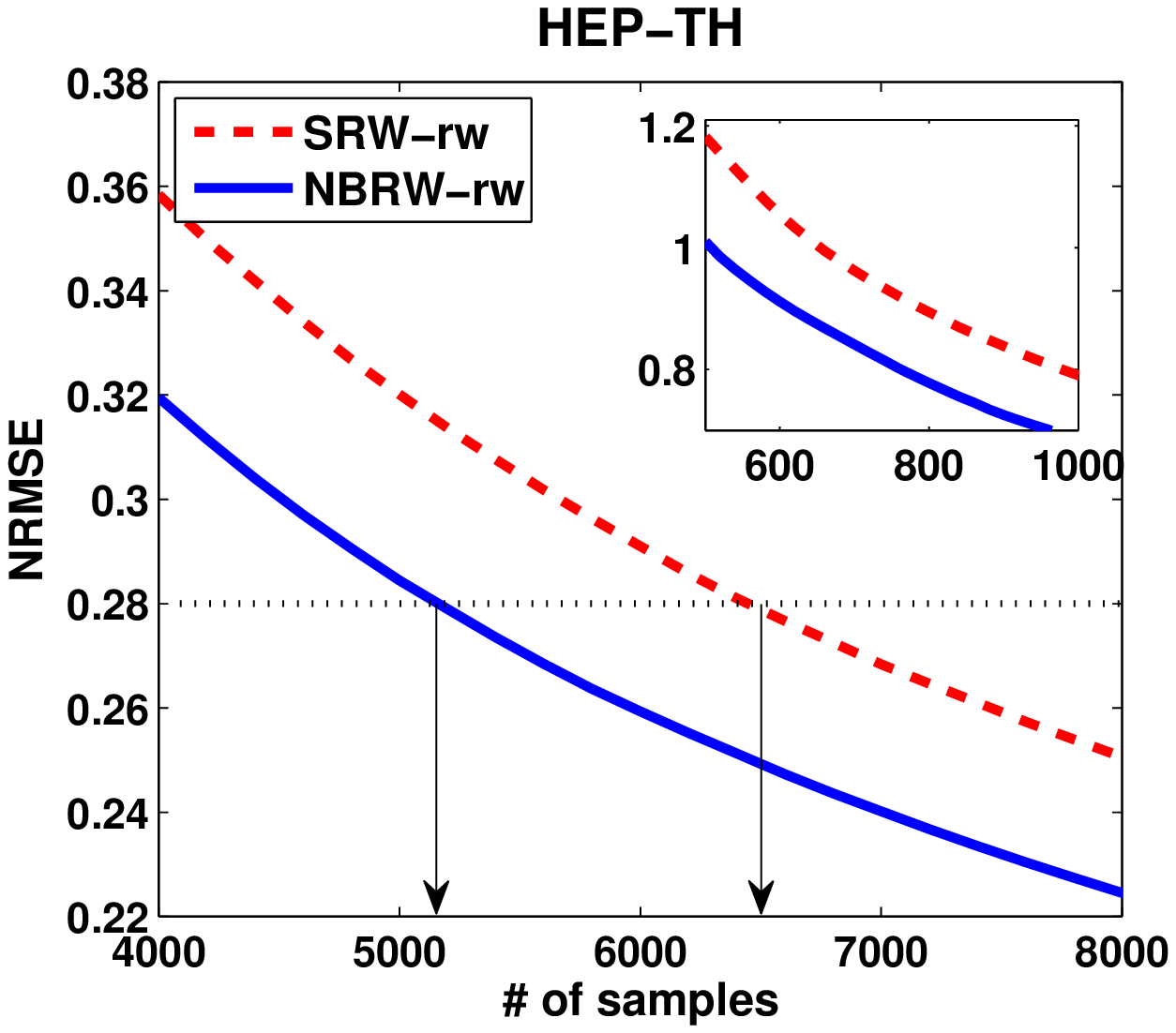}}
    \hspace{-0mm}\subfigure[MHRW vs. MHRW-DA]{\includegraphics[width=3in,height=2.3in]{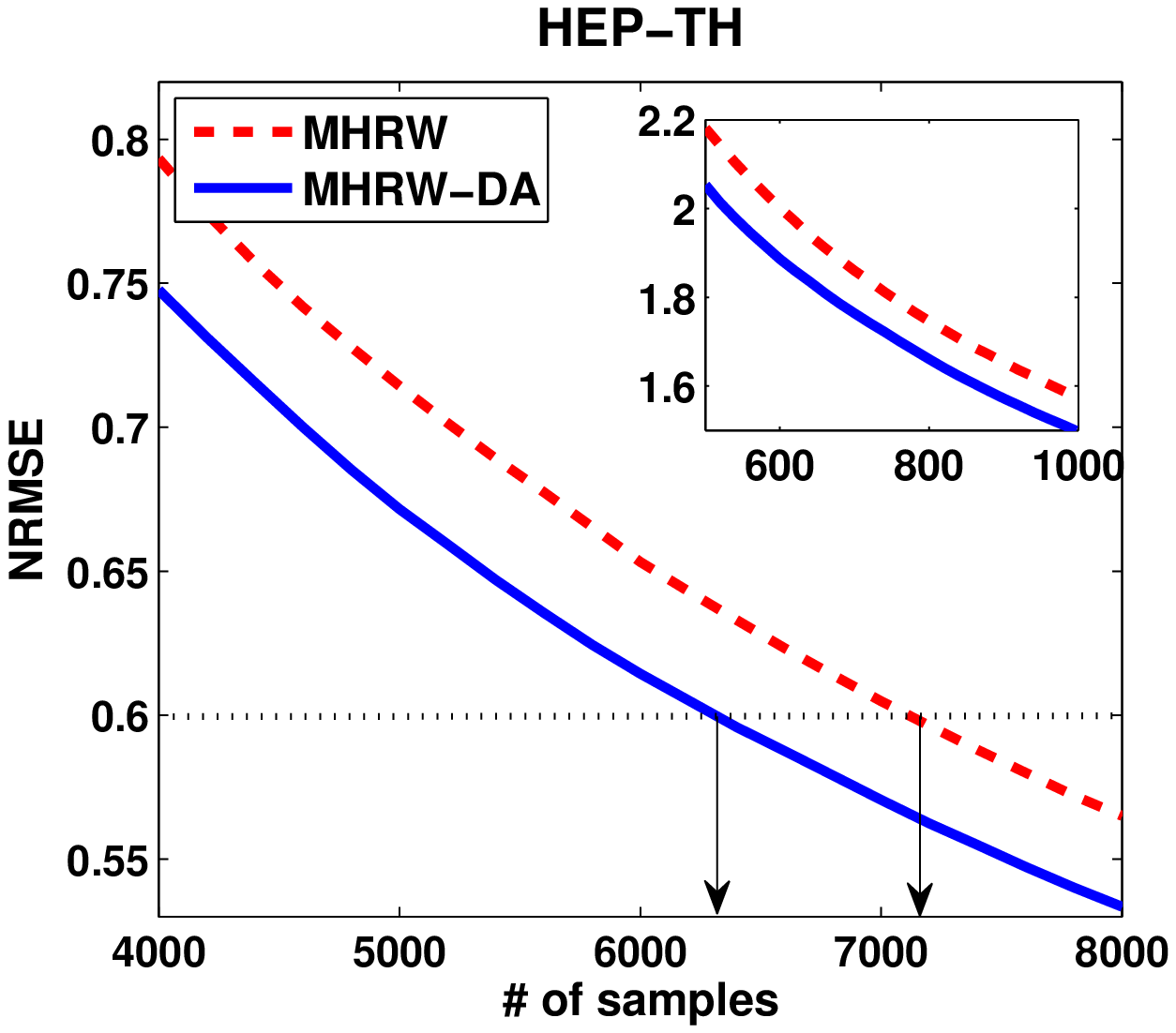}}
    \hspace{-0mm}
    \vspace{-0mm}
    \caption{HEP-TH graph. NRMSE (averaged over all possible $d$) of the estimator of $\pr\{D_\G \!=\! d\}$ with different number of samples; the insets are for smaller number of samples.} \label{fig:nrmse-hep}
    \vspace{-0mm}
\end{figure}
\begin{figure}[t!]
    \centering
    \vspace{-0mm}
    \hspace{-0mm}\subfigure[SRW-rw vs. NBRW-rw]{\includegraphics[width=3in,height=2.3in]{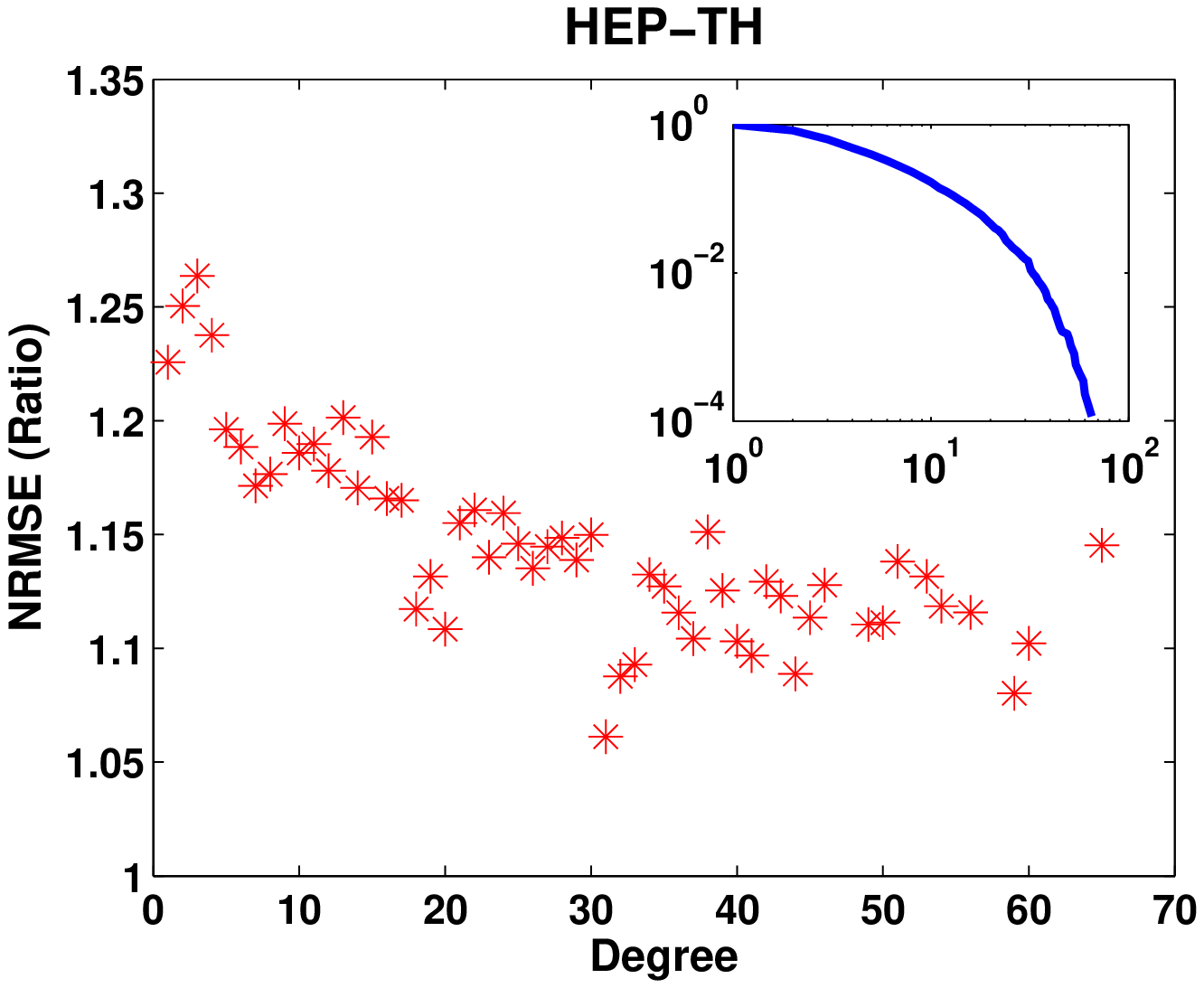}}
    \hspace{-0mm}\subfigure[MHRW vs. MHRW-DA]{\includegraphics[width=3in,height=2.3in]{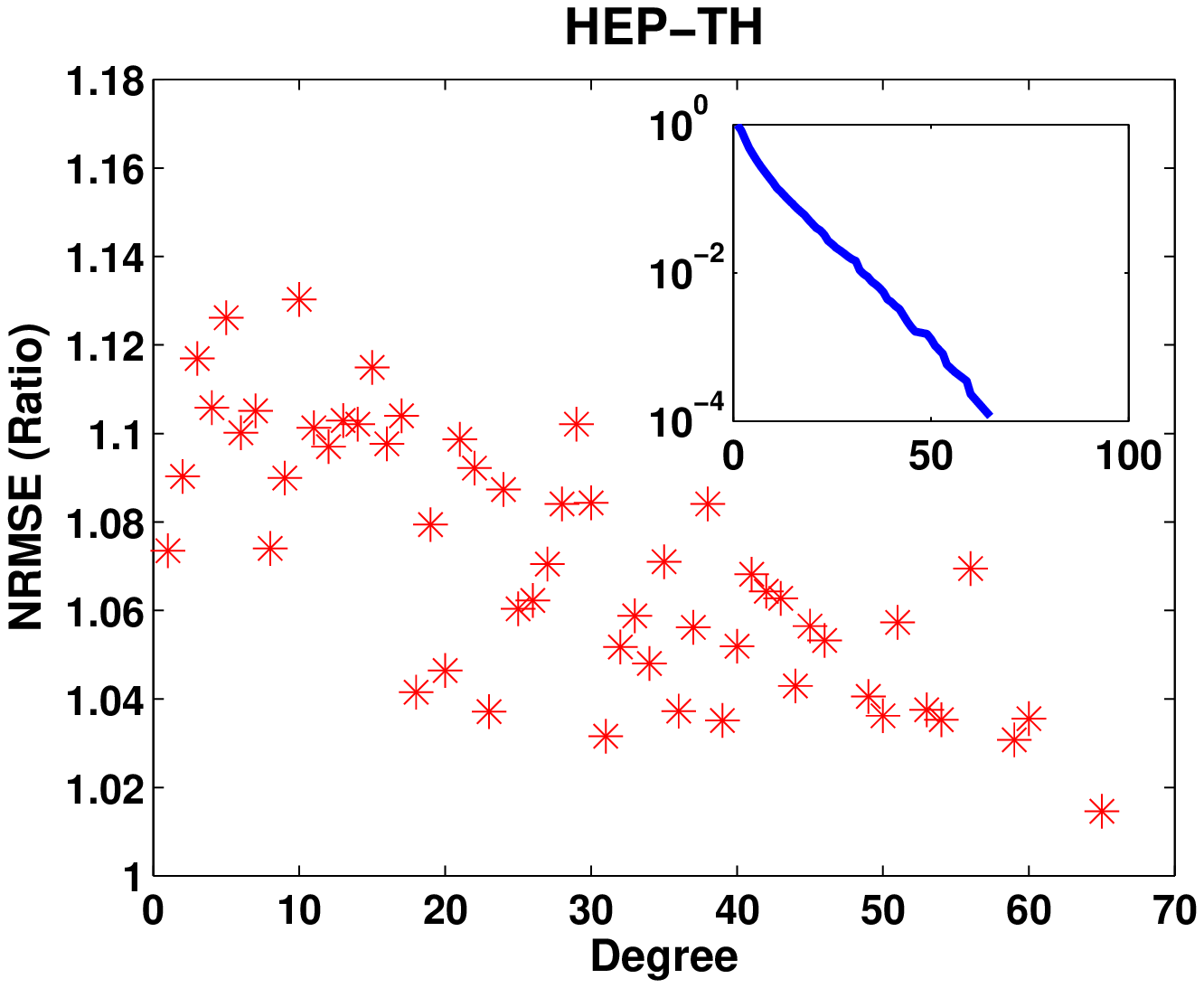}}
    \hspace{-0mm}
    \vspace{-0mm}
    \caption{HEP-TH graph. NRMSE ratio (per degree $d$) for the estimation of $\pr\{D_\G \!=\! d\}$ with $10^4$ samples; the insets represent the `actual' degree distribution (ccdf) in (a) log-log scale, (b) semi-log scale.} \label{fig:perdegree-pdf-hep}
    \vspace{-0mm}
\end{figure}
\begin{figure}[t!]
    \centering
    \vspace{-0mm}
    \hspace{-0mm}\subfigure[SRW-rw vs. NBRW-rw]{\includegraphics[width=3in,height=2.3in]{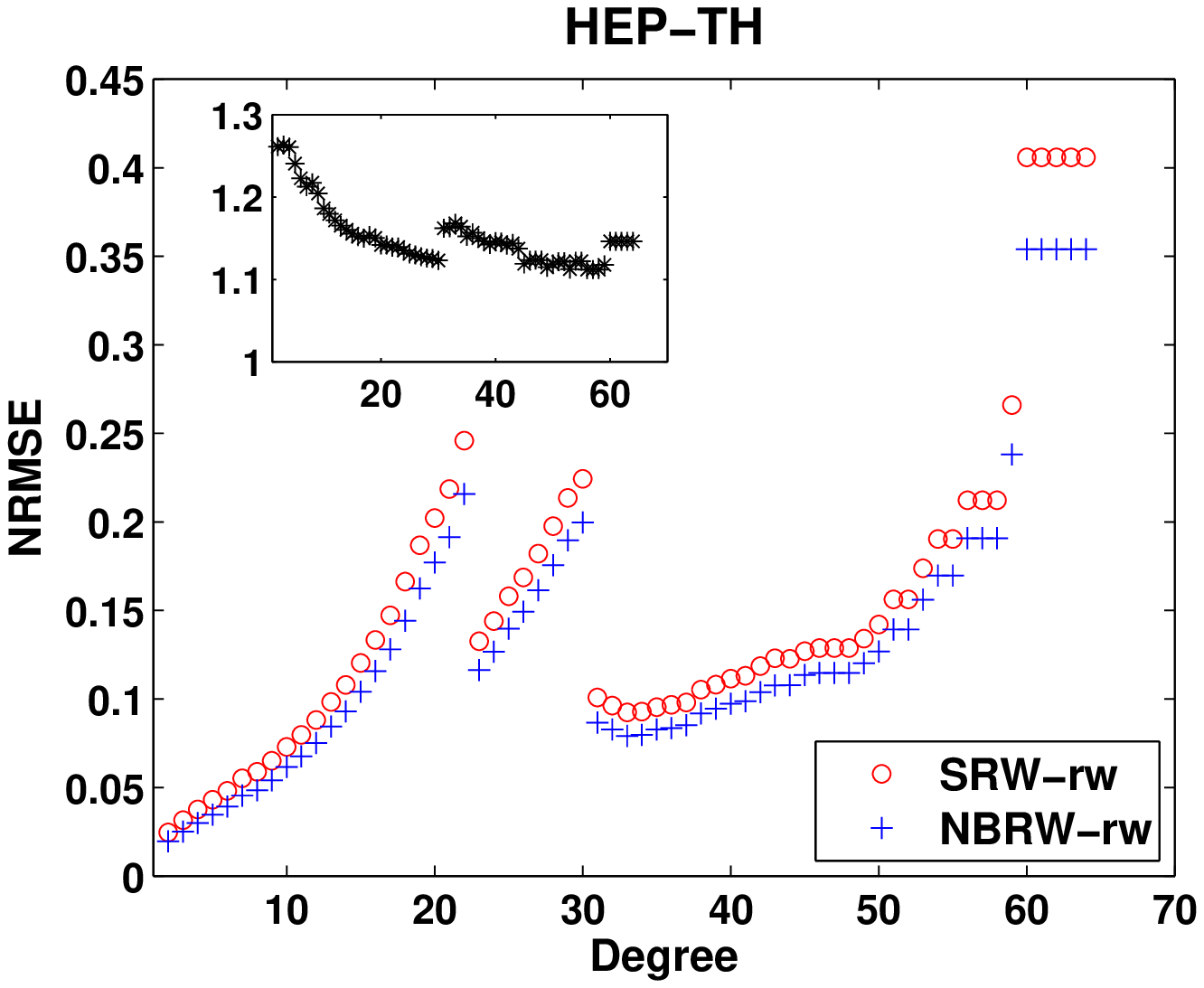}}
    \hspace{-0mm}\subfigure[MHRW vs. MHRW-DA]{\includegraphics[width=3in,height=2.3in]{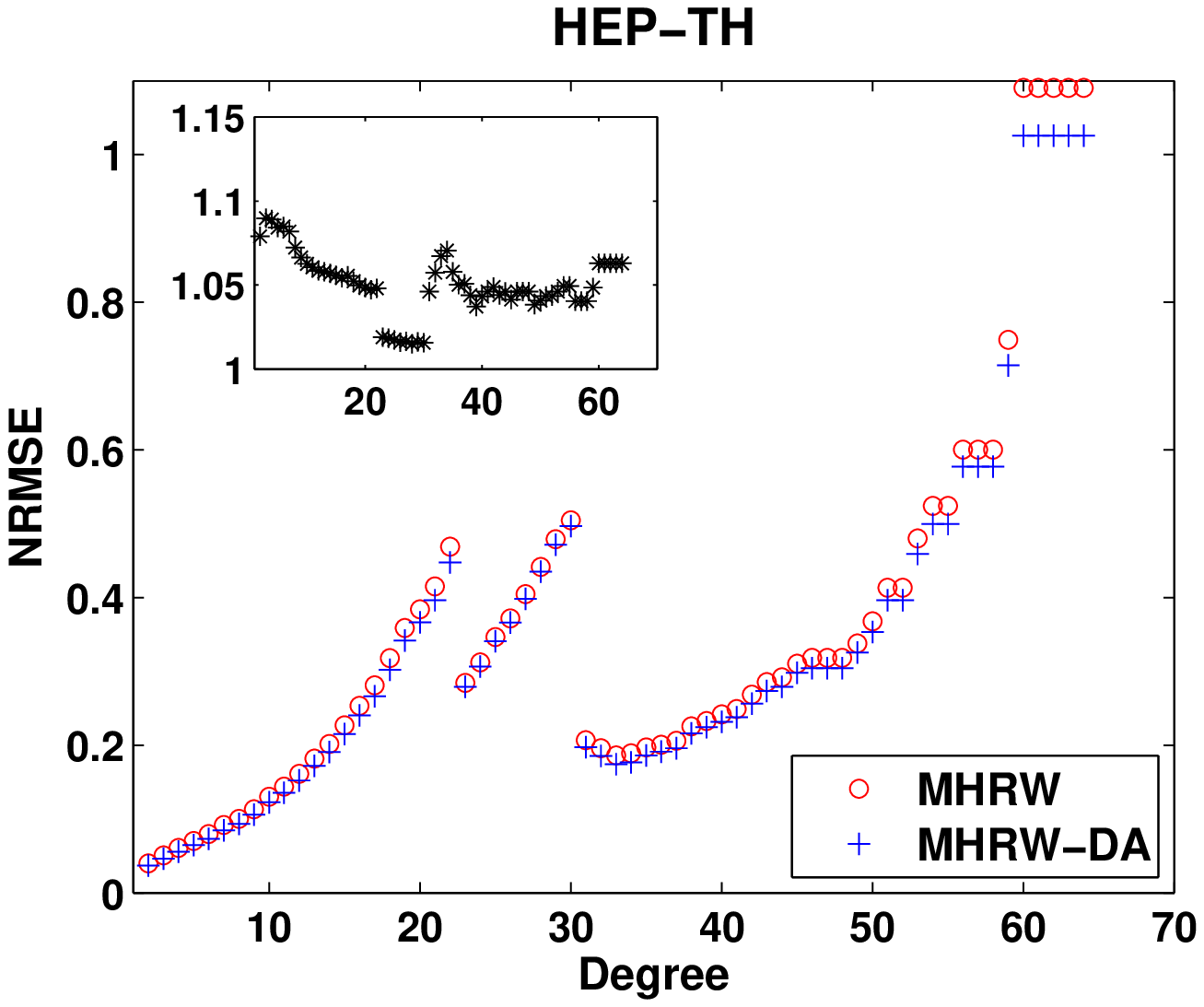}}
    \hspace{-0mm}
    \vspace{-0mm}
    \caption{HEP-TH graph. NRMSE (per degree $d$) when estimating $\pr\{D_\G \!>\! d\}$ with $10^4$ samples; the insets show NRMSE ratio.} \label{fig:perdegree-ccdf-hep}
    \vspace{-0mm}
\end{figure}
\begin{figure}[t!]
    \centering
    \vspace{-0mm}
    \hspace{-0mm}\subfigure[SRW-rw vs. NBRW-rw]{\includegraphics[width=3in,height=2.3in]{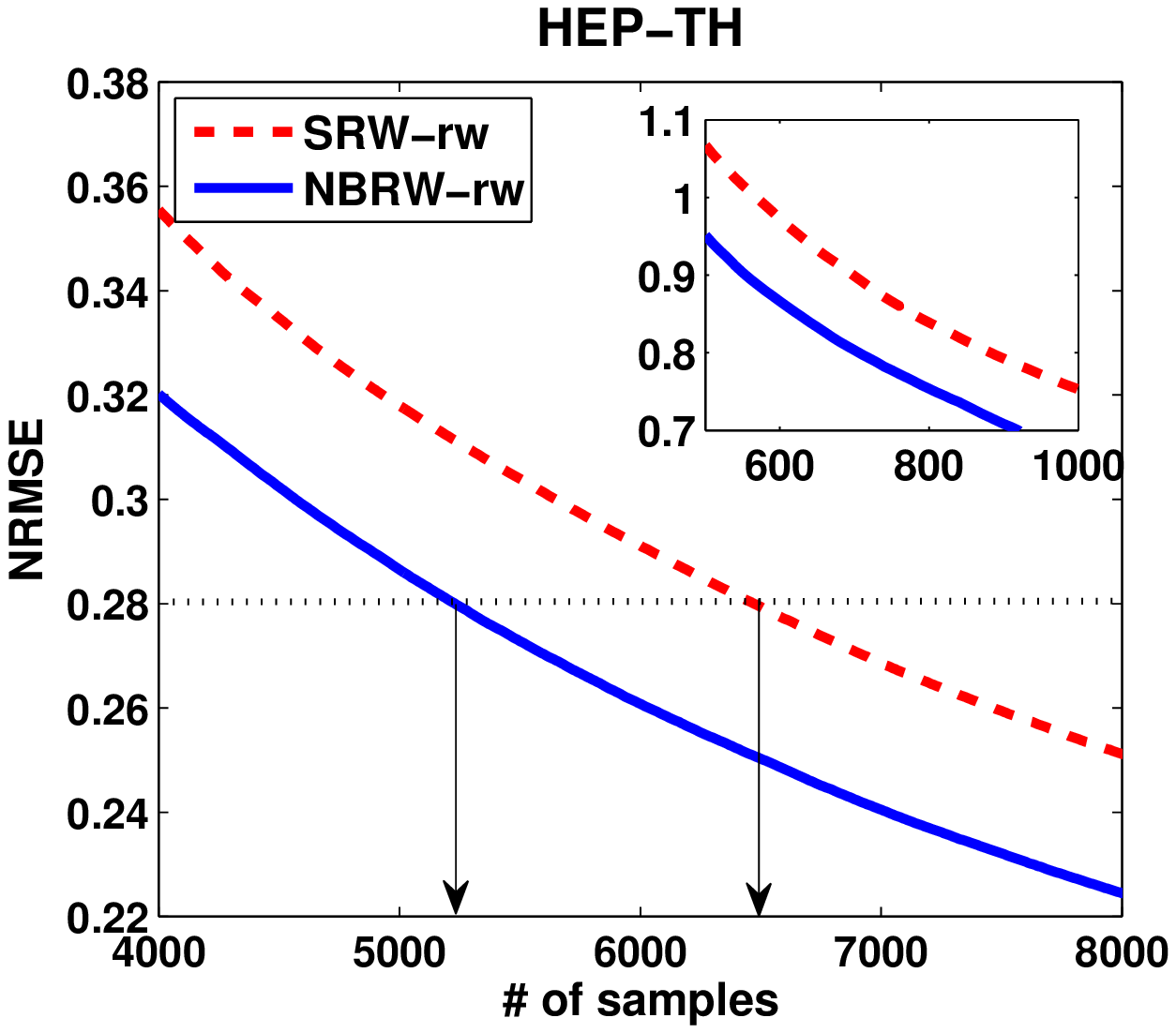}}
    \hspace{-0mm}\subfigure[MHRW vs. MHRW-DA]{\includegraphics[width=3in,height=2.3in]{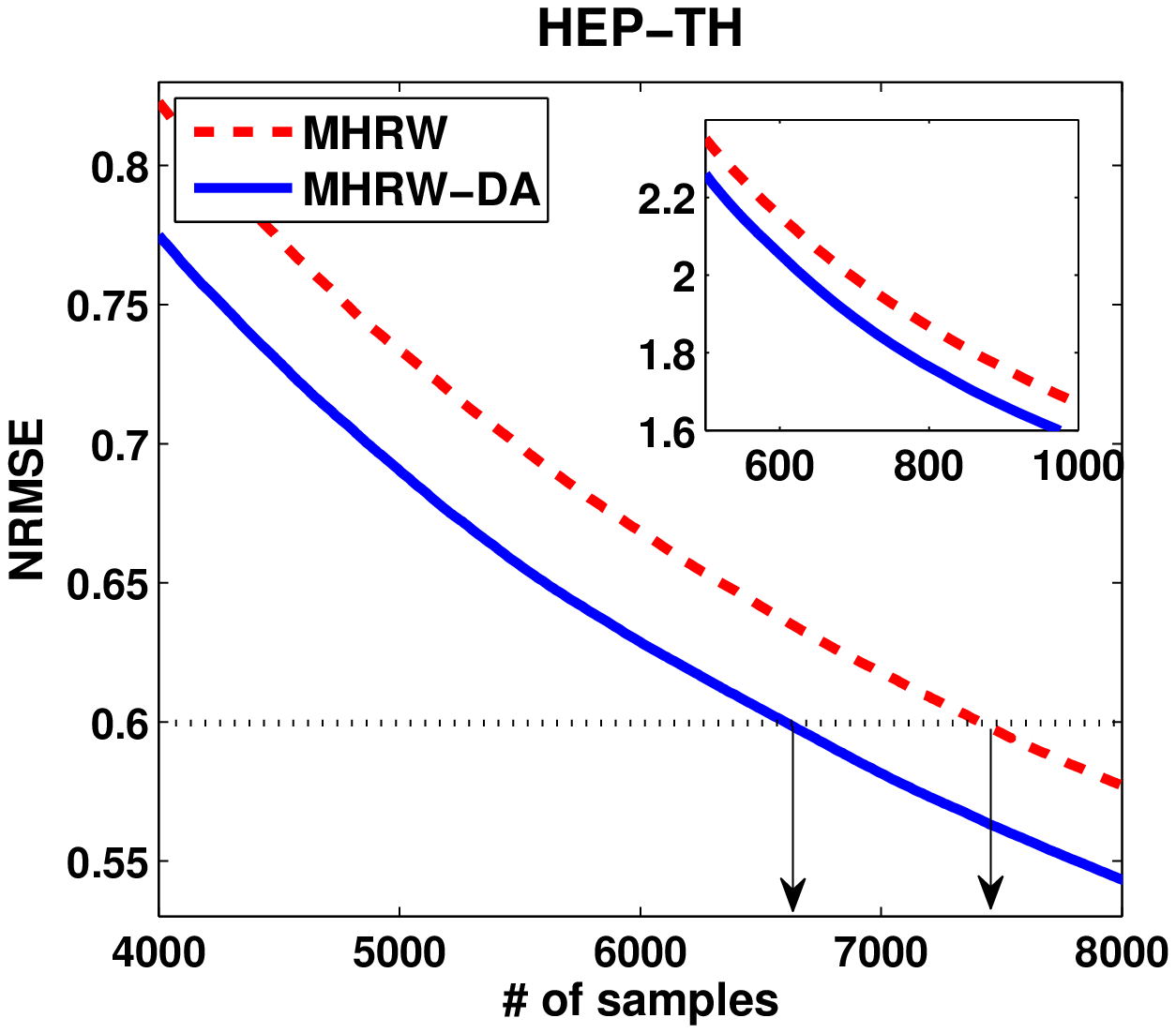}}
    \hspace{-0mm}
    \vspace{-0mm}
    \caption{HEP-TH graph. NRMSE (averaged over all possible $d$) of the estimator of $\pr\{D_\G \!=\! d\}$, when each random walk does not start in the steady-state.} \label{fig:nrmse-nonstationary-hep}
    \vspace{-0mm}
\end{figure}

We next provide the simulation results for HEP-TH graph whose actual degree distribution is close to exponential as depicted in Figure~\ref{fig:perdegree-pdf-hep} (insets). As before, Figure~\ref{fig:nrmse-hep} demonstrates that NBRW-rw and MHRW-DA surpass SRW-rw and MHRW for the estimation of $\pr\{D_\G \!=\! d\}$, respectively. Specifically, the NBRW-rw (resp. MHRW-DA) saves, on average, about 22\% (resp. 12\%) of the required number of samples to attain the same level of estimation accuracy, which compared to the SRW-rw (resp. MHRW). Also, Figure~\ref{fig:perdegree-pdf-hep} shows the NRMSE ratio of SRW-rw (resp. MHRW) to the case of NBRW-rw (resp. MHRW-DA) for every degree $d$ for the estimation of
$\pr\{D_\G \!=\! d\}$ with $10^4$ samples, while Figure~\ref{fig:perdegree-ccdf-hep} depicts the NRMSE curve (with its ratio) when estimating $\pr\{D_\G \!>\! d\}$ with
$10^4$ samples. Both results are again in good agreement with our theoretical results. Moreover, after repeating the same experiment for the non-stationary start as above, we observe that the improvement from NBRW-rw and MHRW-DA remains preserved, as shown in Figure~\ref{fig:nrmse-nonstationary-hep}.

\begin{figure}[t!]
    \centering
    \vspace{-0mm}
    \hspace{-0mm}\subfigure[SRW-rw vs. NBRW-rw]{\includegraphics[width=3in,height=2.3in]{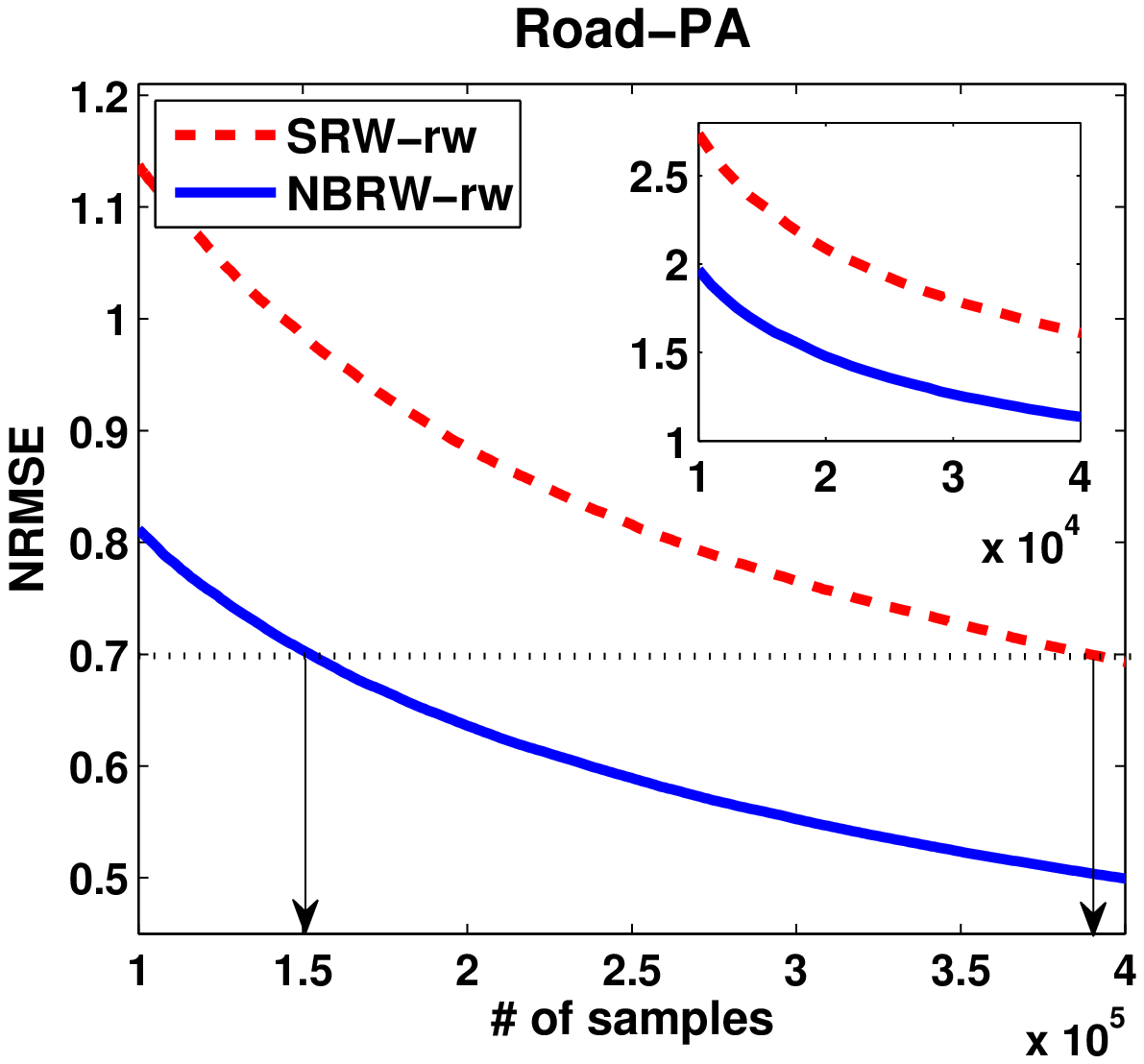}}
    \hspace{-0mm}\subfigure[MHRW vs. MHRW-DA]{\includegraphics[width=3in,height=2.3in]{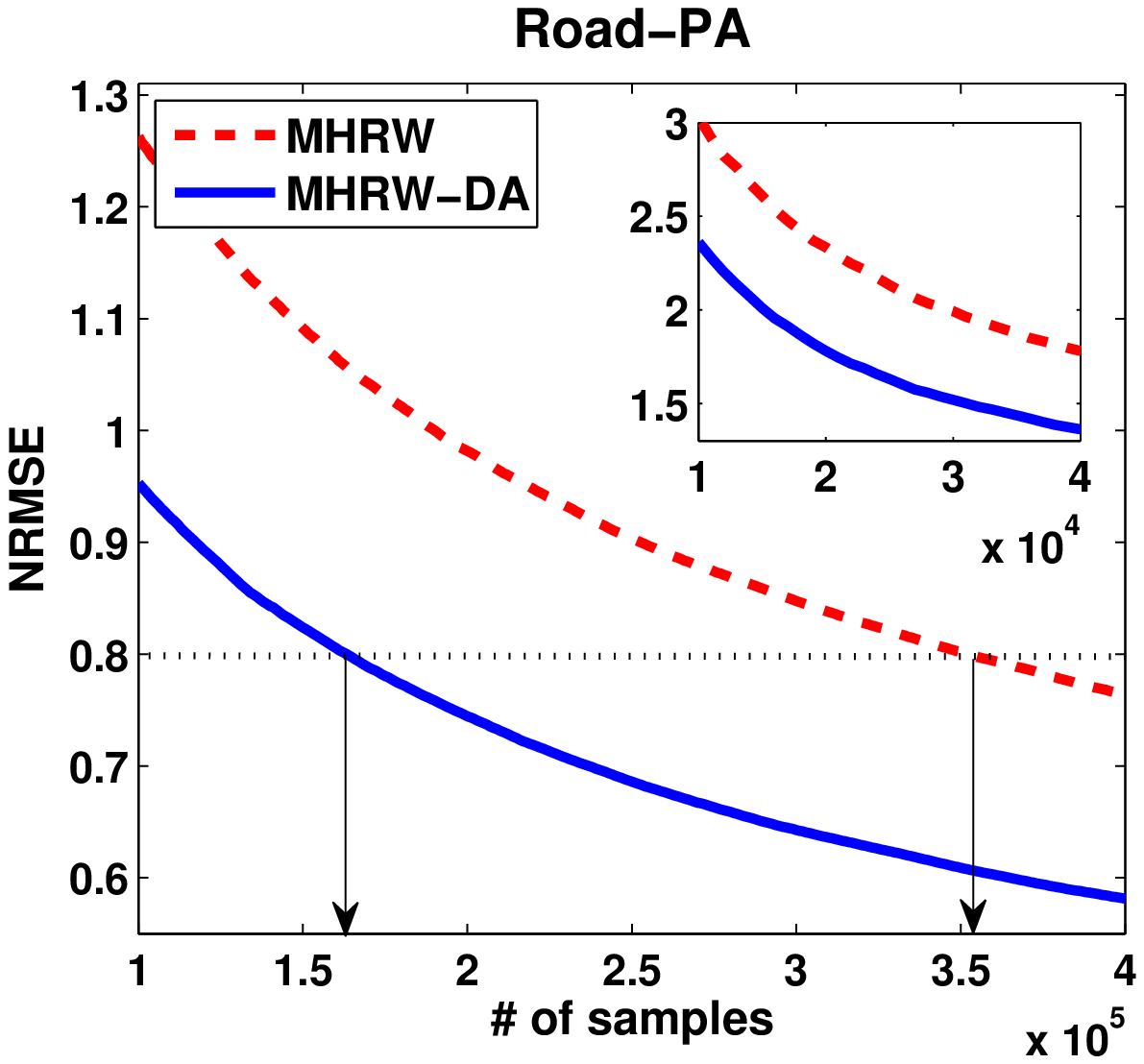}}
    \hspace{-0mm}
    \vspace{-0mm}
    \caption{Road-PA graph. NRMSE (averaged over all possible $d$) of the estimator of $\pr\{D_\G \!=\! d\}$, while varying the number of samples; the insets are for smaller number of samples.} \label{fig:nrmse-road}
    \vspace{-0mm}
\end{figure}

\begin{figure}[t!]
    \centering
    \vspace{-0mm}
    \hspace{-0mm}\subfigure[SRW-rw vs. NBRW-rw]{\includegraphics[width=3in,height=2.3in]{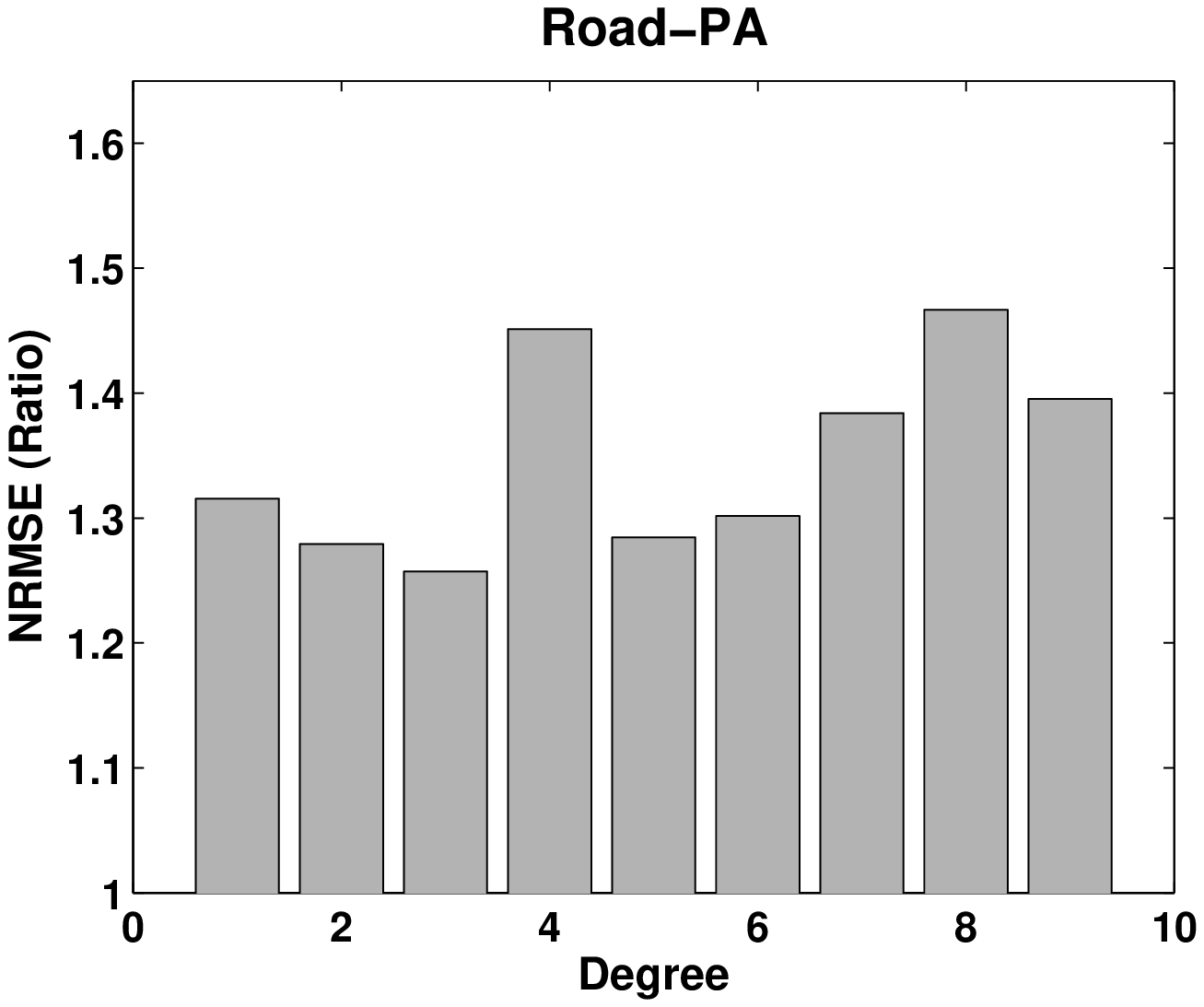}}
    \hspace{-0mm}\subfigure[MHRW vs. MHRW-DA]{\includegraphics[width=3in,height=2.3in]{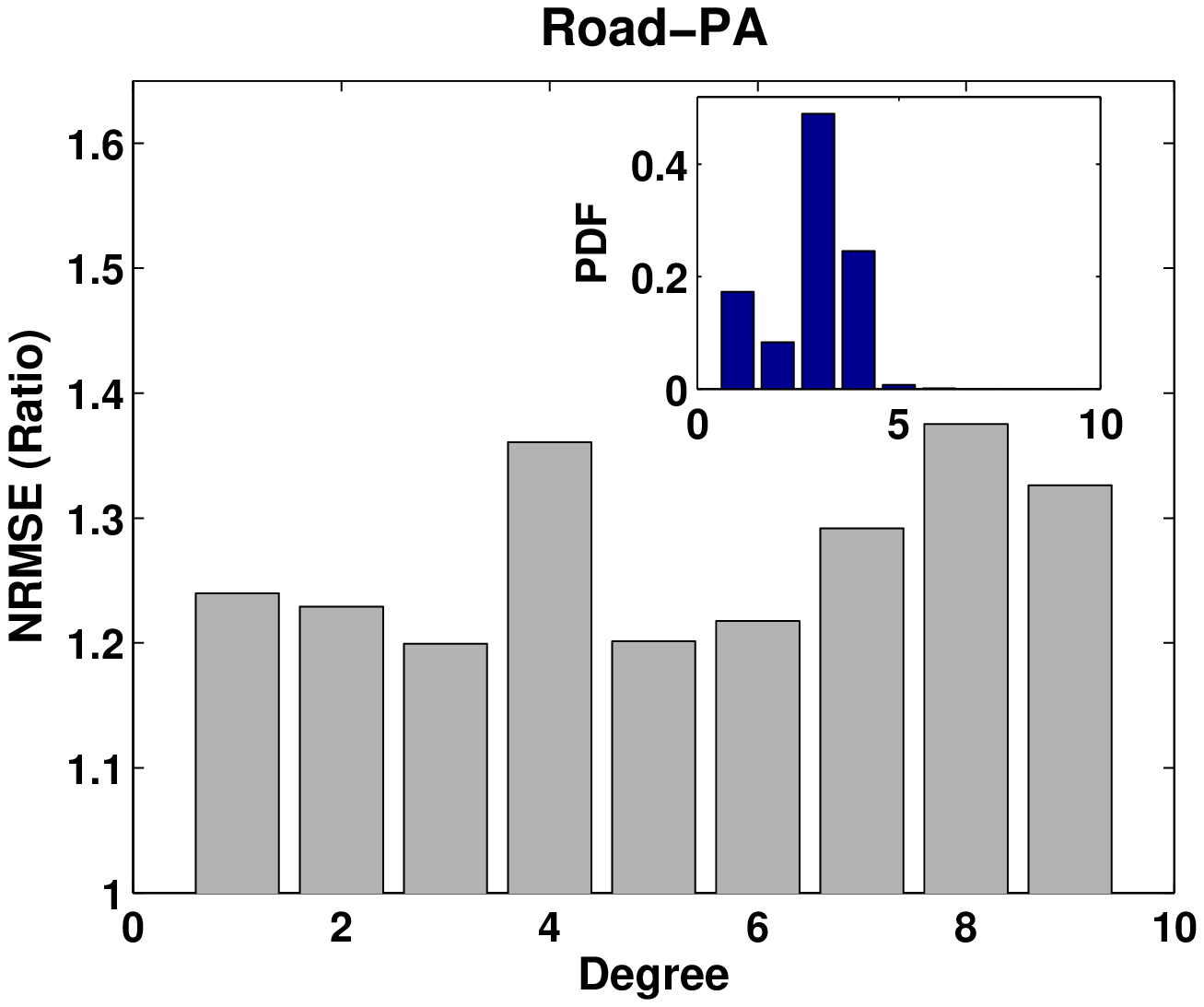}}
    \hspace{-0mm}
    \vspace{-0mm}
    \caption{Road-PA graph. NRMSE ratio (per degree $d$) when estimating $\pr\{D_\G \!=\! d\}$ with $5\cdot10^5$ samples; the inset represents the `actual' degree distribution (pdf).} \label{fig:perdegree-pdf-road}
    \vspace{-0mm}
\end{figure}

\begin{figure}[t!]
    \centering
    \vspace{-0mm}
    \hspace{-0mm}\subfigure[SRW-rw vs. NBRW-rw]{\includegraphics[width=3in,height=2.3in]{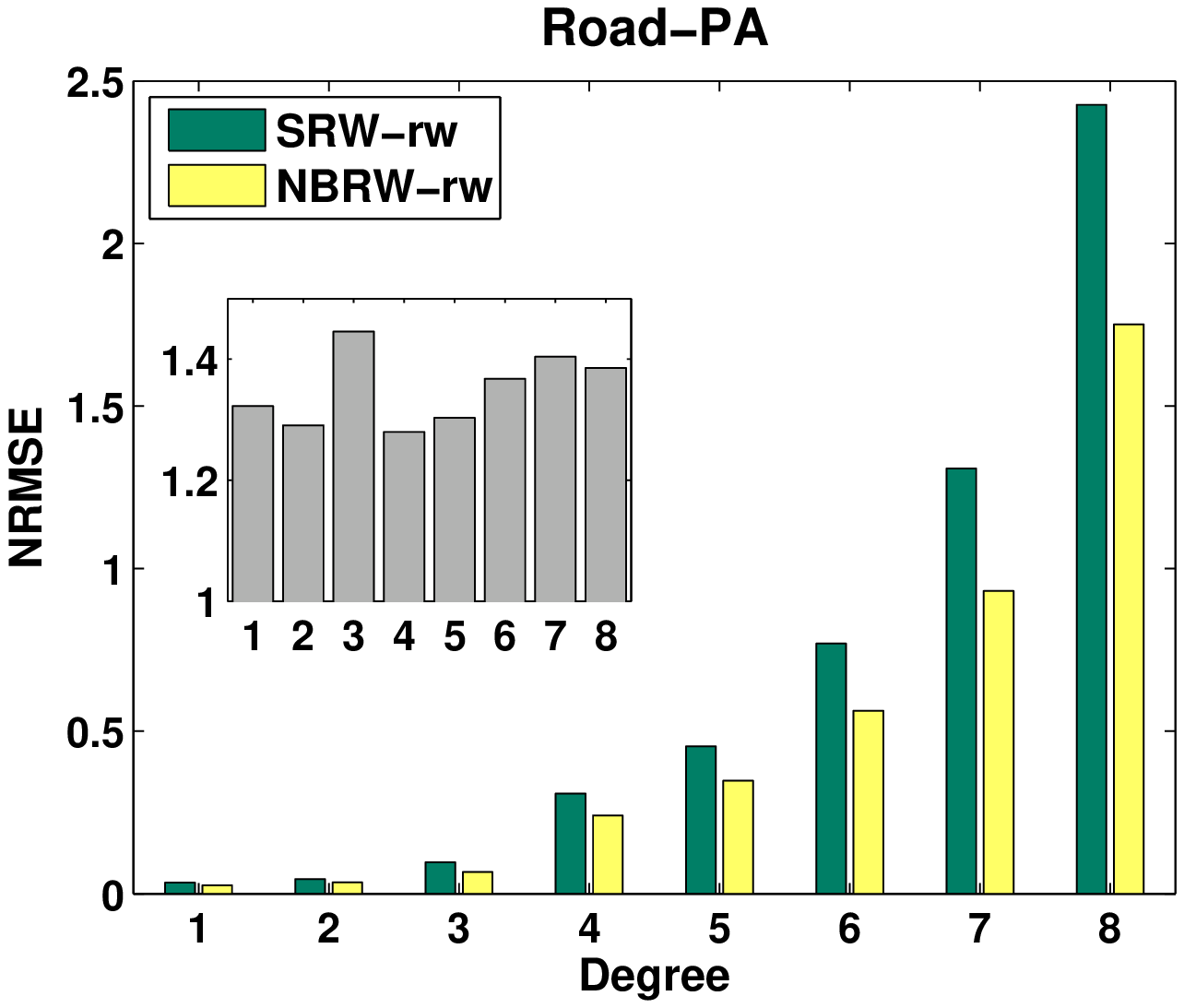}}
    \hspace{-0mm}\subfigure[MHRW vs. MHRW-DA]{\includegraphics[width=3in,height=2.3in]{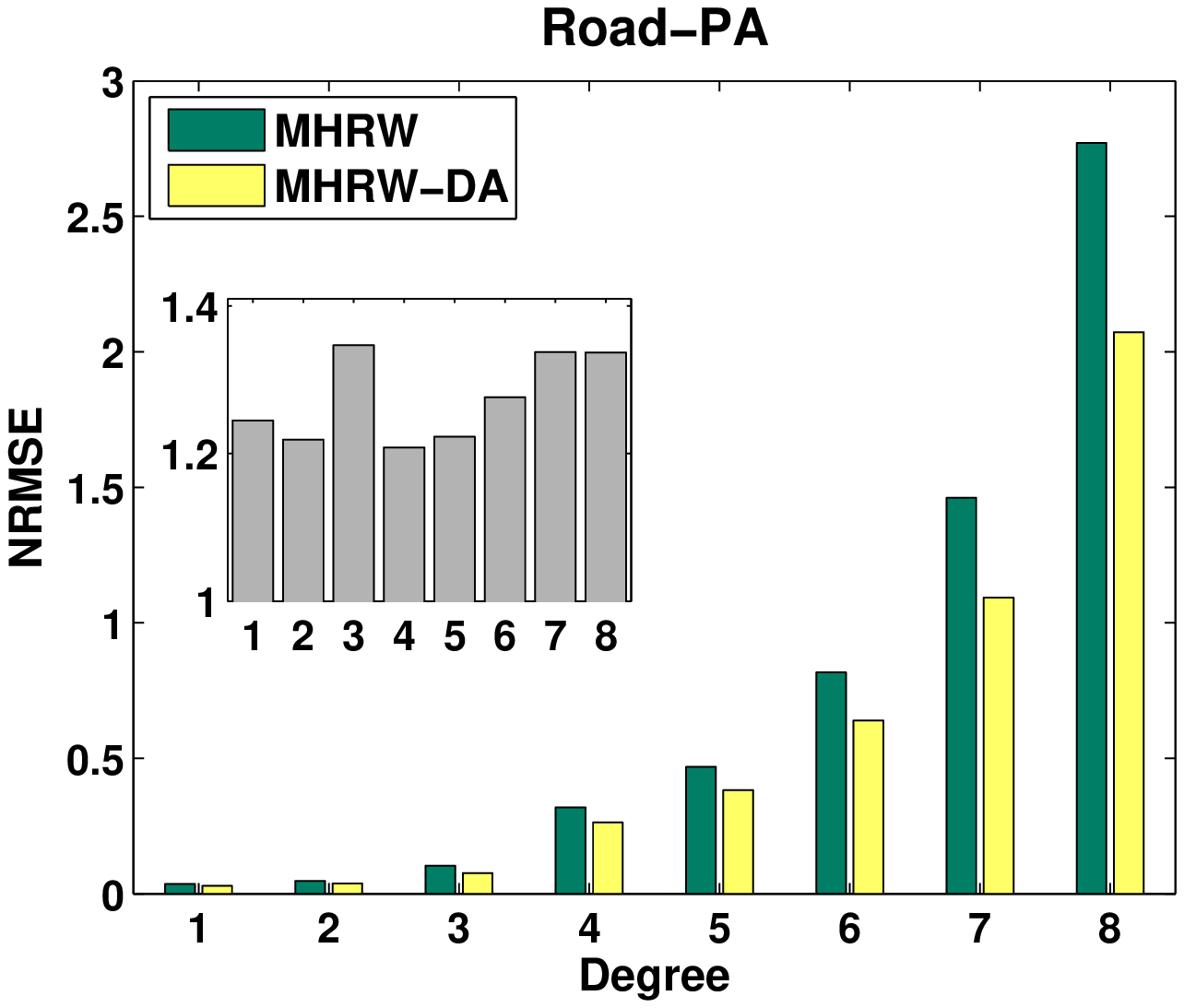}}
    \hspace{-0mm}
    \vspace{-0mm}
    \caption{Road-PA graph. NRMSE (per degree $d$) for the estimation of $\pr\{D_\G \!>\! d\}$ with $5\cdot10^5$ samples; the insets show NRMSE ratio.} \label{fig:perdegree-ccdf-road}
    \vspace{-0mm}
\end{figure}

\begin{figure}[t!]
    \centering
    \vspace{-0mm}
    \hspace{-0mm}\subfigure[SRW-rw vs. NBRW-rw]{\includegraphics[width=3in,height=2.3in]{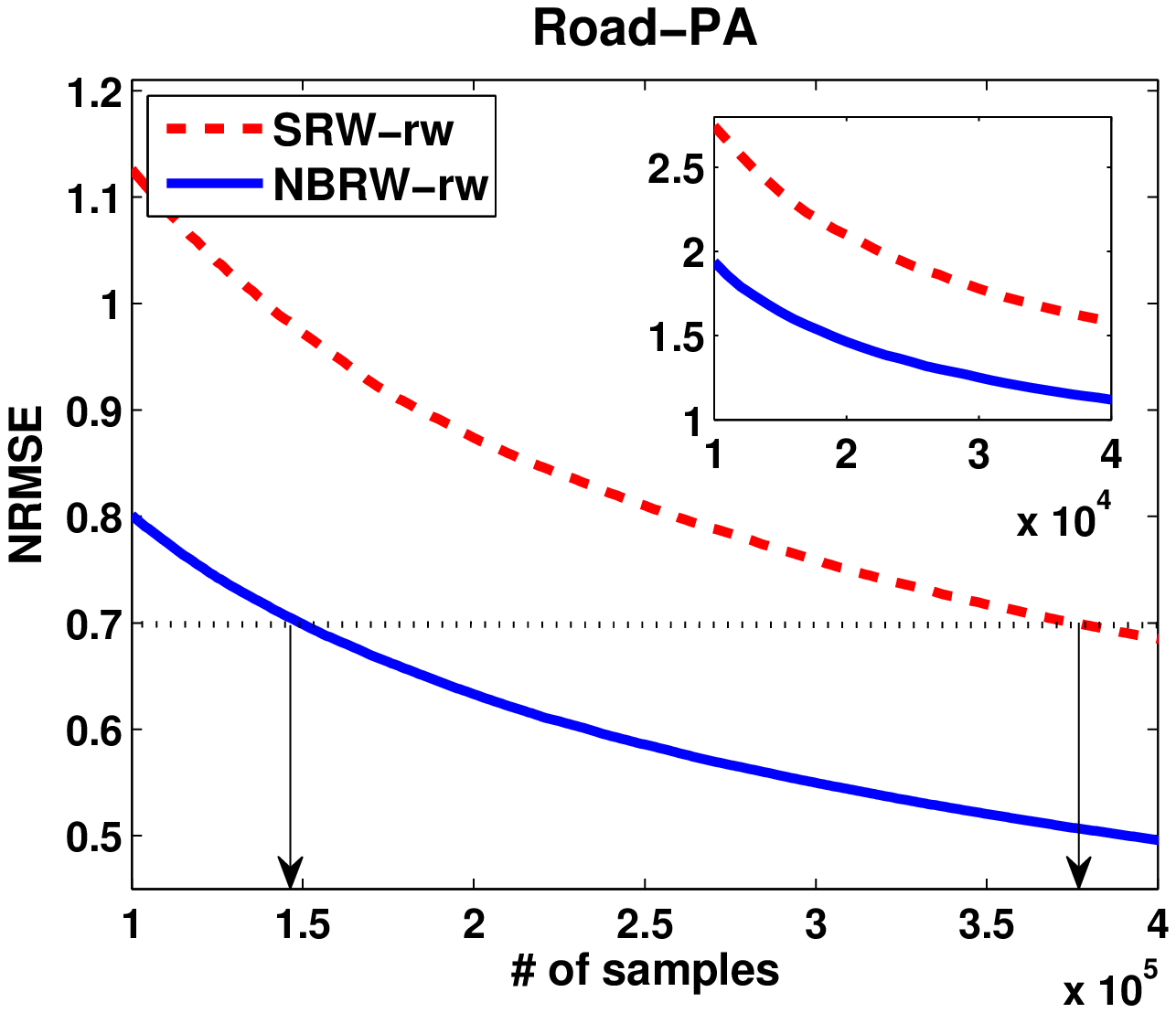}}
    \hspace{-0mm}\subfigure[MHRW vs. MHRW-DA]{\includegraphics[width=3in,height=2.3in]{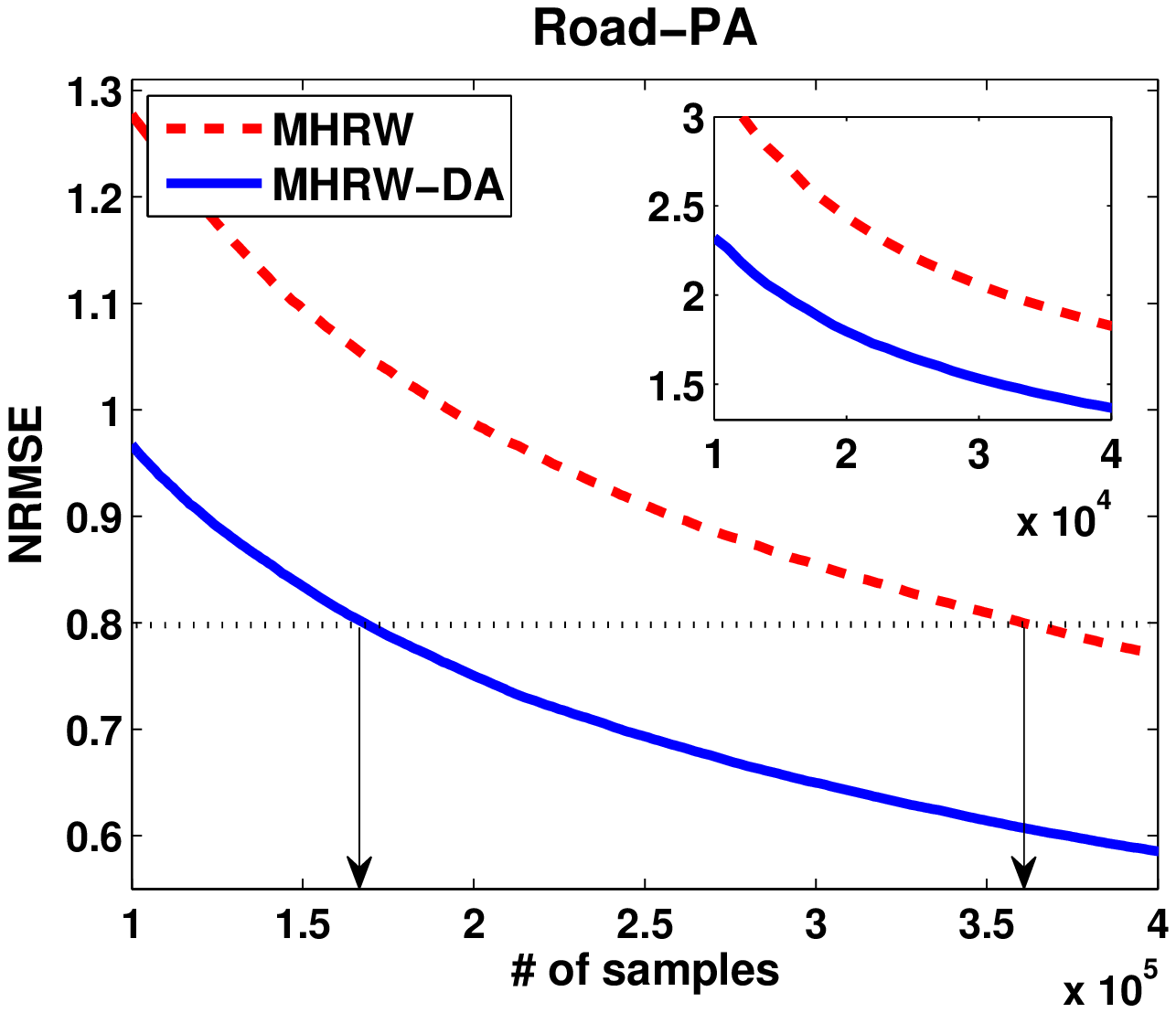}}
    \hspace{-0mm}
    \vspace{-0mm}
    \caption{Road-PA graph. NRMSE (averaged over all possible $d$) of the estimator of $\pr\{D_\G \!=\! d\}$, when an initial position of each random walk is not drawn from its stationary distribution.} \label{fig:nrmse-nonstationary-road}
    \vspace{-0mm}
\end{figure}

We also present the simulation results for Road-PA graph in which every node has small degree, ranging from 1 to 9, and the actual degree distribution (pdf) is given in Figure~\ref{fig:perdegree-pdf-road} (inset). As seen from Figure~\ref{fig:nrmse-road}, SRW-rw (resp. MHRW) requires \emph{more than twice larger samples} than the case of NBRW-rw (resp.
MHRW-DA) to attain the same level
of accuracy for the estimation of $\pr\{D_\G \!=\! d\}$.
Specifically, the
NBRW-rw and MHRW-DA leads to about 60\% and 54\% cost saving on average. Also, as before, Figure~\ref{fig:perdegree-pdf-road} shows the NRMSE ratio of SRW-rw (resp. MHRW) to the case of NBRW-rw (resp. MHRW-DA) for every degree $d$ when estimating
$\pr\{D_\G \!=\! d\}$ with $5\cdot10^5$ samples, and Figure~\ref{fig:perdegree-ccdf-road} depicts the NRMSE curve (with its ratio) for the estimation of $\pr\{D_\G \!>\! d\}$ with
$5\cdot10^5$ samples, which are all in good agreement with our theoretical findings. In addition, Figure~\ref{fig:nrmse-nonstationary-road} demonstrates that such considerable performance improvement of NBRW-rw (resp. MHRW-DA) over SRW-rw (resp. MHRW) still prevails for the case of non-stationary start. We observe that the NBRW-rw and MHRW-DA are remarkably effective for Road-PA graph, as the graph structure with small node degrees makes the less-backtracking feature more favorable.

\begin{figure}[t!]
    \centering
    \vspace{-0mm}
    \hspace{-0mm}\subfigure[SRW-rw vs. NBRW-rw]{\includegraphics[width=3in,height=2.3in]{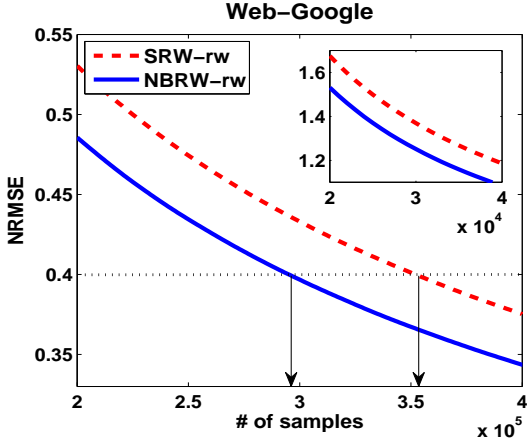}}
    \hspace{-0mm}\subfigure[MHRW vs. MHRW-DA]{\includegraphics[width=3in,height=2.3in]{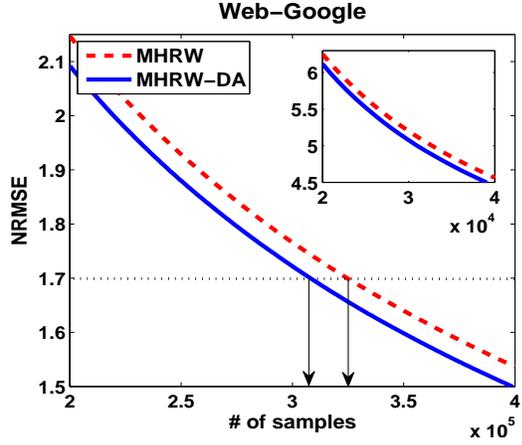}}
    \hspace{-0mm}
    \vspace{-0mm}
    \caption{Web-Google graph. NRMSE (averaged over all possible $d$) of the estimator of $\pr\{D_\G \!=\! d\}$ with different number of samples; the insets are for smaller number of samples.} \label{fig:nrmse-web}
    \vspace{-0mm}
\end{figure}

\begin{figure}[t!]
    \centering
    \vspace{-0mm}
    \hspace{-0mm}\subfigure[SRW-rw vs. NBRW-rw]{\includegraphics[width=3in,height=2.3in]{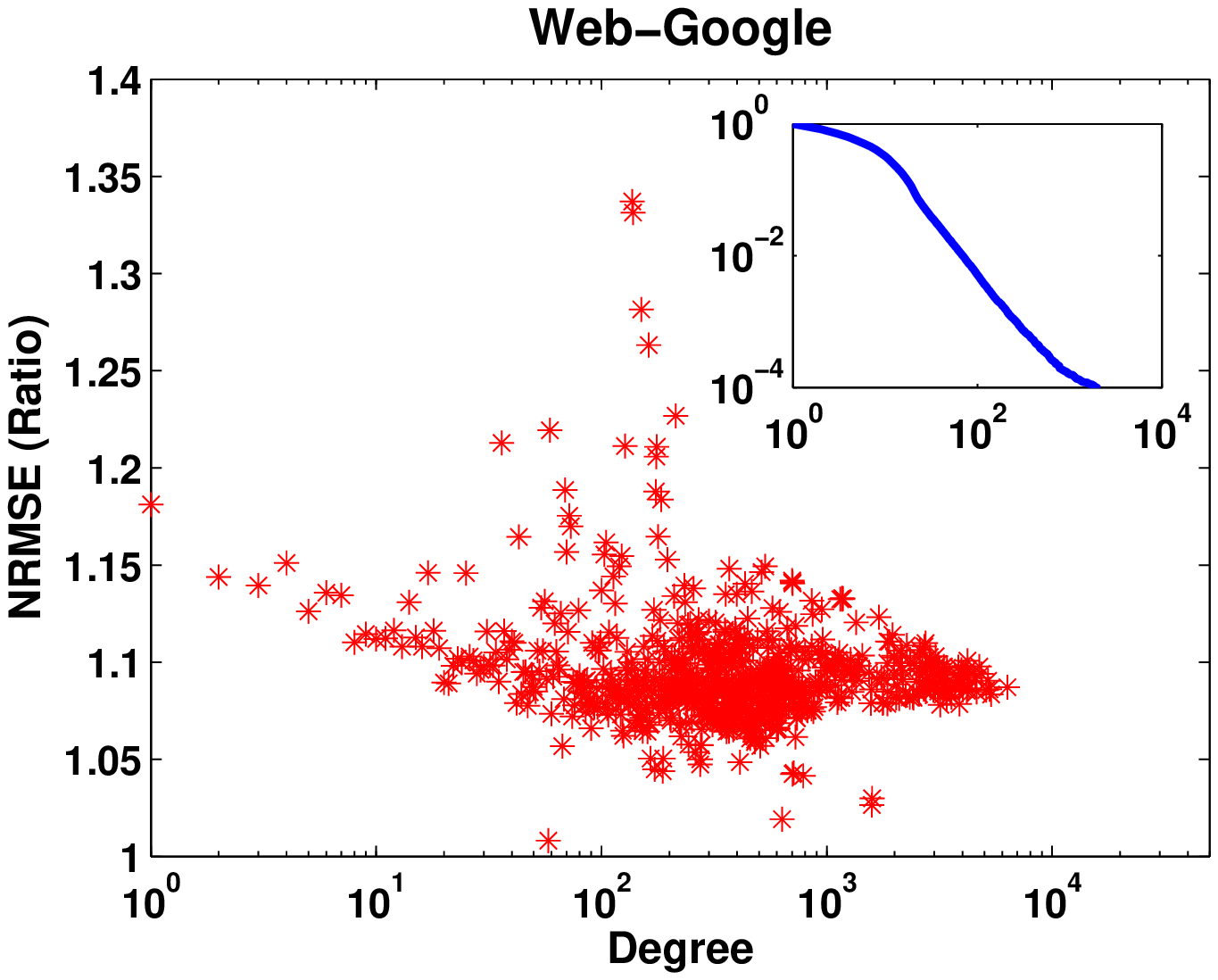}}
    \hspace{-0mm}\subfigure[MHRW vs. MHRW-DA]{\includegraphics[width=3in,height=2.3in]{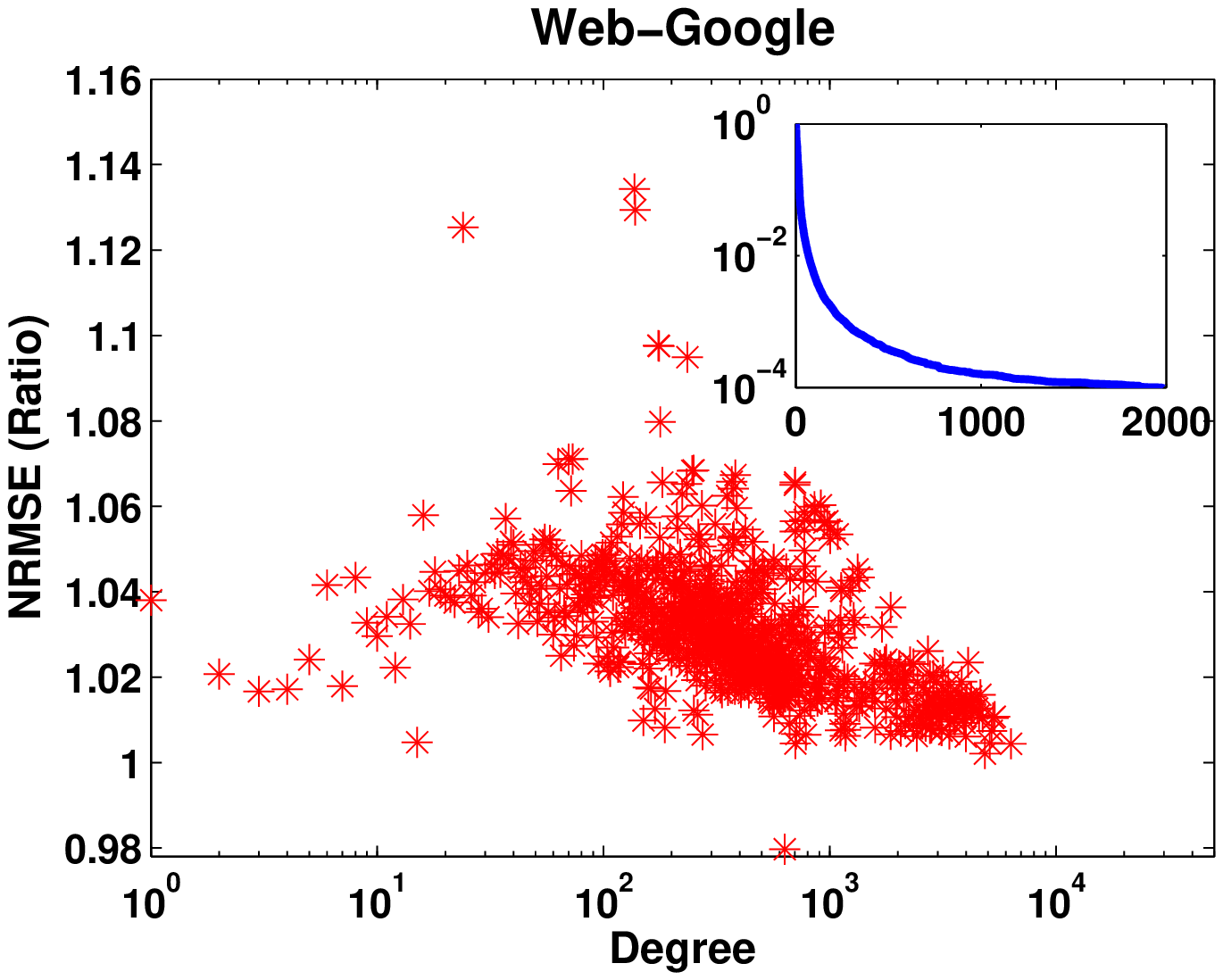}}
    \hspace{-0mm}
    \vspace{-0mm}
    \caption{Web-Google graph. NRMSE ratio (per degree $d$) when estimating $\pr\{D_\G \!=\! d\}$ with $5\cdot10^5$ samples; the insets represent the `actual' degree distribution (ccdf) in (a) log-log scale, (b) semi-log scale.} \label{fig:perdegree-pdf-web}
    \vspace{-0mm}
\end{figure}

\begin{figure}[t!]
    \centering
    \vspace{-0mm}
    \hspace{-0mm}\subfigure[SRW-rw vs. NBRW-rw]{\includegraphics[width=3in,height=2.3in]{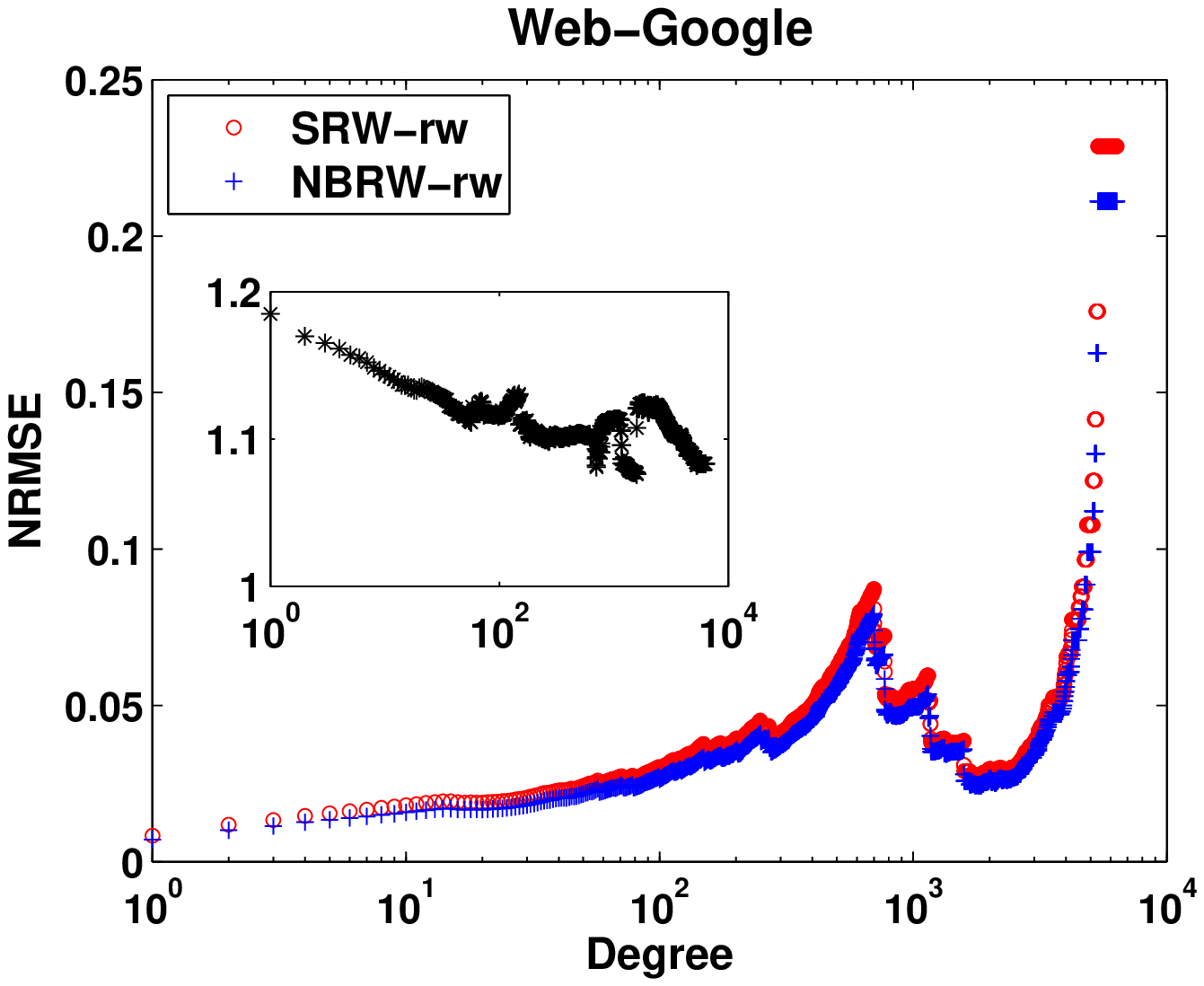}}
    \hspace{-0mm}\subfigure[MHRW vs. MHRW-DA]{\includegraphics[width=3in,height=2.3in]{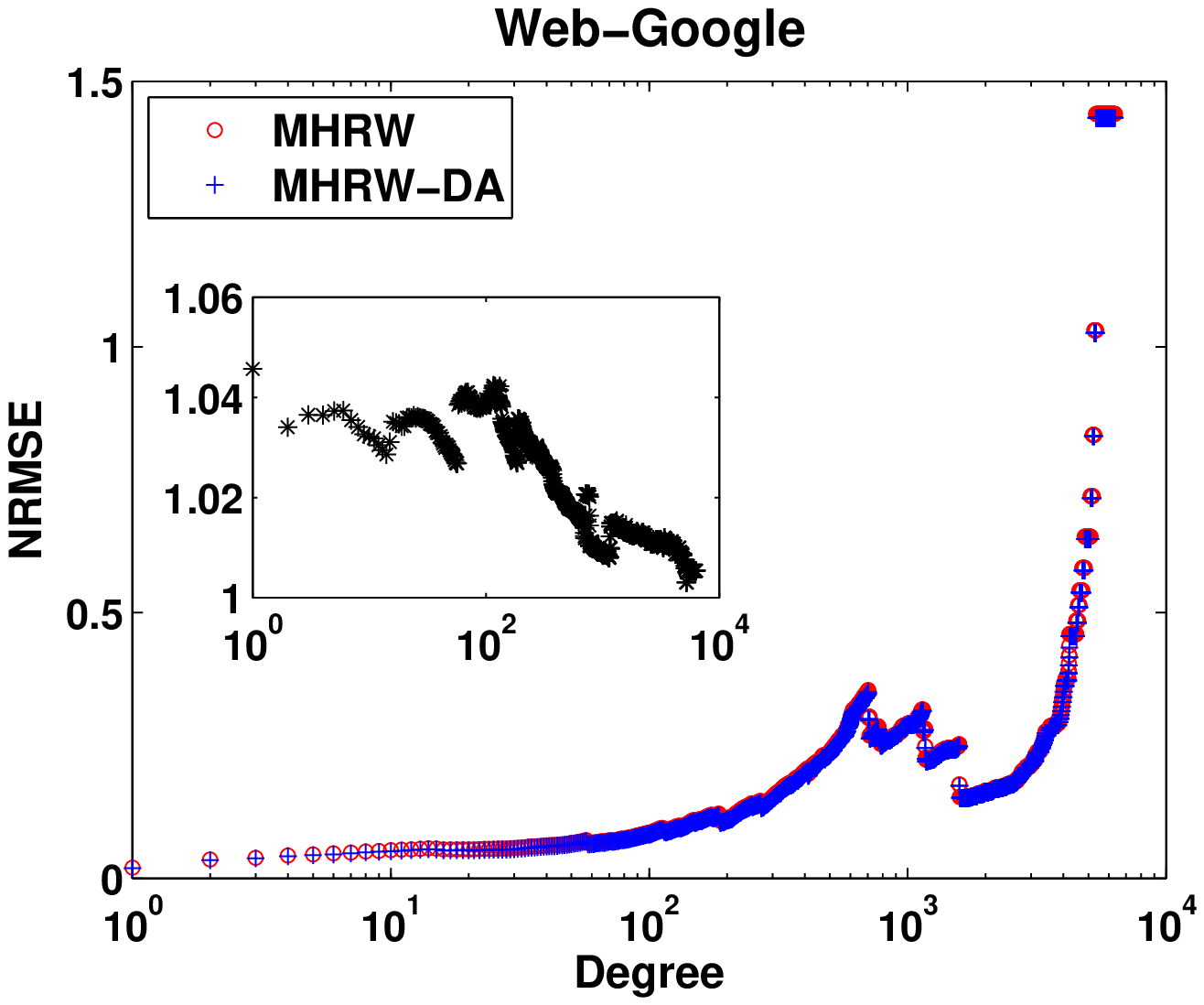}}
    \hspace{-0mm}
    \vspace{-0mm}
    \caption{Web-Google graph. NRMSE (per degree $d$) when estimating $\pr\{D_\G \!>\! d\}$ with $5\cdot10^5$ samples; the insets show NRMSE ratio.} \label{fig:perdegree-ccdf-web}
    \vspace{-0mm}
\end{figure}

\begin{figure}[t!]
    \centering
    \vspace{-0mm}
    \hspace{-0mm}\subfigure[SRW-rw vs. NBRW-rw]{\includegraphics[width=3in,height=2.3in]{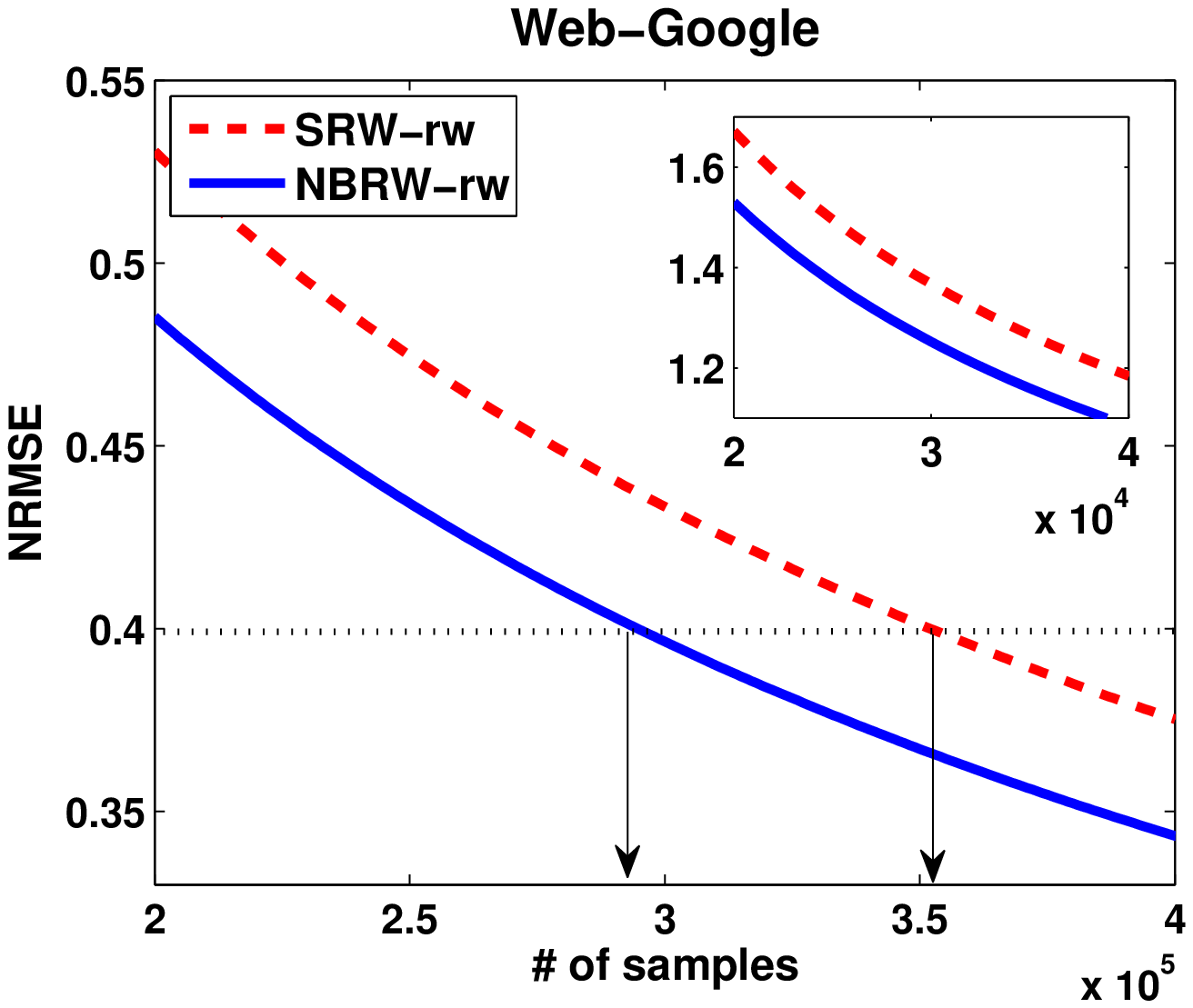}}
    \hspace{-0mm}\subfigure[MHRW vs. MHRW-DA]{\includegraphics[width=3in,height=2.3in]{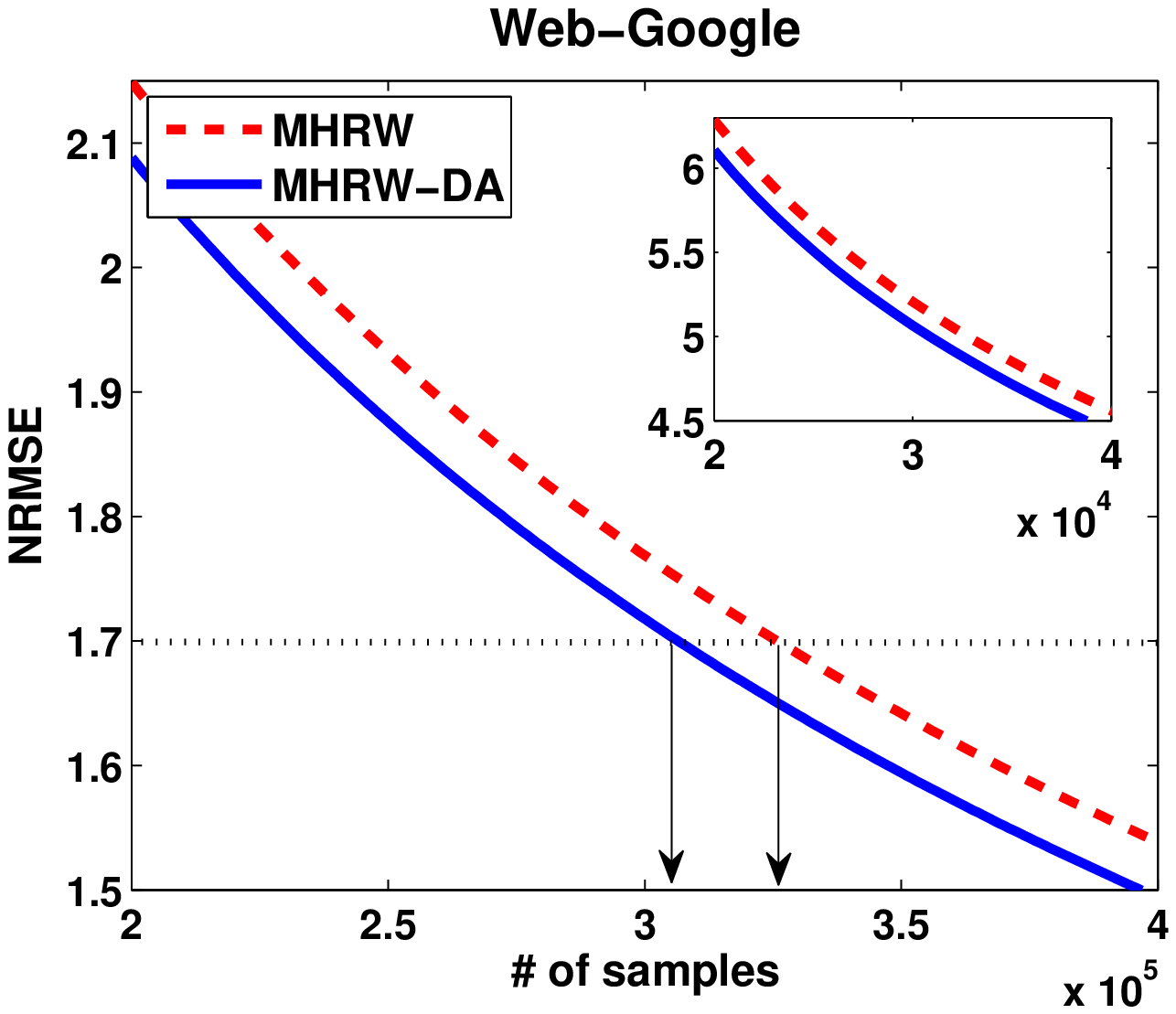}}
    \hspace{-0mm}
    \vspace{-0mm}
    \caption{Web-Google graph. NRMSE (averaged over all possible $d$) of the estimator of $\pr\{D_\G \!=\! d\}$, when each random walk does not start in the stationary regime.} \label{fig:nrmse-nonstationary-web}
    \vspace{-0mm}
\end{figure}

We finally provide the simulation results for Web-Google graph whose actual degree distribution is more like a power-law as shown in Figure~\ref{fig:perdegree-pdf-web} (insets). Figure~\ref{fig:nrmse-web} demonstrates that NBRW-rw (resp. MHRW-DA) surpasses SRW-rw (resp. MHRW) overall for the estimation of $\pr\{D_\G \!=\! d\}$, although their improvements are not as large as before. Again, Figure~\ref{fig:perdegree-pdf-web} shows the NRMSE ratio of SRW-rw (resp. MHRW) to the case of NBRW-rw (resp. MHRW-DA) for every degree $d$ in estimating $\pr\{D_\G \!=\! d\}$ with $5\cdot10^5$ samples, and Figure~\ref{fig:perdegree-ccdf-web} depicts the NRMSE curve (with its ratio) for the estimation of $\pr\{D_\G \!>\! d\}$ with
$5\cdot10^5$ samples. Clearly, NBRW-rw performs better than SRW-rw for each degree $d$ (all data points are above one), as expected from our theoretical results. We also observe similar results when comparing MHRW-DA and MHRW. There is, however, just one data point below one (in the ratio) for the estimation of $\pr\{D_\G \!=\! d\}$. We admit that such an `outlier' may be possible, since our theoretical results hold in the asymptotic sense. Nonetheless, MHRW-DA leads to an overall performance improvement (over all possible $d$). In addition, we observe that NBRW-rw (resp. MHRW-DA) remains effective in achieving higher sampling accuracy than SRW-rw (resp. MHRW), even when each random walk does not start in the stationary regime, as shown in Figure~\ref{fig:nrmse-nonstationary-web}.

It is also worth noting that a direct comparison between SRW-rw
(or NBRW-rw) and MH algorithm (or MHDA) may not
be appropriate. Recently, \cite{GjokaJSAC11-2} numerically shows a
counter-example that MHRW can be more efficient, although
SRW-rw has been shown to be better than the MHRW over
several numerical
simulations~\cite{WillingerInfocom09,GjokaJSAC11-2}. In addition,
\cite{Bassetti06} proved, through several examples, that there is
no clear winner between the importance sampling for reversible
Markov chains (whose special case is the SRW-rw)
and the MH algorithm. The MH algorithm also is valuable because it can be used to construct a reversible chain with any given stationary
distribution. We thus have focused on improving each of the SRW-rw
and the MH algorithm separately.

\section{Concluding Remarks} We demonstrated, in theory and
simulation, that our proposed NBRW-rw and MHDA
guarantee unbiased graph sampling, while also achieving higher
sampling efficiency than SRW-rw and MH
algorithm, respectively. While the focus of this paper was on the
unbiased graph sampling, we cannot stress enough the versatile
applicability of the MHDA for \emph{any} non-uniform node sampling
(e.g., intentionally creating a known bias toward preferable
nodes), not to mention its improvement over the famous MH
algorithm in sampling efficiency. We expect that the MHDA can
be applied to many other problems beyond the unbiased graph
sampling.

\bibliographystyle{abbrv}
\bibliography{chlee-ref}

\clearpage
\appendix

\allowdisplaybreaks

\begin{center}
\textbf{\LARGE Appendix}
\end{center}

\section{Proof of Proposition 1}\label{append2}

Observe that the condition in (\ref{condition1}) can be written
as, for all $e_{ij},e_{ji},e_{jk},e_{kj} \in \Omega$ with $i \ne
k$,
\begin{align}
&\tilde{\pi}(j) \tilde{P}(j,i) \tilde{P}'(e_{ij},e_{jk}) = \tilde{\pi}(j) \tilde{P}(j,k) \tilde{P}'(e_{kj},e_{ji})\nonumber\\
\Longrightarrow~ &\tilde{\pi}(i) \tilde{P}(i,j)
\tilde{P}'(e_{ij},e_{jk}) = \tilde{\pi}(k) \tilde{P}(k,j)
\tilde{P}'(e_{kj},e_{ji}) \nonumber\\
\Longrightarrow~ &~~~~~\tilde{\pi}'(e_{ij})
\tilde{P}'(e_{ij},e_{jk}) = \tilde{\pi}'(e_{kj})
\tilde{P}'(e_{kj},e_{ji}),\label{flow}
\end{align}
which is from the reversibility of the embedded chain
$\{\tilde{X}_m\}$ and $\tilde{\pi}'(e_{ij}) =
\tilde{\pi}(i)\tilde{P}(i,j)$, $e_{ij} \in \Omega$. Also,
(\ref{flow}) trivially holds for $i = k$.
Applying the form of $\tilde{P}'(e_{ij},e_{jk})$ in
(\ref{new-embedded}) to the condition in (\ref{flow}) yields
\begin{align}
&\!\!\tilde{\pi}'(e_{ij}) \tilde{P}(j,k) +
\tilde{\pi}'(e_{ij})\tilde{P}(j,i)Q'(e_{ij},e_{jk})A'(e_{ij},e_{jk})\nonumber\\
&\!\!\!\!= \tilde{\pi}'(e_{kj}) \tilde{P}(j,i) +
\tilde{\pi}'(e_{kj})\tilde{P}(j,k)Q'(e_{kj},e_{ji})A'(e_{kj},e_{ji}).\!\!\label{condition}
\end{align}
Again from the reversibility of the chain $\{\tilde{X}_m\}$ and
$\tilde{\pi}'(e_{ij}) = \tilde{\pi}(i)\tilde{P}(i,j)$, it is not
difficult to see that $\tilde{\pi}'(e_{ij})\tilde{P}(j,k) \!=\!
\tilde{\pi}'(e_{kj})\tilde{P}(j,i)$,
$\tilde{\pi}'(e_{ij})\tilde{P}(j,i) \!=\!
\tilde{\pi}(j)\tilde{P}^2(j,i)$, and
$\tilde{\pi}'(e_{kj})\tilde{P}(j,k) \!=\!
\tilde{\pi}(j)\tilde{P}^2(j,k)$.
Then, observe that (\ref{condition}) holds if and only if
\begin{align}
A'(e_{ij},e_{jk}) & =
\frac{\tilde{P}^2(j,k)Q'(e_{kj},e_{ji})}{\tilde{P}^2(j,i)Q'(e_{ij},e_{jk})}
A'(e_{kj},e_{ji})\nonumber\\
& =
\frac{P^2(j,k)Q'(e_{kj},e_{ji})}{P^2(j,i)Q'(e_{ij},e_{jk})}A'(e_{kj},e_{ji}) \nonumber \\
& = T(e_{kj},e_{ji})A'(e_{kj},e_{ji}), \label{new-condition}
\end{align}
where the second equality is from $\tilde{P}(j,k) \!=\!
P(j,k)/(1\!-\!P(j,j))$ ($j \ne k$) and
\begin{equation}
T(e_{kj},e_{ji}) ~\triangleq~
\frac{P^2(j,k)Q'(e_{kj},e_{ji})}{P^2(j,i)Q'(e_{ij},e_{jk})}.
\label{asdtt}
\end{equation}
Hence, from (\ref{new-embedded}) and (\ref{flow}), we see that,
for any given $\{Q'(e_{ij},e_{lk})\}$, \emph{any} acceptance
probability $A'(e_{ij},e_{jk})$ satisfying (\ref{new-condition})
will make the resulting transition matrix $\tilde{\p}'$ (in
relation to $\tilde{\p}$) also satisfy the two conditions in
(\ref{condition1})--(\ref{condition2}).

By nothing that (\ref{asdtt}) asserts $T(e_{kj},e_{ji}) =
1/T(e_{ij},e_{jk})$, from (\ref{new-condition}), we know that the
acceptance probability $A'(e_{ij},e_{jk})$ is generally in the
form of $F(T(e_{ij},e_{jk}))$, where $0 \leq F \leq 1$ is any
arbitrary function satisfying $F(x) = F(1/x)/x$ for all $x$. Among
infinitely many possible choices, we choose $F(x) = \min\{1, x\}$,
yielding
\begin{equation}
A'(e_{ij},e_{jk}) = \min\lt\{1,T(e_{kj},e_{ji})\rt\},
\end{equation}
which gives rise to (\ref{accept}). This completes the proof.
\hfill $\Box$

\section{Proof of Theorem 6}\label{append3}

\noindent \textbf{(I) Proof for almost sure convergence:} Define a
function $\gamma': \Omega \!\to\! \Rn$ such that $\gamma'(e_{ij})
\!=\! \gamma(j)$. See (\ref{gamma}) for $\gamma(\cdot)$. We also
define another sequence $\{\xi''_m\}_{m \geq 1}$ which depends on
$\tilde{Z}'_m \!=\! (\tilde{X}'_{m-1},\tilde{X}'_m) \!\in\!
\Omega$ and is geometrically distributed with parameter
$\gamma'(\tilde{Z}'_m) \!=\! \gamma(\tilde{X}'_m)$ such that
$\xi''_m \! \equiv\! \xi'_m$. Now, consider a sequence of the
pairs $(\tilde{Z}'_m,\xi''_m)$. For any $f: \N \!\to\! \Rn$, choose another function $g: \Omega
\!\to\! \Rn$ such that $g(e_{ij}) = f(j)$. Then, by noting that
\begin{equation}
\hat{\mu}'_{m, \textsf{MHDA}}(f) = \frac{\sum^m_{l=1} \xi'_l
f(\tilde{X}'_l)}{\sum^m_{l=1} \xi'_l} = \frac{\sum^m_{l=1} \xi''_l
g(\tilde{Z}'_l)}{\sum^m_{l=1} \xi''_l},\label{ttt}
\end{equation}
it suffices to show that, as $m \to \infty$,
\begin{equation*}
\frac{\sum^m_{l=1} \xi''_l g(\tilde{Z}'_l)}{\sum^m_{l=1} \xi''_l}
~\to~ \Ex_{\bm{\pi}}(f)~~\text{a.s.}
\end{equation*}

First, we define
\begin{equation*}
S_m(e_{ij}) \triangleq \sum^m_{l=1}\idc_{\{\tilde{Z}'_l =
e_{ij}\}}, ~~ e_{ij} \in \Omega,
\end{equation*}
to indicate the number of visits to state $e_{ij}$ during the
first $m$ transitions of the (non-reversible) Markov chain
$\{\tilde{Z}'_l\}$ over $\Omega$. Also, let $J_k(e_{ij})$, $k \geq 1$, be the geometrically
distributed time duration associated with the $k^{th}$ visit of
the chain $\{\tilde{Z}'_l\}$ to state $e_{ij} \in \Omega$. Observe
that
\begin{equation}
\frac{\sum^m_{l=1} \xi''_l g(\tilde{Z}'_l)}{\sum^m_{l=1}
\xi''_l} = \frac{\sum_{e_{ij} \in \Omega}\sum^{S_m(e_{ij})}_{l =
1}J_l(e_{ij})g(e_{ij})}{\sum_{e_{ij} \in
\Omega}\sum^{S_m(e_{ij})}_{l = 1}J_l(e_{ij})} = \frac{\sum_{e_{ij}}\frac{S_m(e_{ij})}{m}\sum^{S_m(e_{ij})}_{l =
1}\frac{J_l(e_{ij})g(e_{ij})}{S_m(e_{ij})}}{\sum_{e_{ij}}\frac{S_m(e_{ij})}{m}
 \sum^{S_m(e_{ij})}_{l = 1}\frac{J_l(e_{ij})}{S_m(e_{ij})}}.
 \label{aaa}
\end{equation}
By the SLLN for $i.i.d.$ random variables, for each $e_{ij} \in
\Omega$, as $m \to \infty$ and so $S_m(e_{ij}) \to \infty$,
\begin{align*}
\frac{1}{S_m(e_{ij})}\sum^{S_m(e_{ij})}_{l =
1}J_l(e_{ij})g(e_{ij}) ~&\to~
g(e_{ij})/\gamma'(e_{ij}) ~~\text{a.s.},\\
\frac{1}{S_m(e_{ij})}\sum^{S_m(e_{ij})}_{l = 1}J_l(e_{ij}) ~&\to~
1/\gamma'(e_{ij}) ~~\text{a.s.}
\end{align*}
Since the return times of the chain $\{\tilde{Z}'_m\}$ to state
$e_{ij}$ (the time intervals between two consecutive visits to
$e_{ij}$) are $i.i.d.$ from the strong Markov property, by
applying the strong law for renewal processes~\cite{Ross96a}, we
also have, as $m \to \infty$,
\begin{equation*}
\frac{S_m(e_{ij})}{m} ~\to~ \tilde{\pi}'(e_{ij}) =
\tilde{\pi}(i)\tilde{P}(i,j) ~~\text{a.s.},~~e_{ij} \in \Omega,
\end{equation*}
where $\bm{\tilde{\pi}'}$ is the unique stationary distribution of
the chain $\{\tilde{Z}'_m\}$. (See Proposition~\ref{proposition1}
and (\ref{stationary2}).) Hence, from (\ref{aaa}), we have, as $m
\to \infty$,
\begin{equation}
\frac{\sum^m_{l=1} \xi''_l g(\tilde{Z}'_l)}{\sum^m_{l=1} \xi''_l}
~\to~ \frac{\sum_{e_{ij}}
\tilde{\pi}'(e_{ij})g(e_{ij})/\gamma'(e_{ij})}{\sum_{e_{ij}}
\tilde{\pi}'(e_{ij})/\gamma'(e_{ij})} ~~\text{a.s.} \label{aaa1}
\end{equation}
Here, the RHS of (\ref{aaa1}) becomes
\begin{align*}
\frac{\sum_{e_{ij}}
\tilde{\pi}'(e_{ij})g(e_{ij})/\gamma'(e_{ij})}{\sum_{e_{ij}}
\tilde{\pi}'(e_{ij})/\gamma'(e_{ij})} &= \frac{\sum_{j \in \N}
\sum_{i \ne j} \tilde{\pi}(i)\tilde{P}(i,j)
f(j)/\gamma(j)}{\sum_{j \in \N} \sum_{i \ne j}
\tilde{\pi}(i)\tilde{P}(i,j)/\gamma(j)}\\
&= \frac{\sum_{j \in \N} f(j) \tilde{\pi}(j)/\gamma(j)}{\sum_{j
\in \N} \tilde{\pi}(j)/\gamma(j)} = \Ex_{\bm{\pi}}(f).
\end{align*}
The first two equalities are from that $\tilde{P}(u,v) \!=\! 0$
for all $(u,v) \not\in \Omega$ (including $\tilde{P}(u,u) \!=\! 0$
for all $u$), $\tilde{\pi}'(e_{ij}) \!=\!
\tilde{\pi}(i)\tilde{P}(i,j)$ (and so $\sum_{j \ne
i}\tilde{\pi}(i)\tilde{P}(i,j) = \tilde{\pi}(j)$), $g(e_{ij})
\!=\! f(j)$, and $\gamma'(e_{ij}) \!=\! \gamma(j)$, $e_{ij}
\!\in\! \Omega$. The last equality follows from $\pi(j) \propto
\tilde{\pi}(j)/\gamma(j)$ for all $j$ in (\ref{semi-stationary}).
Therefore, for any function $f$, $\hat{\mu}'_{m,\textsf{MH}}(f)$
converges almost surely to $\Ex_{\bm{\pi}}(f)$, as $m$ goes to
infinity.

\vspace{3mm} \noindent \textbf{(II) Proof for asymptotic
variance:} We next prove that the asymptotic variance of
$\hat{\mu}'_{m, \textsf{MHDA}}(f)$ is no larger than that of
$\hat{\mu}_{m, \textsf{MH}}(f)$, i.e.,
${\sigma'}^2_{\textsf{MHDA}}(f) \!\leq\!
{\sigma}^2_{\textsf{MH}}(f)$. Consider a sequence of the pairs
$(\tilde{X}_m,\xi_m)$ by the MH algorithm. For a function $f: \N
\!\to\! \Rn$, we define by $\Gamma(f)$ the asymptotic variance of
the following estimator
\begin{equation}
\frac{\hat{\mu}_m(f/\gamma)}{\hat{\mu}_m(1/\gamma)} =
\frac{\sum^m_{l=1}f(\tilde{X}_l)/\gamma(\tilde{X}_l)}{\sum^m_{l=1}1/\gamma(\tilde{X}_l)}.
\label{oooo}
\end{equation}
It then follows from a special case of Theorem 1 in~\cite{Douc11} that, for any function $f: \N \!\to\! \Rn$, as $m\to\infty$,
\begin{align}
\sqrt{m}\cdot[\hat{\mu}_{m, \textsf{MH}}(f) - \Ex_{\bm{\pi}}(f)]
&\stackrel{d}{\Longrightarrow}
\mathrm{N}(0,{\sigma}^2_{\textsf{MH}}(f)) = \mathrm{N}(0,\Gamma(f) + \Delta(f)), \label{theorem-douc}
\end{align}
where
\begin{equation*}
\Delta(f) = \Ex_{\bm{\pi}}(\gamma)\Ex_{\bm{\pi}}\{\var\{\xi |
X\}[f(X) - \Ex_{\bm{\pi}}(f)]^2 \gamma(X)\},
\end{equation*}
and $\Ex_{\bm{\pi}}(\gamma) \!=\! \sum_{i \in \N}\gamma(i)\pi(i)$. Here, the expectation is with respect to $X \sim \bm{\pi}$, and
$\xi$ is geometrically distributed with parameter $\gamma(X)$.
We notice that this result is obtained from the sequence of the
pairs $(\tilde{X}_m,\xi_m)$ (not from the fact that $\{X_t\}$ is
an irreducible Markov chain) with the stationary distribution
$\bm{\pi}$ given by $\pi(i) \propto \tilde{\pi}(i)/\gamma(i)$, $i
\!\in\! \N$. See Theorem 1 in~\cite{Douc11} (in addition to Lemma
1 therein) for more details. Thus, we can similarly apply the result in
(\ref{theorem-douc}) for the sequence of the pairs
$(\tilde{Z}'_m,\xi''_m)$ defined earlier in \textbf{(I)}. Our
proof strategy is to show that the first term $\Gamma(f)$ for
${\sigma}^2_{\textsf{MH}}(f)$ is no smaller than its corresponding
term for ${\sigma'}^2_{\textsf{MHDA}}(f)$, while $\Delta(f)$
remains the same for both ${\sigma}^2_{\textsf{MH}}(f)$ and
${\sigma'}^2_{\textsf{MHDA}}(f)$. We start by showing that the
latter is true.

For a given $f$, again consider another function $g: \Omega
\!\to\! \Rn$ such that $g(e_{ij}) \!=\! f(j)$, and recall that
$\gamma'(e_{ij}) = \gamma(j)$, $e_{ij} \!\in\! \Omega$. We define
by $\Gamma'(f)$ the asymptotic variance of the following estimator
\begin{equation*}
\frac{\hat{\mu}'_m(f/\gamma)}{\hat{\mu}'_m(1/\gamma)} =
\frac{\sum^m_{l=1}f(\tilde{X}'_l)/\gamma(\tilde{X}'_l)}{\sum^m_{l=1}1/\gamma(\tilde{X}'_l)}
=
\frac{\sum^m_{l=1}g(\tilde{Z}'_l)/\gamma'(\tilde{Z}'_l)}{\sum^m_{l=1}1/\gamma'(\tilde{Z}'_l)},
\end{equation*}
where the RHS of the second equality (the ratio estimator defined
based on $\{\tilde{Z}'_l\}$) corresponds to (\ref{oooo}). In
addition, if one finds a semi-Markov chain associated with the
sequence $(\tilde{Z}'_m,\xi''_m)$ as was done for the MH
algorithm, then its stationary distribution, denoted as
$\bm{\lambda}$, should be (e.g.,~\cite{Ross96a})
\begin{equation*}
\lambda(e_{ij}) \propto \tilde{\pi}'(e_{ij})/\gamma'(e_{ij}),
~~e_{ij} \in \Omega.
\end{equation*}
Together with this, the facts that $\tilde{P}(u,v) \!=\! 0$ for
all $(u,v) \not\in \Omega$ (including $\tilde{P}(u,u) \!=\! 0$ for
all $u$), $\gamma'(e_{ij}) \!=\! \gamma(j)$, $\tilde{\pi}'(e_{ij})
\!=\! \tilde{\pi}(i)\tilde{P}(i,j)$, and $\pi(i) \propto
\tilde{\pi}(i)/\gamma(i)$ give
\begin{align}
\Ex_{\bm{\lambda}}(\gamma') & = \sum_{e_{ij} \in
\Omega}\gamma'(e_{ij}) \lambda(e_{ij}) = \frac{\sum_{j \in
\N}\sum_{i \ne j}\tilde{\pi}(i)\tilde{P}(i,j)}{\sum_{j \in
\N}\sum_{i \ne j}\tilde{\pi}(i)\tilde{P}(i,j)/\gamma(j)} \nonumber\\
& = \frac{1}{\sum_{j \in
\N}\tilde{\pi}(j)/\gamma(j)} = \sum_{i \in \N}\gamma(i)
\frac{\tilde{\pi}(i)/\gamma(i)}{\sum_{j \in
\N}\tilde{\pi}(j)/\gamma(j)} = \Ex_{\bm{\pi}}(\gamma),
\label{bbb}
\end{align}
where the denominators in the RHS of the second and fourth
equalities are the normalizing constants of $\lambda(e_{ij})$ and
$\pi(i)$, respectively. Following the same lines, we similarly
have $\Ex_{\bm{\lambda}}(g) \!=\! \Ex_{\bm{\pi}}(f)$ for any $g$
such that $g(e_{ij}) \!=\! f(j)$. Then, if we define a random
variable $Y \!\sim\! \bm{\lambda}$ (defined on $\Omega$) and a
geometric random variable $\xi''$ with parameter $\gamma'(Y)$,
following the similar arguments above, we obtain
\begin{align}
&\Ex_{\bm{\lambda}}\{\var\{\xi'' | Y\}[g(Y) -
\Ex_{\bm{\lambda}}(g)]^2
\gamma'(Y)\}\nonumber\\
&~~~= \sum_{e_{ij} \in \Omega} \var\{\xi'' | Y = e_{ij}\}[g(e_{ij}) -
\Ex_{\bm{\lambda}}(g)]^2\gamma'(e_{ij})
\lambda(e_{ij}) \nonumber\\
&~~~= \frac{\sum_{j \in \N}\sum_{i \ne j}\var\{\xi | X = j\} [f(j) -
\Ex_{\bm{\pi}}(f)]^2\tilde{\pi}(i)\tilde{P}(i,j)}{\sum_{j \in
\N}\sum_{i \ne j}\tilde{\pi}(i)\tilde{P}(i,j)/\gamma(j)}
\nonumber\\
&~~~= \sum_{j \in \N}\var\{\xi | X = j\} [f(j) -
\Ex_{\bm{\pi}}(f)]^2\gamma(j)\cdot\frac{\tilde{\pi}(j)/\gamma(j)}{\sum_{j
\in
\N}\tilde{\pi}(j)/\gamma(j)} \nonumber\\
&~~~= \Ex_{\bm{\pi}}\{\var\{\xi | X\}[f(X) - \Ex_{\bm{\pi}}(f)]^2
\gamma(X)\},\label{bbb1}
\end{align}
where $\var\{\xi'' | Y = e_{ij}\} = \var\{\xi | X =
j\}$ follows from $\gamma'(e_{ij}) = \gamma(j)$, $e_{ij} \in
\Omega$. Hence, by applying the result in (\ref{theorem-douc}) for the
sequence $(\tilde{Z}'_m,\xi''_m)$ with (\ref{ttt}) and
(\ref{bbb})--(\ref{bbb1}), we have
\begin{align*}
\!\!\sqrt{m}\lt[\frac{\sum^m_{l=1} \xi''_l
g(\tilde{Z}'_l)}{\sum^m_{l=1} \xi''_l} \!-\!
\Ex_{\bm{\lambda}}(g)\rt] &~=
\sqrt{m}\cdot[\hat{\mu}'_{m,\textsf{MHDA}}(f) \!-\! \Ex_{\bm{\pi}}(f)] \nonumber\\
&\stackrel{d}{\Longrightarrow}~
\mathrm{N}(0,{\sigma'}^2_{\textsf{MHDA}}(f)),
\end{align*}
with ${\sigma'}^2_{\textsf{MHDA}}(f) \!=\! \Gamma'(f) \!+\!
\Delta(f)$. Thus, if $\Gamma'(f) \!\leq\! \Gamma(f)$, then
${\sigma'}^2_{\textsf{MHDA}}(f) \!\leq\!
{\sigma}^2_{\textsf{MH}}(f)$. We below show that this is indeed
true.

Observe that
\begin{align*}
\sqrt{m}\lt[\frac{\hat{\mu}_m(f/\gamma)}{\hat{\mu}_m(1/\gamma)} -
\Ex_{\bm{\pi}}(f)\rt]
& = \sqrt{m}\lt[\frac{\sum^m_{l=1}\frac{f(\tilde{X}_l)}{\gamma(\tilde{X}_l)}}{\sum^m_{l=1}\frac{1}{\gamma(\tilde{X}_l)}} - \Ex_{\bm{\pi}}(f)\rt] \nonumber\\
& =
\frac{m}{\sum^m_{l=1}1/\gamma(\tilde{X}_l)}\sqrt{m}\lt[\frac{\sum^m_{l=1}[f(\tilde{X}_l)
- \Ex_{\bm{\pi}}(f)]/\gamma(\tilde{X}_l)}{m}\rt].
\end{align*}
Define another function $h: \N \to \Rn$ such that
\begin{equation*}
h(i) \triangleq  [f(i) - \Ex_{\bm{\pi}}(f)]/\gamma(i),~~~ i \in
\N,
\end{equation*}
implying $\Ex_{\bm{\tilde{\pi}}}(h) \!=\! \sum_{i \in
\N}h(i)\tilde{\pi}(i) \!=\! 0$, which can be seen from that
$\pi(i) \propto \tilde{\pi}(i)/\gamma(i)$, $i \!\in\! \N$. Then,
Theorems~\ref{ergodic} and \ref{clt} say that, as $m \to \infty$,
\begin{align*}
\frac{1}{m}\sum^m_{l=1}1/\gamma(\tilde{X}_l) ~&\to~ \Ex_{\bm{\tilde{\pi}}}(1/\gamma)~~\text{a.s.}, \\
\sqrt{m}\lt[\frac{1}{m}\sum^m_{l=1}h(\tilde{X}_l)\rt]
~&\stackrel{d}{\Longrightarrow}~ \mathrm{N}(0,\sigma^2(h)),
\end{align*}
and thus, by Slutsky's theorem (and almost sure convergence
implies convergence in probability), we have, as $m \to \infty$,
\begin{equation*}
\sqrt{m}\lt[\frac{\hat{\mu}_m(f/\gamma)}{\hat{\mu}_m(1/\gamma)} -
\Ex_{\bm{\pi}}(f)\rt] ~\stackrel{d}{\Longrightarrow}~
\frac{1}{\Ex_{\bm{\tilde{\pi}}}(1/\gamma)}
\mathrm{N}(0,\sigma^2(h)).
\end{equation*}
Since the process $\{\tilde{X}'_m\}$ has the same stationary
distribution $\bm{\tilde{\pi}}$, together with
(\ref{ergodic-nonrev}) and
(\ref{clt-nonrev})\footnote{Specifically, we mean
(\ref{ergodic-nonrev}) and (\ref{clt-nonrev}) where $\{Z'_t\}$,
$\{X'_t\}$, $\bm{\pi'}$, and $\bm{\pi}$ are replaced by
$\{\tilde{Z}'_m\}$, $\{\tilde{X}'_m\}$, $\bm{\tilde{\pi}'}$, and
$\bm{\tilde{\pi}}$, respectively.}, following the same lines as
above, we similarly have, as $m \to \infty$,
\begin{equation*}
\sqrt{m}\lt[\frac{\hat{\mu}'_m(f/\gamma)}{\hat{\mu}'_m(1/\gamma)}
- \Ex_{\bm{\pi}}(f)\rt] ~\stackrel{d}{\Longrightarrow}~
\frac{1}{\Ex_{\bm{\tilde{\pi}}}(1/\gamma)}
\mathrm{N}(0,{\sigma'}^2(h)).
\end{equation*}
Hence, from Theorem~\ref{nonbacktrack} and
Proposition~\ref{proposition1}, for any function $f$, the
asymptotic variance of $\hat{\mu}'_m(f)$ (based on
$\{\tilde{X}'_m\}$) is no larger than that of $\hat{\mu}_m(f)$
(obtained from $\{\tilde{X}_m\}$), i.e., ${\sigma'}^2(f) \leq
\sigma^2(f)$, so is ${\sigma'}^2(h) \leq \sigma^2(h)$. Finally,
\begin{equation*}
\Gamma'(f) = {\sigma'}^2(h)/(\Ex_{\bm{\tilde{\pi}}}(1/\gamma))^2
\leq  {\sigma}^2(h)/(\Ex_{\bm{\tilde{\pi}}}(1/\gamma))^2 =
\Gamma(f),
\end{equation*}
implying that ${\sigma'}^2_{\textsf{MHDA}}(f) \!\leq\!
{\sigma}^2_{\textsf{MH}}(f)$, and we are done.
\hfill $\Box$

\end{document}